\documentclass[11pt, prd]{revtex4}

\usepackage{amsmath}
\usepackage{amssymb}
\usepackage[dvips]{graphicx,color}

\begin{document}

\title{
Non-BPS exact solutions and 
their relation to bions in 
${\mathbb C}P^{N-1}$ models
}

\author{Tatsuhiro Misumi}
\email{misumi(at)phys.akita-u.ac.jp}
\affiliation{Department of Mathematical Science, Akita 
University, 1-1 Tegata Gakuen-machi,
Akita 010-8502, Japan\\
Research and Education Center for Natural Sciences, 
Keio University, 4-1-1 Hiyoshi, Yokohama, Kanagawa 223-8521, Japan
}

\author{Muneto Nitta}
\email{nitta(at)phys-h.keio.ac.jp}
\affiliation{Department of Physics, and Research and 
Education Center for Natural Sciences, 
Keio University, 4-1-1 Hiyoshi, Yokohama, Kanagawa 223-8521, Japan}

\author{Norisuke Sakai}
\email{norisuke.sakai(at)gmail.com}
\affiliation{Department of Physics, and Research and 
Education Center for Natural Sciences, 
Keio University, 4-1-1 Hiyoshi, Yokohama, Kanagawa 223-8521, Japan}

\begin{abstract}
We investigate non-BPS exact solutions in ${\mathbb C}P^{N-1}$ 
sigma models on ${\mathbb R}^1 \times S^{1}$ with twisted boundary 
conditions, by using the Din-Zakrzewski projection method. 
We focus on the relation of the non-BPS solutions to
the ansatz of multi-instanton (bion) configurations and discuss their 
significance in the context of the resurgence theory.
We find that the transition between seemingly 
distinct configurations of multi-instantons occur as moduli changes 
in the non-BPS solutions, 
and the simplest non-BPS exact solution corresponds to
multi-bion configurations with fully-compressed 
double fractional instantons in the middle. 
It indicates that the non-BPS solutions make small but nonzero 
contribution to the resurgent trans-series as special cases of the multi-bion
configurations.
We observe a generic pattern of transitions between 
distinct multi-bion configurations (flipping partners), 
leading to the three essential properties of the non-BPS exact solution:
(i) opposite sign for terms corresponding to the left and right infinities,
(ii) symmetric location of fractional instantons,
and (iii) the transition between distinct bion configurations.
By studying the balance of forces, we show that 
the relative phases between the instanton constituents play 
decisive roles in stability and instability of the muli-instanton 
configurations.
We discuss local and global instabilities of the 
solutions such as negative modes and the flow to the other 
saddle points, by considering the deformations of the 
non-BPS exact solutions within our multi-instanton ansatz. 
We also briefly discuss some classes of the non-BPS exact solutions in 
Grassmann sigma models.
\end{abstract}

\maketitle

\newpage


\section{Introduction}
\label{sec:Intro}

Recent intensive study on the resurgence theory and the 
complexified path integral in quantum field theories and 
quantum mechanics has revealed the significance of  
topological soliton molecules, which are composed of 
(fractional) instantons and anti-instantons 
\cite{Unsal:2007vu, Unsal:2007jx, Shifman:2008ja, 
Poppitz:2009uq, Anber:2011de, Poppitz:2012sw,Argyres:2012vv, 
Argyres:2012ka, Dunne:2012ae, Dunne:2012zk, Dabrowski:2013kba, 
Dunne:2013ada, Cherman:2013yfa, Basar:2013eka, Dunne:2014bca, 
Cherman:2014ofa, Behtash:2015kna, Bolognesi:2013tya, 
Misumi:2014jua,Misumi:2015dua, Misumi:2014rsa, Misumi:2014raa, 
Misumi:2014bsa, Nitta:2015tua,Nitta:2014vpa,Shermer:2014wxa, 
Dunne:2015ywa, Behtash:2015kva, Behtash:2015zha, Behtash:2015loa, 
Dunne:2015eaa, Gahramanov:2015yxk, Dunne:2016nmc, Dunne:2016qix}. 
Imaginary ambiguities arising in amplitudes of topologically 
neutral configurations of composite solitons 
can cancel out those arising in non-Borel-summable perturbative 
series in certain quantum theories 
\cite{Argyres:2012vv, Argyres:2012ka, Dunne:2012ae, Dunne:2012zk, 
Dabrowski:2013kba, Dunne:2013ada, Cherman:2013yfa, Basar:2013eka, 
Dunne:2014bca, Cherman:2014ofa,Bolognesi:2013tya,Misumi:2014jua, 
Misumi:2015dua, Misumi:2014rsa, Misumi:2014raa, Misumi:2014bsa, 
'tHooft:1977am, Fateev:1994ai, Fateev:1994dp}.
In particular, in field theories on compactified spacetime, 
these objects are termed as ``bions" \cite{Argyres:2012vv, 
Argyres:2012ka, Dunne:2012ae, Dunne:2012zk}.
It is expected that a full semi-classical expansion in 
perturbative and non-perturbative sectors as bions, which is 
called a ``resurgent" trans-series \cite{Ec1, Marino:2007te, 
Marino:2008ya, Marino:2008vx, Pasquetti:2009jg, Drukker:2010nc,
Aniceto:2011nu, Marino:2012zq, Hatsuda:2013gj, Schiappa:2013opa, 
Hatsuda:2013oxa, Aniceto:2013fka, Santamaria:2013rua, 
Kallen:2013qla, Honda:2014ica,Grassi:2014cla, Sauzin, Kallen:2014lsa,
Couso-Santamaria:2014iia, Honda:2014bza, Aniceto:2014hoa, 
Couso-Santamaria:2015wga, Honda:2015ewa, Hatsuda:2015owa,
Aniceto:2015rua, Dorigoni:2015dha, Honda:2016mvg}, leads to unambiguous and 
self-consistent definition of field theories in the same manner 
as the conjecture in quantum mechanics \cite{Bogomolny:1980ur, 
ZinnJustin:1981dx, ZinnJustin:1982td, ZinnJustin:1983nr, 
ZinnJustin:2004ib, ZinnJustin:2004cg, Jentschura:2010zza, Jentschura:2011zza}.
It is also discussed that the complexification of variables and parameters in path integrals 
play significant roles in the resurgence theory \cite{Behtash:2015zha, Behtash:2015loa}. 

In order to reach deeper understanding on bions and 
the related physics, it is of great importance to study 
examples in the field theory models in low-dimensions such as 
${\mathbb C}P^{N-1}$ models \cite{Dunne:2012ae, Dunne:2012zk, 
Misumi:2014jua, Misumi:2015dua, Misumi:2014rsa, Misumi:2014raa, 
Misumi:2014bsa}, 
principal chiral models \cite{Cherman:2013yfa, Basar:2013eka, 
Dunne:2014bca} and quantum mechanics \cite{Dunne:2014bca, 
Cherman:2014ofa, Behtash:2015kna, Misumi:2015dua}. 
In particular, the ${\mathbb C}P^{N-1}$ model in 1+1 dimensions has 
been studied for a long time as a toy model 
of the Yang-Mills theory in 3+1 dimensions, 
because of similarities between them such as 
dynamical mass gap, asymptotic freedom and the existence of 
instantons. The ${\mathbb C}P^{N-1}$ model  
on ${\mathbb R}^1 \times S^1$ with 
 ${\mathbb Z}_{N}$-twisted 
boundary conditions admits fractional instantons 
(domain-wall instantons) as Bogomol'nyi-Prasad-Sommerfield (BPS) 
solutions with the minimal topological charge ($Q=1/N$) 
\cite{Eto:2004rz,Eto:2006mz} (see also 
Refs.~\cite{Bruckmann:2007zh}). 
A part of the nonperturbative contributions in resurgence theory comes 
from bion configurations, consisting of a fractional instanton 
and anti-instanton. 
They are most conveniently obtained by means of an ansatz 
which reduces to solutions of field equations asymptotically 
at large separations of constituent fractional (anti-)instantons 
\cite{Misumi:2014jua,  Misumi:2014rsa, Misumi:2015dua, Misumi:2014bsa}. 
It has been found that the two-body forces change its character 
from attractive to repulsive as the relative phase of these 
fractional (anti-)instantons varies 
\cite{Misumi:2014jua}.

Non-BPS exact solutions in the ${\mathbb C}P^{N-1}$ model 
on ${\mathbb R}^1 \times S^1$ with a ${\mathbb Z}_{N}$-twisted 
boundary condition are found \cite{Dabrowski:2013kba}, 
by means of
the Din-Zakrzewski projection method \cite{Din:1980jg,Din:1981bx,Din:1983fj} 
generating a tower of non-BPS solutions from a BPS solution. 
The simplest non-BPS solution that they found corresponds to
the configuration composed of one doubly-compressed fractional 
instanton ($Q=2/N$) and two fractional anti-instantons ($Q=-1/N$) 
for $N\geq3$. 
It is not yet clear how the balance of forces is 
achieved in the non-BPS exact solutions, in contrast to the 
naive configuration composed of the double fractional instanton 
and two fractional anti-instantons, where the attractive force 
exists among the nearest constituents \cite{Misumi:2014jua}. 
On the other hand, the non-BPS exact solutions have been known to have unstable 
directions (negative mode) \cite{Din:1980jg}. 
However, the local and global instabilities such as negative 
modes and the flow to the other saddle points have not been 
explicitly worked out for the non-BPS exact solutions. 
It is also observed recently that the non-BPS exact solutions 
exhibit unexpected properties with respect to the clustering 
\cite{Bolognesi:2013tya}.

The purpose of this work is to investigate the properties of 
non-BPS exact solutions \cite{Dabrowski:2013kba} by using the 
ansatz of multi-instanton configurations, which is used to 
obtain the nonperturbative contributions needed for the resurgence 
\cite{Misumi:2014jua,  Misumi:2015dua, Misumi:2014rsa, Misumi:2014bsa}. 
We find that the non-BPS exact solutions are contained as 
subspaces of the parameter space of the ansatz, giving (a part 
of) the multi-instanton contributions, and hence they play a 
role in resurgence.
Relative phases between constituent fractional instantons are 
found to play a decisive role to achieve the balance of forces. 
We also find that the transition between seemingly distinct 
configurations occurs as moduli parameters change in the non-BPS 
solutions, and the simplest non-BPS solution contains a multi-bion 
configuration with fully-compressed double fractional instantons. 
This fully-compressed double fractional instanton eliminates 
the strong phase dependence of two-body forces between a
fractional instanton and anti-instanton \cite{Misumi:2014jua}, and 
helps to achieve the balance of forces in the non-BPS exact 
solution. 
We obtain a generic pattern of transitions among distinct 
multi-bion configurations (flipping partners), which helps to 
list various flipping partners in the non-BPS exact solutions. 
We observe the three essential properties of the non-BPS exact 
solutions : (i) opposite sign for terms corresponding to the left and right infinities,
(ii) the symmetric location of instantons,
and (iii) the transition between flipping partners. 
The balance of forces in the non-BPS exact solution is examined 
step by step, and many-body forces related to the relative phases 
are found to be important. 
We study the local and global instabilities of the non-BPS 
exact solution and find physical meaning of the negative modes 
and the flow to the other saddle points. 
We also briefly discuss some classes of the non-BPS exact solutions in 
Grassmann sigma models. 

This paper is organized as follows. 
In Sec.~\ref{sec:CPN}, we review the multi-bion ansatz and 
introduce the simplest non-BPS solution in the 
${\mathbb C}P^{N-1}$ model on ${\mathbb R}^1 \times S^1$ with 
a ${\mathbb Z}_{N}$-twisted boundary condition.
In Sec.~\ref{sec:FP} we discuss the transition between seemingly 
distinct configurations in the non-BPS solutions, and conjecture 
a generic structure of non-BPS solutions diagrammatically.
In Sec.~\ref{sec:balance} we show how the balance of force 
exists in the non-BPS solutions, and discuss essential properties 
of the solutions.
In Sec.~\ref{sec:NM} we investigate local and global instabilities 
of the non-BPS solutions by deforming them within the multi-bion 
ansatz, and discuss the number of negative modes.
In Sec.~\ref{sec:Gr} we briefly discuss the simplest classes 
of non-BPS solutions in Grassmann sigma models.
Section \ref{sec:SD} is devoted to a summary and discussion.



\section{Non-BPS solutions in ${\mathbb C}P^{N-1}$ models}
\label{sec:CPN}

\subsection{The multi-instanton ansatz}
\label{sc:ansatz}

Action density $s(x)$ and topological charge density $q(x)$ 
of the ${\mathbb C}P^{N-1}$ model in euclidean $2$-dimensions 
are given in terms of a 
normalized row vector $n(x)$ of an $N$-component complex scalar 
fields $\omega(x)$ as 
\begin{eqnarray}
s(x)={1\over{2\pi g^{2}}}  (D_{\mu} n) (D_{\mu}n)^{\dag}
={1\over{2\pi g^{2}}}( \partial_\mu n \partial_{\mu}n^{\dagger} 
+\partial_{\mu} n n^\dagger \partial_\mu n n^\dagger)
\,,
\label{eq:action}
\end{eqnarray}
\begin{eqnarray}
q(x)=
 \frac{-1}{2\pi}\epsilon_{\mu\nu} \partial_{\mu} A_{\nu} 
=\frac{-i}{2\pi} \epsilon_{\mu\nu} 
\partial_{\mu}(n \partial_{\nu}n^\dagger )
\,,
\label{eq:top_charge}
\end{eqnarray}
\begin{equation}
n(x)=\omega(x)/|\omega(x)|, \quad |\omega|
=\sqrt{\omega \omega^\dagger }, 
\quad A_{\mu}=i n \partial_\mu n^\dagger. 
\label{eq:normalized_vector}
\end{equation}
The unnormalized vector field $\omega$ is also called a moduli 
matrix, since it contains entire informations of moduli for 
BPS solutions \cite{Eto:2006pg}. 
We denote the total action as $S$ and the total topological 
charge as $Q$. 
The covariant derivative $D_{\mu}=\partial_{\mu}+iA_{\mu}$ 
is defined in terms of the composite gauge field $A_{\mu}$. 
One should note that the unnormalized vector field $\omega(x)$ 
multiplied by arbitrary nonsingular functions should be 
identified, since it gives the same physical field $n(x)$. 
We compactify the $x_2$ direction as $x_2+L\sim x_2$, and 
impose the ${\mathbb Z}_{N}$-twisted boundary 
conditions on $n$ and $\omega$ 
\begin{equation}
\omega(x_1, x_2+L)=\Omega \; \omega(x_1,x_2), \quad 
\Omega={\rm diag.}\left[1, e^{2\pi i/N}, e^{4\pi i/N},
\cdots, e^{2(N-1)\pi i/N} \right]. 
\label{eq:twisted_bc}
\end{equation}
The twisted boundary condition introduces effectively a 
potential which lifts the vacuum moduli of 
${\mathbb C}P^{N-1}$ model, leaving only $N$ discrete vacua :
$n=(1,\cdots,0), \cdots, (0,\cdots,1)$.

With $z=x_1+ix_2, \bar z=x_1-ix_2$, the holomorphic fields 
$\omega(z)$ solve the BPS equation $D_{\bar z}n=0$, which 
automatically satisfies the field equation. 
The simplest BPS solution with the topological charge $Q=1/N$ 
is the fractional instanton given by the following unnormalized 
$N$-component fields $\omega(x)$ 
\begin{equation}
\omega=(0, \cdots, a e^{i\theta_{a}} e^{-\frac{2\pi}{NL}z}, 1, 
\cdots, 0), 
\label{eq:fractional_instanton}
\end{equation}
with moduli parameters $a \in {\mathbb R}$ and $-\pi<\theta \leq \pi$. 
One should note that the twisted boundary condition introduces 
the nontrivial $x_1$ dependence automatically, leading to the kink-like 
behavior as follows. 
For $x_{1}\ll 0$, the term $a e^{i\theta_{a}} e^{-{2\pi\over{NL}}z}$ 
is dominant, while the term $1$ is dominant for $x_{1}\gg 0$. 
Whenever a single component dominates, the normalized vector 
$n$ becomes a vacuum configuration as $(0,\cdots, 1, \cdots,0)$. 
\begin{figure}[htbp]
\begin{center}
\includegraphics[width=0.4\textwidth]{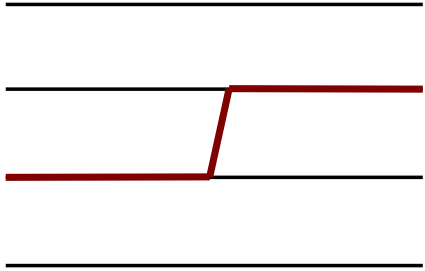}
\end{center}
\caption{An example of BPS fractional instantons in the 
${\mathbb C}P^{N-1}$ model.
Black lines stand for some of $N$ discrete vacua while
a kinky red line indicates a transition between different vacua.
}
\label{fracInst}
\end{figure}
Consequently the action density $s(x)$ and topological charge 
density $q(x)$ are localized in a region of width $NL$ around 
the transition point between the two vacua 
\begin{equation}
\begin{pmatrix}
...,\,\,&
1,   \,\,&
0, \,\,&
...
\end{pmatrix}\,\to\,
\begin{pmatrix}
...,\,\,&
0,\,\,&
1, \,\,&
...
\end{pmatrix}\, ,
\end{equation}
as depicted in Fig.~\ref{fracInst}. 
This kinky configuration has been understood as a D-brane 
configuration \cite{Eto:2004vy}. 
The condition $a e^{-{2\pi\over{NL}}x_{1} } = 1$ gives the 
transition point at $x_1=\frac{NL}{2\pi}\log a$, which is 
defined as the position of the fractional instanton. 
Because of the translational invariance, integrated total action 
does not depend on the position of the fractional instanton. 
Moreover, the action density $s(x)$ and topological charge 
density $q(x)$ do not depend on the phase angle $\theta_{a}$. 

It is important to note that both $s(x)$ and $q(x)$ for 
multi-fractional instantons ($|Q|<1$) are independent of $x_{2}$. 
Even if we have many fractional instantons at varius locations, 
$x_{2}$ appears only as the phase factor $e^{-i\frac{2\pi}{NL}x_{2}}$ 
common to all terms in the same component due to the twisted 
boundary conditions (\ref{eq:twisted_bc}). 
Then $s(x)$ and $q(x)$ do not depend on $x_2$. 
In such a situation, we can take the compactification limit 
$L\to 0$, and obtain the domain wall situation that has been 
studied extensively \cite{Isozumi:2004jc, Isozumi:2004va}. 
If and only if (more than) $N$ fractional instantons coexist 
(not far apart), an additional term with a distinct $x_2$ dependent 
phase (by an additional amount $e^{-i\frac{2\pi}{L} x_{2}}$) coexits 
in the same component. 
In such a situation with $|Q|\ge1$, $x_2$ dependence of $s(x)$ 
and $q(x)$ emerges.

From now on, we take the compactification scale $L=1$ 
and the coupling constant $g^2=1$, for simplicity. 
Before discussing the non-BPS exact solutions in ${\mathbb C}P^{N-1}$ 
models with the ${\mathbb Z}_{N}$-twisted boundary conditions 
(\ref{eq:twisted_bc}), we introduce an ansatz 
\cite{Misumi:2014jua, Misumi:2014raa, Misumi:2014bsa} 
for multi fractional instanton configurations that reduces to 
solutions of field equations for asymptotically large 
separations of constituent fractional instantons, and carries 
the correct number of moduli for each individual constituent: its 
phase and position in $x_1$. 
Therefore the ansatz serves as a basis to obtain nonperturbative 
contributions by integrating over the (quasi)-moduli of the 
multi fractional instantons. 
Among them, we have bions (one-fractional-instanton $+$ 
one-fractional-anti-instanton) that play a vital role to 
achieve a resurgent trans-series.

\begin{figure}[htbp]
\begin{center}
\includegraphics[width=0.4\textwidth]{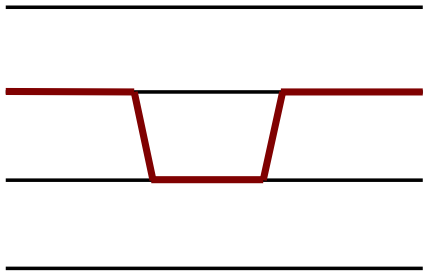}
\end{center}
\caption{An example of bion configurations in the 
${\mathbb C}P^{N-1}$ model.
For simplicity, we depict only four vacua lines among $N$ here.}
\label{bion}
\end{figure}

Let us first construct the ansatz for a bion as a non-BPS 
configuration as depicted in Fig.~\ref{bion}, by adding a 
fractional anti-instanton to the left of the fractional 
instanton (in Fig.~\ref{fracInst}). 
To return to the second vacuum for large negative $x_1$, 
we need to add a term proportional to $e^{-\frac{2\pi}{N}(z+\bar z)}$ 
in the second component in order to satisfy the twisted boundary 
condition (\ref{eq:twisted_bc}). 
Thus the ansatz $\omega(x)$ for a bion in ${\mathbb C}P^{N-1}$ 
models is given by 
\begin{equation}
\omega = 
(0, \cdots, a e^{i\theta_{a}} e^{-{2\pi\over{N}}z} ,
 b e^{i\theta_{b}} e^{-{2\pi\over{N}}(z+\bar{z})}+
1, \cdots,\cdots, 0)\,,
\label{eq:bion}
\end{equation}
with $a,\,b\, \in {\mathbb R}$, $-\pi<\theta_{a},\theta_{b} \leq \pi$. 
As $x_{1}$ varies from negative to positive values, 
the normalized vector $n$ makes transitions between two different 
vacua as 
\begin{equation}
\begin{pmatrix}
...,\,\,&
0,   \,\,&
1, \,\,&
...
\end{pmatrix}\,\to\,
\begin{pmatrix}
...,\,\,&
1,\,\,&
0, \,\,&
...
\end{pmatrix}\,\to\,
\begin{pmatrix}
...,\,\,&
0,   \,\,&
1, \,\,&
...
\end{pmatrix}\,.
\end{equation}
The first transition point gives the position of the left 
fractional anti-instanton 
\begin{equation}
x_{\bar{\cal I}} \,=\, {N\over{2\pi}} \log \frac{b}{a}\,, 
\end{equation}
and the second transition point gives the position of the 
right fractional instanton 
\begin{equation}
x_{\cal I} \,=\, {N\over{2\pi}} \log {a}\,.
\end{equation}
Thus, the separation $R$ between these instanton constituents is
\begin{equation}
R \,=\, x_{\cal I}-x_{\bar{\cal I}} \,
=\,{N\over{2\pi}} \log {a^2\over{b}} \,,
\end{equation}
whereas the center of mass position is given by 
$x_{\rm cm}={N\over{4\pi}} \log {b}$. 
The configuration associated with $\omega$ in Eq.(\ref{eq:bion}) 
for $R>1$ is depicted in Fig.~\ref{bion}.

We note that action density $s(x)$ and topological charge 
density $q(x)$ depend on the (relative) phase $\theta_b$ 
of two terms in a single component, 
but are independent of the phase $\theta_a$. 
Because of translational invariance, the total energy is 
independent of the center of mass position $x_{\rm cm}$. 
Therefore we consider densities $s(x)$ and $q(x)$ as functions 
of two relevant parameters : 
{\it the separation $R$ between the fractional instanton and 
anti-instanton} and {\it the relative phase $\theta_b$ 
between them}. 
We also note that densities $s(x)$ and $q(x)$ of this 
configuration are independent of the coordinate $x_{2}$ of the 
compactified direction, since no instantons ($|Q|\ge1$) is 
invloved in any region of $x_1$.

As shown in \cite{Misumi:2014jua, Misumi:2014raa, Misumi:2014bsa},
by decreasing this separation $R$ from positive to negative 
values, the value of the total action changes from $S=2/N$ to $S=0$. 
We can define the effective interaction potential 
between the two constituents as 
$
V_{\rm eff}(R, \theta_b) \,=\, S - S_{\cal I} - S_{\bar {\cal I}} 
= S - \frac{2}{N}$, representing the potential for the static 
two-body force between the consitituent fractional instanton 
and fractional anti-instanton. 
For large separations $R$, we have found 
\cite{Misumi:2014jua, Misumi:2014raa, Misumi:2014bsa} 
that 
\begin{equation}
V_{\rm eff}(R, \theta_b) \propto -\cos \theta_b e^{-\xi R}, \qquad 
\xi =2\pi/N\, . 
\label{eq:interation_pot}
\end{equation}
The notable point is that the interaction is attractive 
(repulsive) for $|\theta_b|\leq \pi/2$ ($\pi/2\leq |\theta_b|$).  
Based on the extended Bogomolnyi--Zinn-Justin prescription 
\cite{Bogomolny:1980ur, ZinnJustin:1981dx}, it has been shown 
that the bion configuration in the ${\mathbb C}P^{N-1}$ model 
produces the imaginary ambiguity \cite{Misumi:2015dua}, which 
is expected to cancel the imaginary ambiguity in the 
non-Borel-summable perturbative contributions similarly to the 
case in the quantum mechanics \cite{Bogomolny:1980ur, ZinnJustin:1981dx}.

\begin{figure}[htbp]
\begin{center}
\includegraphics[width=0.99\textwidth]{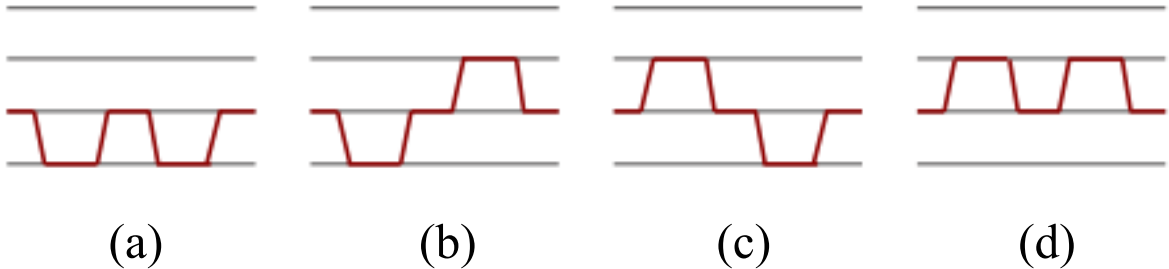}
\end{center}
\caption{Generic two-bion configurations in Eq.~(\ref{eq:2bion}).}
\label{G2bion}
\end{figure}

Let us next introduce an ansatz for two bions. 
For simplicity, we assume $N \ge 3$ 
\footnote{
If $N=2$, the upper and lower lines are identified in 
Fig.~\ref{G2bion} (b) (and (c)), and the corresponding ansatz 
becomes more complicated, since it involves an (anti-)instanton 
with $|Q|=1$ in the middle. 
}. 
In order to add one more bion to the left of the single bion in 
Eq.(\ref{eq:bion}) as in Fig.~\ref{G2bion}(a), 
we need to add a term $c e^{-\frac{2\pi}{N}(2z+\bar z)}$ 
(with a complex constant $c$) in the first component, and 
another term $d e^{-\frac{4\pi}{N}(z+\bar z)}$ to the second 
component. 
We will call this diagram in Fig.~\ref{G2bion}(a) as 
$\bar {\cal I}{\cal I} \bar{\cal I} {\cal I}$.   
If we wish to consider the diagram with ordering 
$\bar {\cal I} {\cal I} {\cal I} \bar{\cal I}$ in 
Fig.~\ref{G2bion}(b), we need to have a term 
$f e^{-{2\pi\over{N}}\bar{z}}$ in the third component. 
If we wish to consider the diagram with ordering 
${\cal I} \bar {\cal I} \bar {\cal I} {\cal I}$ in 
Fig.~\ref{G2bion}(c), we need to have a term 
$g e^{-{4\pi\over{N}}(z+2\bar{z})}$ in the third component. 
With these terms added, we obtain the ansatz for the most 
general configuration for two bions (with no (anti-)instanton 
with $|Q|=1$ in any region of $x_1$) 
\begin{equation}
\omega = 
\begin{pmatrix}
\cdots,\,&
a e^{-{2\pi\over{N}}z} + c e^{-{2\pi\over{N}}(2z+\bar{z})},\, &
1+ b e^{-{2\pi\over{N}}(z+\bar{z})} + d e^{-{4\pi\over{N}}(z+\bar{z})},\,&
f e^{-{2\pi\over{N}}\bar{z}} + g e^{-{2\pi\over{N}}(z+2\bar{z})},
\cdots
\end{pmatrix},
\label{eq:2bion}
\end{equation}
with seven complex parameters $a,b,c,d,f,g, \in {\mathbb C}$. 
One should note that one more diagram with the ordering 
${\cal I} \bar {\cal I} \bar {\cal I} {\cal I}$ in 
Fig.~\ref{G2bion}(d) is automatically included in the ansatz. 
The action and topological charge densities are independent of 
a phase common to terms in the same component. 
Therefore only four phases are relevant among six phase parameters 
: relative phases $(a, c)$, 
and $(f, g)$ besides phases of $b$ and $d$. 
Although six modulus correspond to six positions of constituent 
fractional (anti-)instantons, total action is independent of 
the center of mass position because of translational invariance. 
Therefore the action and topological charge densities can be 
considered as functions of only nine relevant parameters: four 
relative phases, five separation parameters corresponding to 
the lengths of bion and anti-bion, and the distance between them. 
This ansatz contains four seemingly-different configurations 
depending on the values of these parameters as shown in 
Fig.~\ref{G2bion}(a)-(d) for large separations. 

Since our ansatz for the most general two-bion configurations 
 in Eq.~(\ref{eq:2bion}) does not have (anti-)instanton with 
$|Q|=1$, all terms in the same component have a common 
$x_2$-dependence as a phase factor such as $e^{-i\frac{2\pi}{N}x_2}$. 
Therefore the action density $s(x)$ and the topological charge 
density $q(x)$ are independent of $x_2$ in our ansatz. 
If (anti-)instanton with $|Q|=1$ would be involved, additional 
terms with distinct $x_2$-dependent phase factor would emerge such as 
$C(x_{1}) e^{-(2\pi/N)ix_{2} }+ C'(x_{1}) e^{-(2\pi/N+2\pi)ix_{2} }$, 
and $s(x), q(x)$ would depend on $x_2$.

Let us first consider the case with $|g| \gg |c|$ and 
$|a| \gg |f|$, where the terms with $c$ and $f$ are superficially 
negligible. 
Then, as $x_{1}$ varies from negative infinity to positive 
infinity, the dominant term changes as 
$d e^{-{4\pi\over{N}}(z+\bar{z})}\,\to\,  
g e^{-{2\pi\over{N}}(z+2\bar{z})}\,\to\,
b e^{-{2\pi\over{N}}(z+\bar{z})}\,\to\,
a e^{-{2\pi\over{N}}{z}}\,\to\,1$ for a choice 
of the parameters 
$|g|^2 \gg |b||d|,\,\,|b|^{2}\gg |a||g|,\,\,|a|^{2}\gg |b|$. 
It means that the vacua undergo the transition 
\begin{equation}
\begin{pmatrix}
0,\,\,&
1, \,\,&
0
\end{pmatrix}\,\to\,
\begin{pmatrix}
0,\,\,&
0,\,\,&
1
\end{pmatrix}\,\to\,
\begin{pmatrix}
0,\,\,&
1,\,\,&
0
\end{pmatrix}\,\to\,
\begin{pmatrix}
1,\,\,&
0,\,\,&
0
\end{pmatrix}\,\to\,
\begin{pmatrix}
0,\,\,&
1,\,\,&
0
\end{pmatrix}\,,
\label{eq:23212}
\end{equation}
as shown in Fig.~\ref{G2bion}(c).
Using a similar reasoning as the bion case, we find the 
four fractional instanton constituents show up successively 
as $x_{1}$ increases from negative infinity to positive infinity 
: the first fractional instanton at 
$R_{1} = {N\over{2\pi}} \log {|d|\over{|g|}}$, 
the first fractional anti-instanton at 
$R_{2} = {N\over{2\pi}} \log {|g|\over{|b|}}$, 
the second fractional anti-instanton at 
$R_{3} = {N\over{2\pi}} \log {|b|\over{|a|}}$,
and the second fractional instanton at 
$R_{4} = {N\over{2\pi}} \log {|a|}$. 
The three separations between these instanton constituents are
\begin{equation}
R_{21} \,=\, {N\over{2\pi}} \log {|g|^{2}\over{|b||d|}}\,,\,\,\,\,\,\,\,\,\,
R_{32} \,=\, {N\over{2\pi}} \log {|b|^{2}\over{|a||g|}}\,,\,\,\,\,\,\,\,\,\,
R_{43} \,=\, {N\over{2\pi}} \log {|a|^{2}\over{|b|}}\,.
\end{equation}
The configuration in Fig.~\ref{G2bion}(c) is visible in the 
parameter region $|d/g| \ll |g/b| \ll |b/a| \ll |a|$. 

On the other hand, we can recognize the configuration in 
Fig.~\ref{G2bion}(b) in another region of the parameter 
space: the case with $|c| \gg |g|$ and $|f| \gg  |a|$, 
where the terms with $g$ and $a$ are superficially negligible.
Then we obtain transitions of vacua as 
\begin{equation}
\begin{pmatrix}
0,\,\,&
1, \,\,&
0
\end{pmatrix}\,\to\,
\begin{pmatrix}
1,\,\,&
0,\,\,&
0
\end{pmatrix}\,\to\,
\begin{pmatrix}
0,\,\,&
1,\,\,&
0
\end{pmatrix}\,\to\,
\begin{pmatrix}
0,\,\,&
0,\,\,&
1
\end{pmatrix}\,\to\,
\begin{pmatrix}
0,\,\,&
1,\,\,&
0
\end{pmatrix}\,, 
\label{eq:21232}
\end{equation}
instead of Eq.~(\ref{eq:23212}). 
As $x_{1}$ increases from negative infinity to positive infinity, 
four fractional instanton constituents show up successively 
: the first fractional anti-instanton at 
$R_{1}' = {N\over{2\pi}} \log {|d|\over{|c|}}$, 
the first fractional instanton at 
$R_{2}' = {N\over{2\pi}} \log {|c|\over{|b|}}$, 
the second fractional instanton at 
$R_{3}' = {N\over{2\pi}} \log {|b|\over{|f|}}$,
and the second fractional anti-instanton at 
$R_{4}' = {N\over{2\pi}} \log {|f|}$. 
Thus, the three separations between these instanton constituents are
\begin{equation}
R_{21}' \,=\, {N\over{2\pi}} \log {|c|^{2}\over{|b||d|}}\,,\,\,\,\,\,\,\,\,\,
R_{32}' \,=\, {N\over{2\pi}} \log {|b|^{2}\over{|c||f|}}\,,\,\,\,\,\,\,\,\,\,
R_{43}' \,=\, {N\over{2\pi}} \log {|f|^{2}\over{|b|}}\,.
\end{equation}
The configuration in Fig.~\ref{G2bion}(b) is visible in the 
parameter region $|d/c| \ll |c/b| \ll |b/f| \ll |f|$. 
We note that there is a relation among six separations 
$R_{21}+2R_{32}+R_{43}=R_{21}'+2R_{32}'+R_{43}'$, leading to 
five independent separation variables. 
In both of these parameter regions corresponding to 
 Fig.~\ref{G2bion}(b) and (c), only two phases of $b$ 
and $d$ are relevant, since only a single term is dominant 
in the first and the third components.

It is interesting to note that the most general ansatz for 
two bions automatically contains two other possible diagrams 
containing the ordering $\bar {\cal I}{\cal I}\bar {\cal I} {\cal I}$ 
 in Fig.~\ref{G2bion}(a), or 
${\cal I}\bar {\cal I}{\cal I}\bar {\cal I}$ in Fig.~\ref{G2bion}(d). 
The configuration of  $\bar {\cal I}{\cal I}\bar {\cal I} {\cal I}$ 
in Fig.~\ref{G2bion}(a) is 
visible in the parameter region 
$|d/c| \ll |c/b| \ll |b/a| \ll |a|$. 
In this parameter region, the relative phases $(a,c)$, 
and phases of $b$ and $d$ are relevant among the four phase 
parameters. 
The configuration of ${\cal I}\bar {\cal I}{\cal I}\bar {\cal I}$ 
in Fig.~\ref{G2bion}(d) 
is visible in the parameter region 
$|d/g| \ll |g/b| \ll |b/f| \ll |f|$, where only three phases 
are relevant : the relative phase $(g,f)$ and phases $b$ and $d$. 
It is interesting to note that a different partial set of 
four relative phases are relevant in each parameter 
region corresponding to four different diagrams in Fig.~\ref{G2bion}. 
One should also note that there are five independent 
length parameters, but only three different combinations of 
them emerge as separations of constituent fractional 
(anti-)instantons in different parameter regions. 
In the next section, we will demonstrate that this 
ansatz (\ref{eq:2bion}) contains the simplest non-BPS exact 
solution. 
This fact implies that the non-BPS exact solution contributes 
to the resurgent expansion as a special configuration of 
two-bion configurations.

\subsection{The simplest non-BPS exact solution}

Based on the procedure of projection operations \cite{Din:1980jg},
the non-BPS exact solutions in ${\mathbb C}P^{N-1}$ model on 
${\mathbb R}^1 \times S^{1}$ with ${\mathbb Z}_{N}$-twisted 
boundary conditions are constructed in \cite{Dabrowski:2013kba}.
We here discuss the properties of the solutions. 
The non-BPS solutions are obtained through the following 
projection applied to any of the BPS solutions $\omega$, 
\begin{equation}
Z_{+}:\,\,\omega\,\to\, Z_{+}\omega\equiv \partial_{z}
\omega-{(\partial_{z}\omega)\omega^{\dag}\over{\omega\omega^{\dag}}}
\omega\,.
\label{proj}
\end{equation}
\begin{figure}[htbp]
\begin{center}
\includegraphics[width=0.4\textwidth]{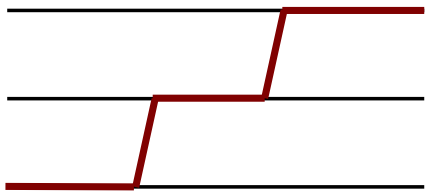}
\end{center}
\caption{BPS solution $\omega_{\cal I I}$ of two fractional instantons in Eq.~(\ref{bps}) 
in ${\mathbb C}P^{2}$ model.}
\label{bpscp2}
\end{figure}
Starting from the BPS solution of two fractional instantons in 
${\mathbb C}P^{2}$ model with ${\mathbb Z}_{N}$-twisted boundary condition 
\begin{equation}
\omega_{\cal I I} = 
\begin{pmatrix}
l_{1}e^{i\theta_{1}}e^{-{4\pi\over{3}}z},&
l_{2}e^{i\theta_{2}}e^{-{2\pi\over{3}}z},&
1
\end{pmatrix}\,,
\label{bps}
\end{equation}
with $S=2/3, \,\,Q=2/3$ shown in Fig.~\ref{bpscp2},
the projection produces the non-BPS exact solution 
\begin{equation}
\omega_{\rm nbps} = 
\begin{pmatrix}
e^{i\theta_{1}}\left(\frac{2l_{1}}{l_2}e^{-{2\pi\over{3}}z}
+l_{1}l_{2}e^{-{2\pi\over{3}}(2z+\bar{z})}\right), &
e^{i\theta_{2}}\left(1
-l_{1}^{2}e^{-{4\pi\over{3}}(z+\bar{z})}\right),&
-l_{2}e^{-{2\pi\over{3}}\bar{z}}
-\frac{2l_{1}^{2}}{l_2}e^{-{2\pi\over{3}}(z+2\bar{z})}
\end{pmatrix}\,.
\label{eq:nonbps}
\end{equation}
The total action of this solution is $S=4/3$ while the total 
topological charge is $Q=0$.
This $\omega$ is obviously a special case of the generic 
two-bion configuration (\ref{eq:2bion}).
We note that the action $s(x)$ and topological $q(x)$ charge 
densities are independent of 
overall phase variables 
$\theta_{1}, \theta_{2}$ in each component.

\begin{figure}[htbp]
\begin{center}
\includegraphics[width=0.4\textwidth]{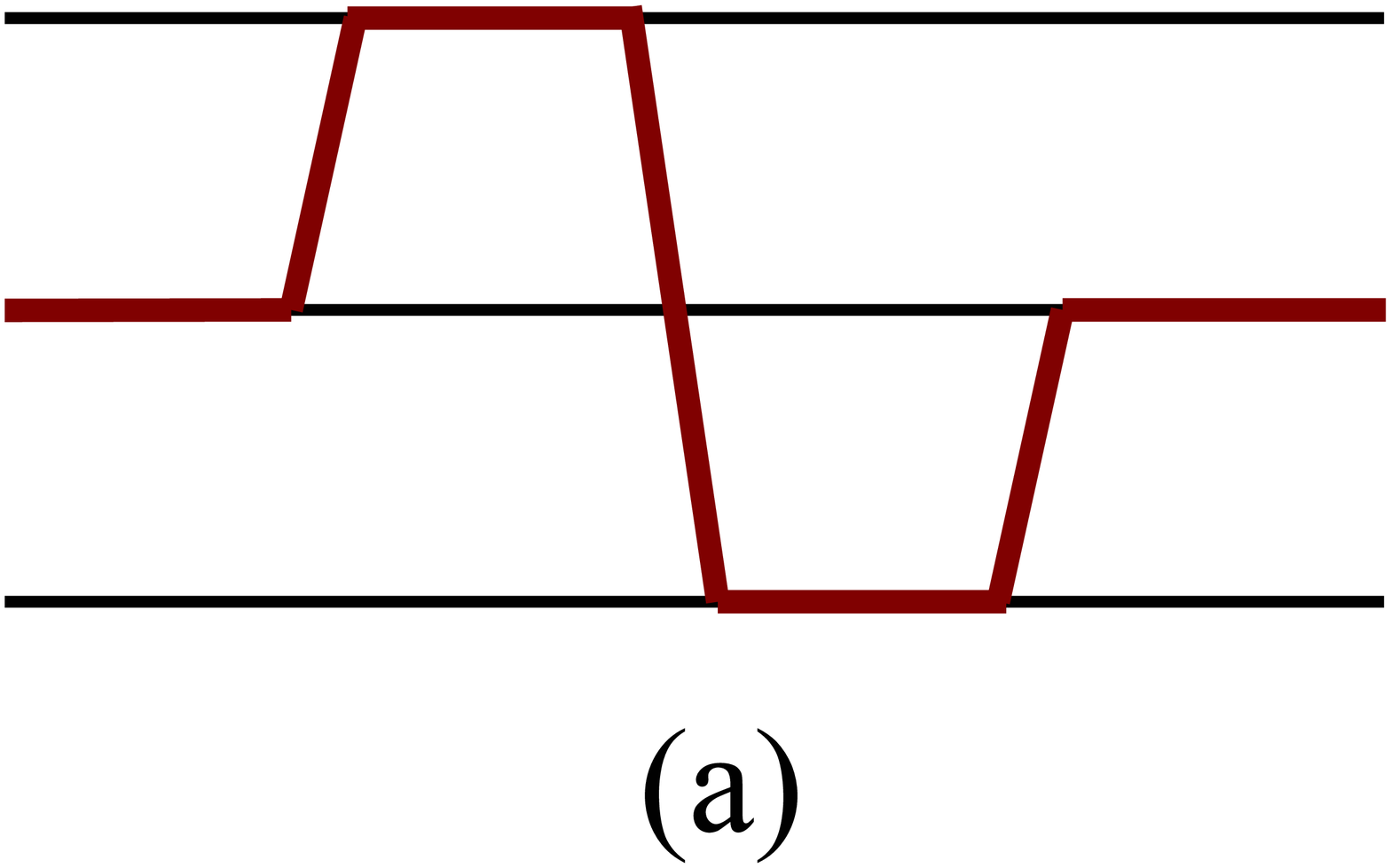}
\includegraphics[width=0.4\textwidth]{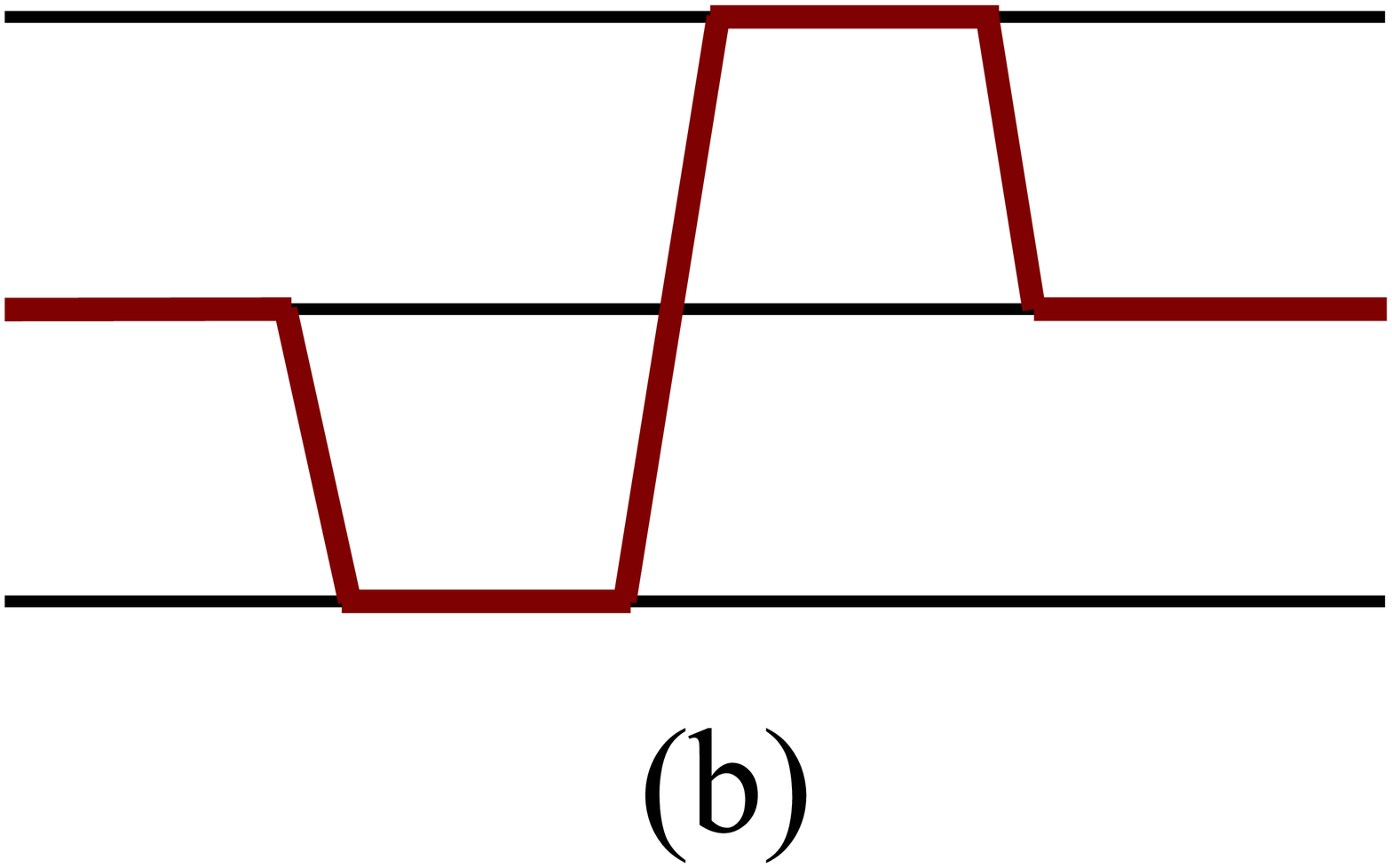}
\end{center}
\caption{The non-BPS exact solution in ${\mathbb C}P^{2}$ 
model makes transition between these two configurations 
(flipping parteners) as a moduli $2l_{1}/l_{2}^2$ varies. }
\label{nonbpscp2}
\end{figure}

\begin{figure}[htbp]
\begin{center}
\includegraphics[width=0.85\textwidth]{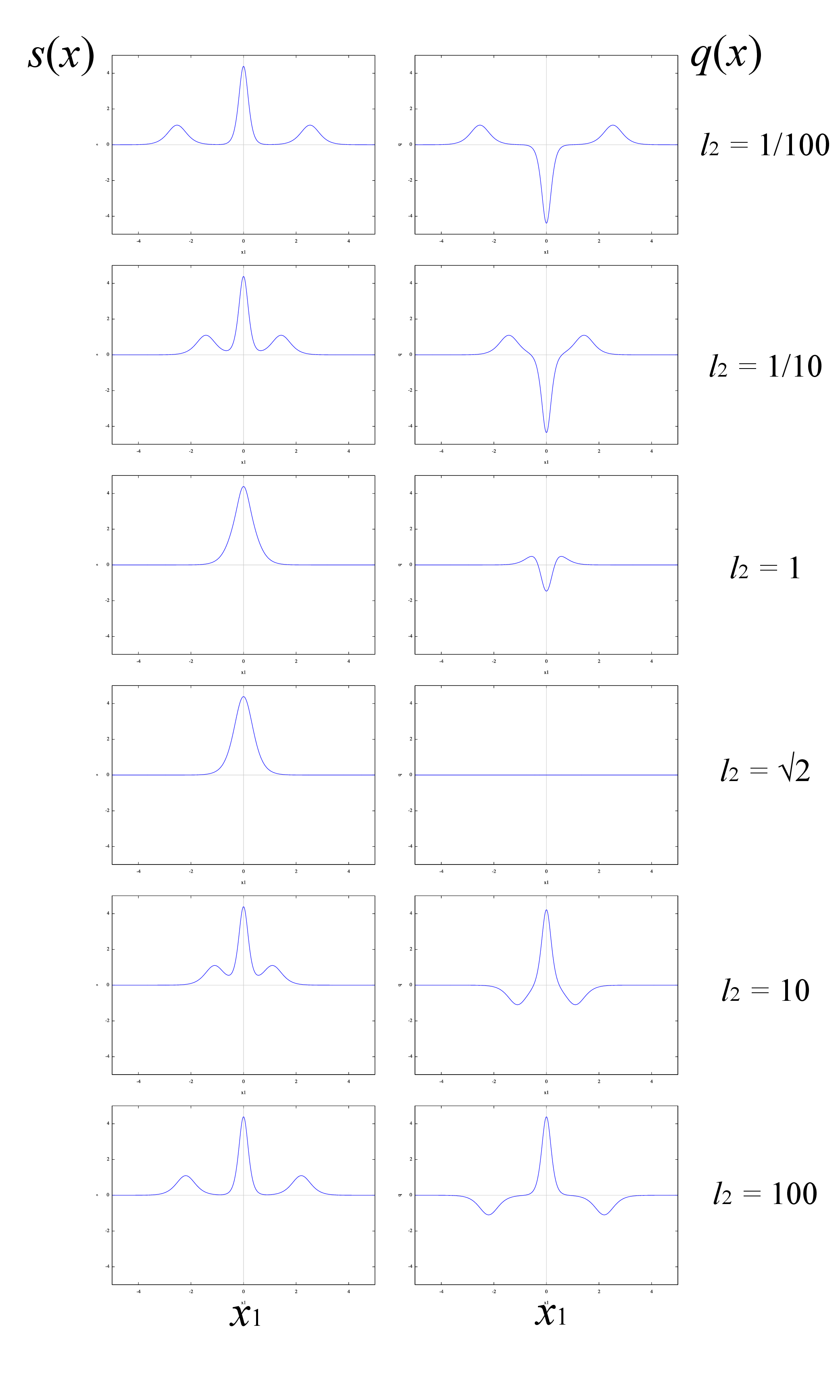}
\end{center}
\caption{Action density $s(x)$(left) and topological charge 
density $q(x)$(right) of the non-BPS solution Eq.(\ref{eq:nonbps}) with 
$l_{2}=1/100,1/10,1,\sqrt{2},10,100$ (from top to bottom) with $l_{1}=1$ fixed.}
\label{densitycp2}
\end{figure}

The exact solution (\ref{eq:nonbps}) is included as a 
subspace of parameters in our most general ansatz in 
Eq.~(\ref{eq:2bion}) as : 
\begin{equation}
a=2l_1/l_2,\,\,c=l_1l_2,\,\, b=0,\,\, d=-l_1^2,\,\, 
f=-l_2,\,\, g=-2l_1^2/l_2\,. 
\label{eq:embedding}
\end{equation}
What is special in this solution is that this solution 
contains in different corners of moduli space 
the two seemingly distinct configurations, each of which 
can be seen as a compressed case of the 
{\it two fractional (anti-)instanton configuration} 
in the middle, because of $b=0$. 
As depicted in Fig.~\ref{nonbpscp2}, the solution can describe 
two types of configurations depending on the ratio of 
$2l_{1}$ and $l_{2}^{2}$: Fig.~\ref{nonbpscp2}(a) arises 
for $2l_1\gg l_2^2$, and contains locally one compressed double 
fractional anti-instanton sandwiched between two fractional 
instantons, whereas Fig.~\ref{nonbpscp2}(b) arises 
for $2l_1\ll l_2^2$, and contains locally one compressed double 
fractional instanton sandwiched between two fractional 
anti-instantons and one double fractional instanton. 
These two configurations are similar to the two-bion 
configuration in Fig.~\ref{G2bion}(c) and (b), 
but the two adjacent fractional (anti-)instantons are 
completely compressed. 
Let us call these two configurations as the flipping partners. 
We show how densities of action $s(x)$ and topological charge 
$q(x)$ vary as $2l_{1}/l_{2}^2$ varies in 
Fig.~\ref{densitycp2}, although the value of the integrated total 
action remains constant. 
We find the action densities of flipping partners are related by 
the transformation $l_{2}/\sqrt{2}\to \sqrt{2}/l_2$. 

We will find that the flipping partners contained in the non-BPS 
exact solution is important to achieve the balance of forces 
between the BPS and anti-BPS constituent fractional instantons. 
One should also note that the two separations $R_{1}$ ($R_{1}'$) 
and $R_{2}$ ($R_{2}'$) from the middle compressed fractional instantons to the 
left and right fractional instantons are identical : 
\begin{equation}
R_1=R_2={3\over{4\pi}}\log(4l_{1}/l_{2}^{2})\,,
\end{equation}
for $2l_{1} \gg  l_{2}^{2}$ and
\begin{equation}
R_{1}'=R_{2}'={3\over{4\pi}}\log(l_{2}^{2}/l_{1})\,,
\end{equation}
for $2l_{1} \ll  l_{2}^{2}$.

\section{Flipping partners 
and generic construction}
\label{sec:FP}

\subsection{Flipping partners 
in the non-BPS exact solution
}

Let us analyze how transitions between different configurations 
occur in the non-BPS exact solution in the ${\mathbb C}P^{2}$ 
model as shown in Fig.~\ref{nonbpscp2} (the extension to 
${\mathbb C}P^{N-1}$ models is straightforward). 
We can re-express the BPS solution with two fractional-instantons 
in Eq.(\ref{bps}) as 
\begin{align}
\omega_{\cal I I
}&=
\begin{pmatrix}
A(z),\,\,&
B(z),\,\,&
1
\end{pmatrix}\,,
\nonumber\\
A(z)&=l_{1}e^{i\theta_{1}}e^{-2\alpha z}\,,\,\,\,\,\,\,
B(z)=l_{2}e^{i\theta_{2}}e^{-\alpha z}\,,
\end{align}
with $\alpha=2\pi/3$.
Then, the projection operation in Rq.(\ref{proj}) gives 
a non-BPS exact solution in Eq.(\ref{eq:nonbps}), which 
is now re-expressed as 
\begin{align}
\omega_{\rm nbps}\,&=\,
\begin{pmatrix}
A(B\bar{B}+2)\,\,&
B(-A\bar{A}+1)\,\,&
2A\bar{A}+B\bar{B}
\end{pmatrix}\,,
\end{align}
where the difference by an overall factor from Eq.(\ref{eq:nonbps}) is irrelevant.

We now explain how various constituents emerge in the non-BPS 
exact solution : 
(a)fractional anti-instantons, (b)fractional instantons, 
(c)double fractional instantons, and (d)double fractional 
anti-instantons.

{\bf (a)Fractional anti-instantons}\\
When $AB\bar{B}$ in the first component and $-BA\bar{A}$ 
in the second component are dominant, 
the unnormalized vector 
field $\omega$ is equivalent to a simpler one as 
\begin{align}
\omega_{\rm nbps}\,\approx\,
\begin{pmatrix}
AB\bar{B}\,\,&
-BA\bar{A}\,\,&
0
\end{pmatrix}
\,\sim\,
\begin{pmatrix}
\bar{B},\,\,&
-\bar{A},\,\,&
0\,\,
\end{pmatrix}\,,
\end{align}
because the division of $\omega$ by a common factor $AB$ to 
all components gives identical physical vector $n$. 
Since $|\bar{B}|= l_{2}e^{-\alpha {x_1}}$ and 
$|\bar{A}|= l_{1}e^{-2\alpha {x_1}}$, 
the vacua undergo transition as $(0,1,0)\to(1,0,0)$, implying 
a fractional anti-instanton at $x_{1}=(1/\alpha)\log (l_{1}/l_{2})$. 
In the same manner, when $B$ in the second component and 
$B\bar{B}$ in the third component are dominant, we obtain 
\begin{align}
\omega_{\rm nbps}\,\approx\,
\begin{pmatrix}
0,\,\,&
B,\,&
B\bar{B}
\end{pmatrix}
\,\sim\,
\begin{pmatrix}
0,\,\,&
1,\,\,&
\bar{B}\,\,
\end{pmatrix}\,,
\end{align}
which implies the vacuum transition $(0,0,1)\to(0,1,0)$ 
and a fractional anti-instanton at $x_{1}=(1/\alpha)\log l_{2}$.
These two fractional anti-instantons correspond to upside-down 
constituents (instanton $\to$ anti-instanton) 
of the original BPS solution in Fig.~\ref{bpscp2}.

{\bf (b)Fractional instantons}\\
When $2A$ in the first component and $B$ in the second component
are dominant, 
we obtain 
\begin{align}
\omega_{\rm nbps}\,\approx\,
\begin{pmatrix}
2A,\,\,&
B,\,\,&
0
\end{pmatrix}\,,
\end{align}
which implies the vacuum transition $(1,0,0)\to(0,1,0)$ 
and a fractional instanton at $x_{1}=(1/\alpha)\log (2l_{1}/l_{2})$. 
When $-BA\bar{A}$ in the second component and $2A\bar{A}$ in 
the third component are dominant, we obtain 
\begin{align}
\omega_{\rm nbps}\,\approx\,
\begin{pmatrix}
0,\,\,&
-BA\bar{A},\,\,&
2A\bar{A}
\end{pmatrix}
\,\sim\,
\begin{pmatrix}
0,\,\,&
-B,\,\,&
2\,\,
\end{pmatrix}\,,
\end{align}
implying the vacuum transition $(0,1,0)\to(0,0,1)$ 
and a fractional instanton at $x_{1}=(1/\alpha)\log (l_{2}/2)$.
These two fractional instantons correspond to the constituents
of the original BPS solution in Fig.~\ref{bpscp2}.

{\bf (c)Double fractional instantons}\\
When $AB\bar{B}$ in the first component and $B\bar{B}$ in the 
third component are dominant, we obtain 
\begin{align}
\omega_{\rm nbps}\,\approx\,
\begin{pmatrix}
AB\bar{B},\,\,&
0,\,\,&
B\bar{B}
\end{pmatrix}
\sim
\begin{pmatrix}
A,\,\,&
0,\,\,&
1
\end{pmatrix}\,,
\end{align}
implying the vacuum transition $(1,0,0)\to(0,0,1)$ 
and a compressed double fractional instanton at 
$x_{1}=(1/(2\alpha))\log l_{1}$.

{\bf (d)Double fractional anti-instantons}\\
When $2A$ in the first component and $2A\bar{A}$ in the third component
re dominant, we obtain 
\begin{align}
\omega_{\rm nbps}\,\approx\,
\begin{pmatrix}
2A,\,\,&
0,\,\,&
2A\bar{A}
\end{pmatrix}
\sim
\begin{pmatrix}
1,\,\,&
0,\,\,&
\bar{A}
\end{pmatrix}\,.
\end{align}
implying the vacuum transition $(0,0,1)\to(1,0,0)$, and a 
compressed double fractional anti-instanton at 
$x_{1}=(1/(2\alpha))\log l_{1}$.

If $2l_{1}>l_{2}^{2}$, the conditions for (b) and (d) are 
satisfied and the configuration in Fig.~\ref{nonbpscp2}(a) 
is clearly visible. 
On the other hand, the conditions for (a) and (c) are satisfied 
and the configuration in Fig.~\ref{nonbpscp2}(b) is visible 
if $2l_{1}<l_{2}^{2}$. 

We also note that the left and right fractional anti-instantons 
in (b) can be regarded as upside-down of fractional instantons 
in the original BPS solution in Fig.~\ref{bpscp2}. 
The configuration in (a) can be obtained by reflecting (b) 
in the middle. 
These observations leads to a generic pattern of non-BPS exact 
solutions constructed from the BPS-solutions and to a graphical 
construction of various configurations contained in the non-BPS 
exact solution in the next subsection.

\subsection{Generic pattern of non-BPS exact solutions}

We have seen how two different configurations emerge in a 
single non-BPS exact solution as moduli varies. 
Since this feature is quite generic in non-BPS exact solutions, 
we now summarize our observation on generic patterns of various 
configurations contained in non-BPS exact solutions. 
Below, we begin with the original BPS solution and show how 
various configurations in the non-BPS exact solution can be 
graphically obtained : 
\begin{enumerate}
\item

Turn all the BPS constituents upside down in the original BPS 
solution (fractional instanton $\leftrightarrow$ fractional 
anti-instanton).

\item

Connect the constituents by multiple instantons (double fractional instanton,
triple fractional instanton etc.). Then we end up with one of configurations of the non-BPS solution.

\item

Non-BPS solutions also contain configurations, in which 
the part with zero-instanton number is reflected in the middle. 

\end{enumerate}

\begin{figure}[htbp]
\begin{center}
\includegraphics[width=0.95\textwidth]{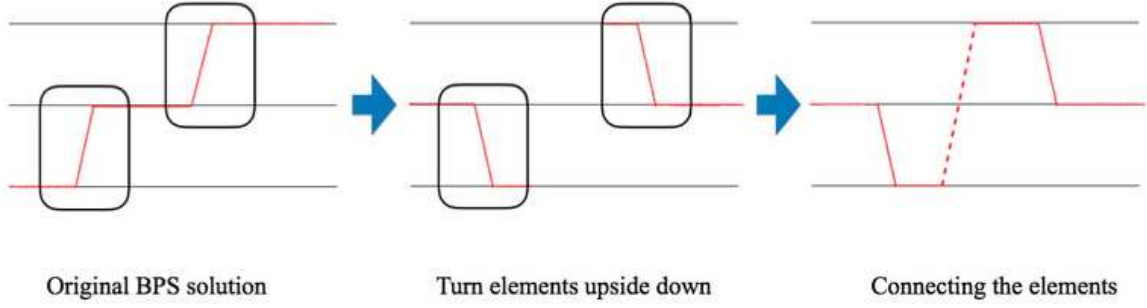}
\end{center}
\caption{Construction of the non-BPS exact solution from 
the BPS solution with two fractional instantons. 
}
\label{nonbps_fig1}
\end{figure}

\begin{figure}[htbp]
\begin{center}
\includegraphics[width=0.95\textwidth]{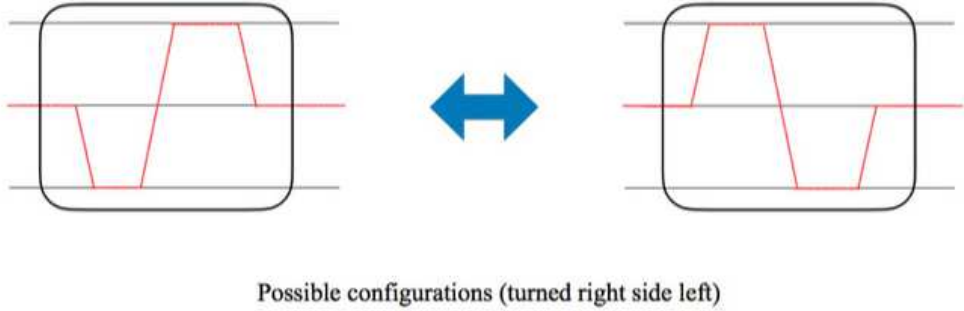}
\end{center}
\caption{Possible configurations are depicted for the non-BPS 
exact solution constructed from the two fractional instantons as 
the starting BPS solution.}
\label{nonbps_fig2}
\end{figure}

In Figs.~\ref{nonbps_fig1} and \ref{nonbps_fig2}, we show 
this construction for the present simplest case.
In this case, the whole configuration is a zero-instanton-number 
part (surrounded by the black lines in \ref{nonbps_fig2}), 
thus the possible configurations include only two types.

We conjecture the above features are generic patterns of non-BPS 
exact solutions in ${\mathbb C}P^{N-1}$ model on 
${\mathbb R}^1\times S^{1}$ with ${\mathbb Z}_{N}$-twisted boundary 
condition. 
To check these features in other examples,
we first begin with a BPS solution with three fractional 
instantons in the ${\mathbb C}P^{3}$ model. 
\begin{align}
\omega_{\cal III}&=
\begin{pmatrix}
l_{1}e^{-3\beta z},\,\,&
l_{2}e^{-2\beta z},\,\,&
l_{3}e^{-\beta z},\,\,&
1
\end{pmatrix}
\nonumber\\
&=
\begin{pmatrix}
C(z),\,\,&
D(z),\,\,&
E(z),\,\,&
1
\end{pmatrix}\,.
\label{eq:3instanton}
\end{align}
with $\beta =2\pi/4=\pi/2$.
The projection operation leads to the following non-BPS solution,
\begin{align}
\omega_{\rm nbps}&=
\begin{pmatrix}
C(D\bar{D}+2E\bar{E}+3),&
D(-C\bar{C}+E\bar{E}+2),&
E(2C\bar{C}+D\bar{D}-1),&
3C\bar{C}+2D\bar{D}+E\bar{E}
\end{pmatrix}\,.
\label{eq:nonBPS_3instanton}
\end{align}
\begin{figure}[htbp]
\begin{center}
\includegraphics[width=0.95\textwidth]{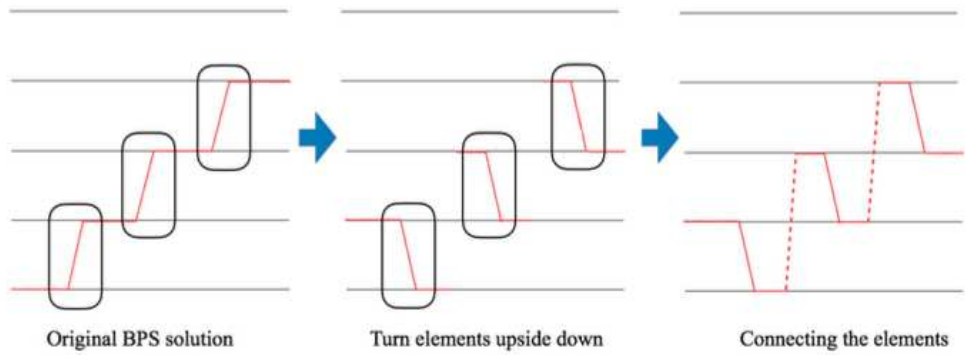}
\end{center}
\caption{Construction of the non-BPS exact solution from the BPS 
three fractional instanton solution ${\mathbb C}P^{4}$ model. 
}
\label{nonbps_fig3}
\end{figure}
\begin{figure}[htbp]
\begin{center}
\includegraphics[width=0.95\textwidth]{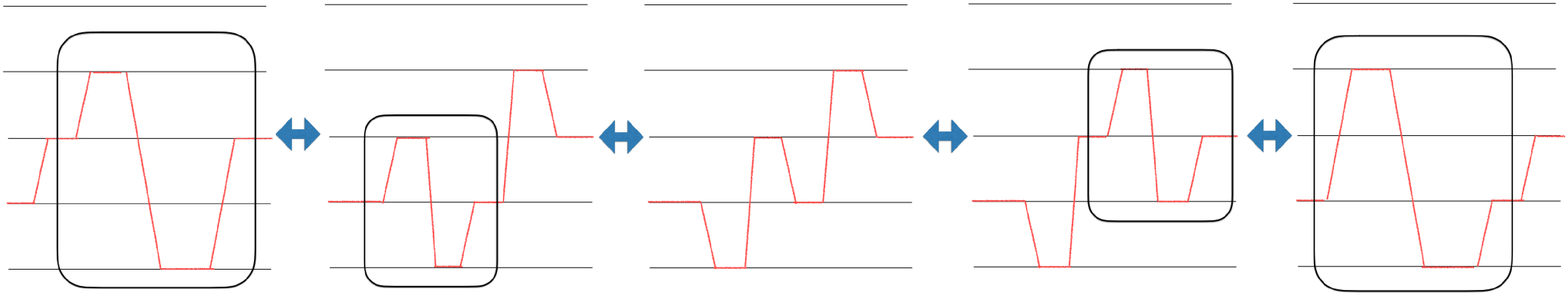}
\includegraphics[width=0.19\textwidth]{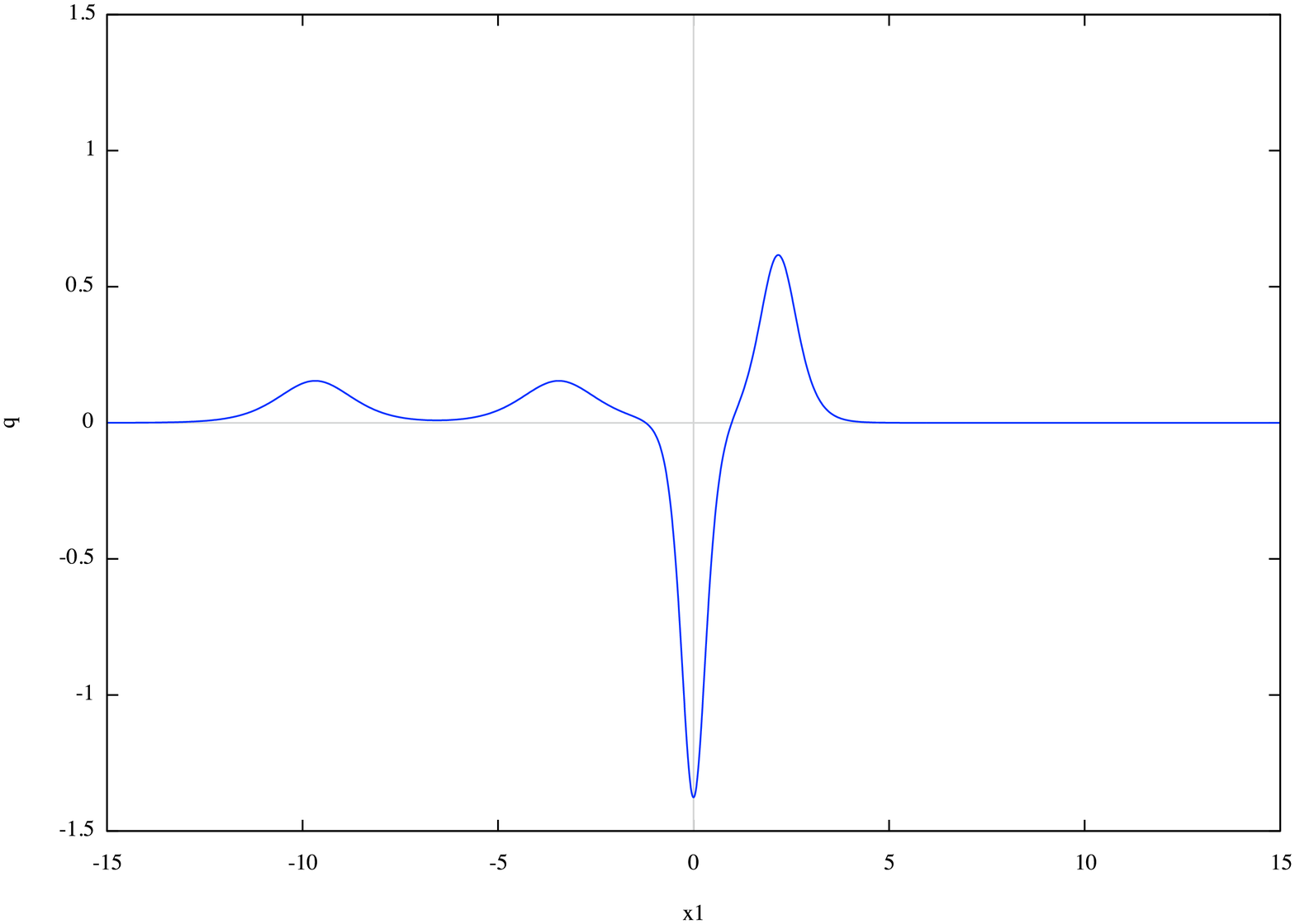}
\includegraphics[width=0.19\textwidth]{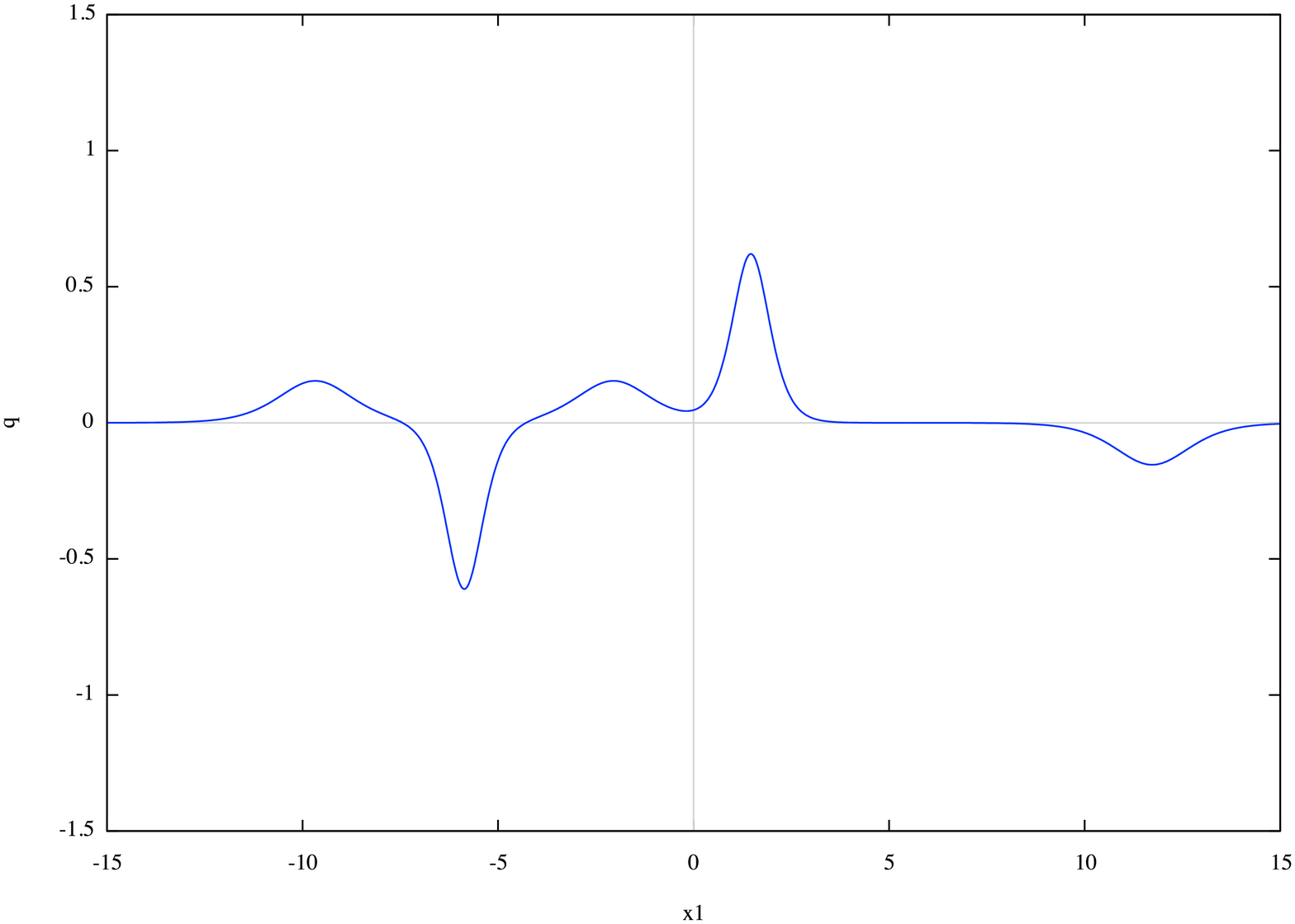}
\includegraphics[width=0.19\textwidth]{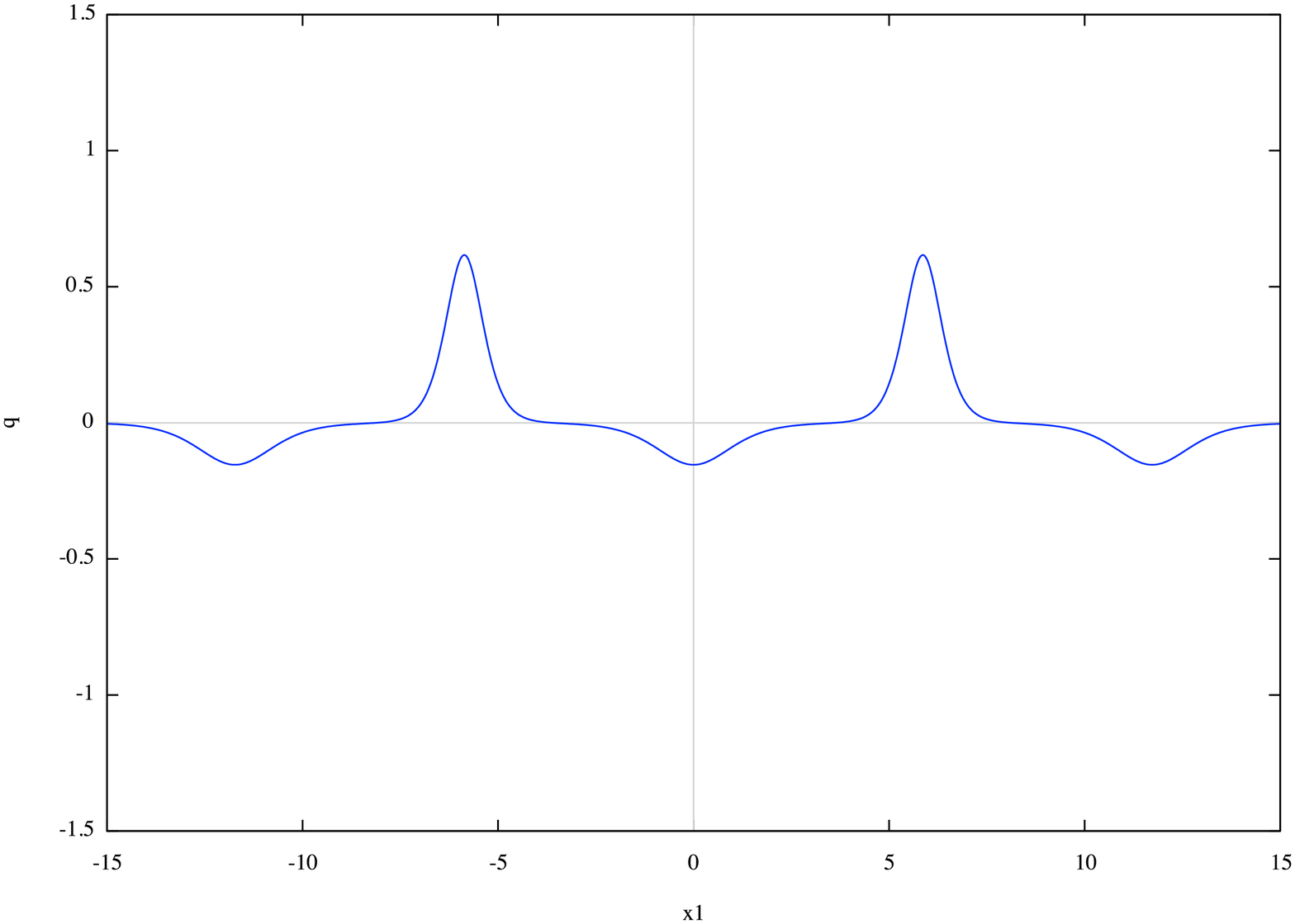}
\includegraphics[width=0.19\textwidth]{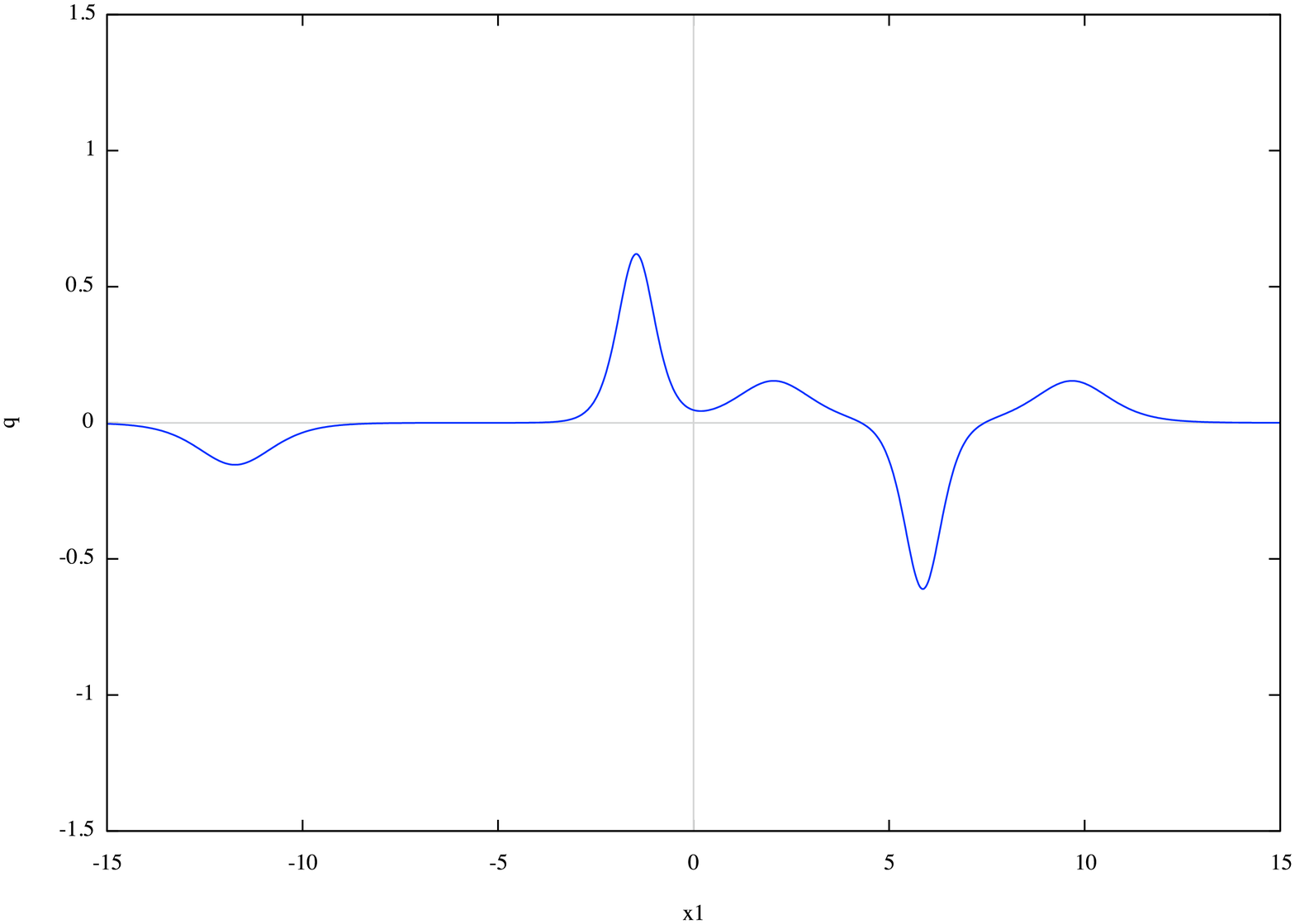}
\includegraphics[width=0.19\textwidth]{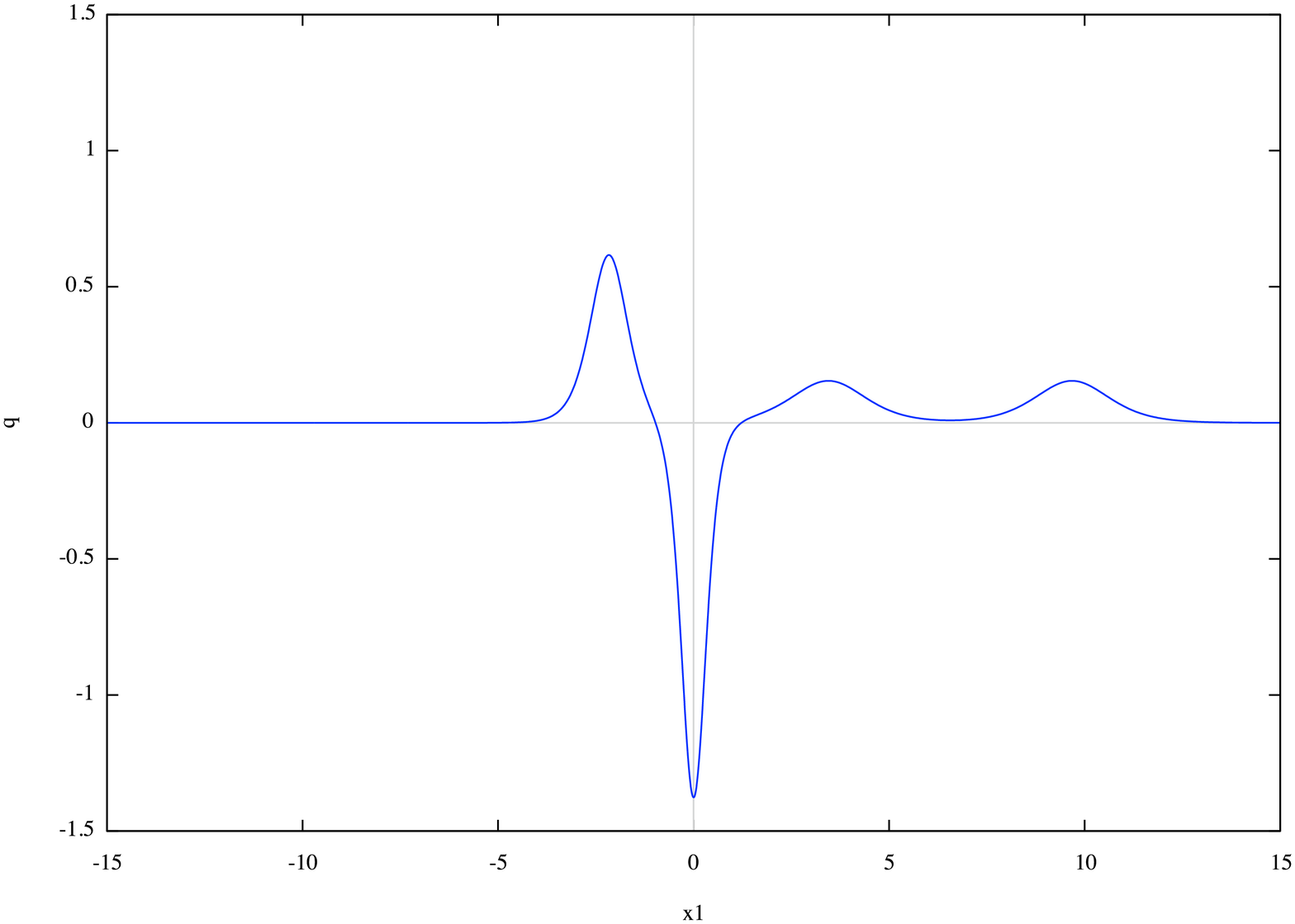}
\end{center}
\caption{Possible configurations are depicted for the non-BPS 
exact solution constructed from the BPS solution with three 
fractional instantons ${\mathbb C}P^{4}$ model. 
}
\label{nonbps_fig4}
\end{figure}
In Figs.~\ref{nonbps_fig3} and \ref{nonbps_fig4}, we show the 
case for three fractional instanton constituents in the starting 
BPS solution. (For generality, we depict the results for 
${\mathbb C}P^{4}$ model with $N=5$.)
In this case, the configuration contains several zero-instanton-number 
parts (surrounded by the black lines in \ref{nonbps_fig4}), 
thus we have five possible configurations as shown in Fig.~\ref{nonbps_fig4}. 
The figures in the lower row are the topological charge densities 
for the corresponding five parameter regions, which are calculated 
from (\ref{eq:nonBPS_3instanton}).

Let us now temporarily think of a similar construction of the 
non-BPS exact solution from the BPS solution of three fractional 
instantons in the ${\mathbb C}P^{2}$ model ($N=3$), instead of 
the $N\ge 4$ case. 
The BPS solution in Figs.~\ref{nonbps_fig3} as well as non-BPS 
solutions in Figs.~\ref{nonbps_fig4} now contain three 
fractional instantons resulting in a single instanton with $Q=1$. 
Hence the action density depends on $x_{2}$ and the instanton 
localizes in the two-dimensional $x_1-x_2$ space when they 
get closer. 
Such a situation was discussed in \cite{Dabrowski:2013kba}.

\begin{figure}[htbp]
\begin{center}
\includegraphics[width=0.95\textwidth]{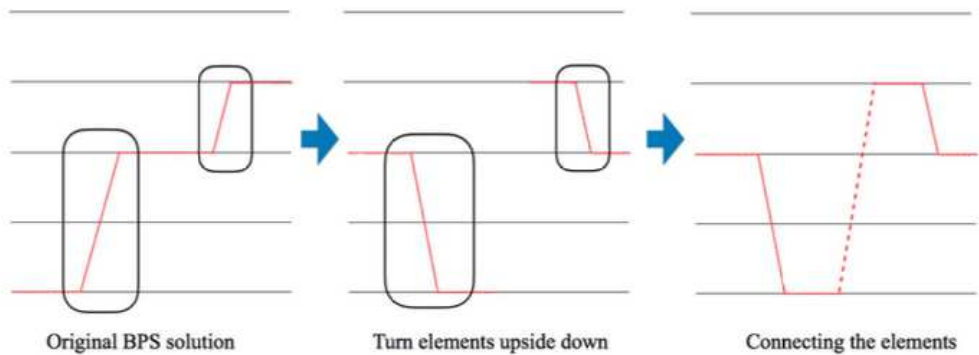}
\end{center}
\caption{Construction of the non-BPS exact solution from the BPS 
solution with a compressed double fractional instanton ${\mathbb C}P^{4}$ model. 
}
\label{nonbps_fig5}
\end{figure}

\begin{figure}[htbp]
\begin{center}
\includegraphics[width=0.95\textwidth]{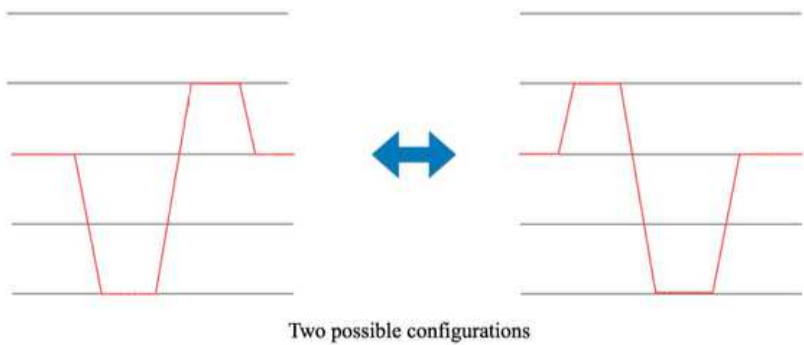}
\end{center}
\caption{Possible configurations are depicted for the non-BPS 
exact solution constructed from the BPS solution with one 
double fractional instanton and one fractional instanton for ${\mathbb C}P^{4}$ model. 
}
\label{nonbps_fig6}
\end{figure}

If two fractional instantons are compressed into a single 
double fractional instanton in the starting BPS solution as 
in the left panel of Fig.~\ref{nonbps_fig5}, we find vanishing 
instanton charge for the non-BPS exact solution, which has only 
two types of configurations as in Fig.~\ref{nonbps_fig6}. 
It is interesting to observe that the compression of the two 
fractional instanton in the starting BPS solution (\ref{eq:3instanton}) 
corresponds to unexpected movement of various constituents 
in the non-BPS solution (\ref{eq:nonBPS_3instanton}) : 
the left-side fractional instanton moves to negative infinity. 
This is the reason why the topological charge in the visible 
(finite) region of $x_1$ changes from unity to zero, resulting 
in Fig.~\ref{nonbps_fig6}. 
This is a manifestation of the unusual clustering property 
of the non-BPS exact solutions compared to the starting BPS 
solutions, that was observed previously \cite{Bolognesi:2013tya}. 
One can explain this unusual property as follows. 
The separation $R$ between constituents in the starting BPS 
solution can be varied by varying the weight of components in 
the BPS solution. 
The separation between constituents in various configurations 
contained in various corners of the moduli space of the non-BPS 
solution are (linearly) related to the separation variables in 
the starting BPS solution. 
It is important to realize that the variable $R$ has a simple 
physical meaning of separation only when it is positive. 
The negative values of the separation variable $R$ between two 
constituents actually means the degree of compression between 
them, since the metric for $R$ in such a region is cigar-like 
\cite{Tong:2002hi}. 
The full-compression of two constituent fractional instantons 
in the starting BPS solution corresponds to large negative values 
$R\to-\infty$. 
The separation variables $R$'s in configurations of the non-BPS 
solution are linearly related to $R$'s in the starting BPS solution. 
Depending on the corners of moduli spaces, they can be positive 
implying the corresponding constituent going to infinity. 
This is observed as the breakdown of clustering property. 
Similar phenomenon occurs between different configurations 
(flipping partners) contained in different corners of moduli 
space.

\begin{figure}[htbp]
\begin{center}
\includegraphics[width=0.95\textwidth]{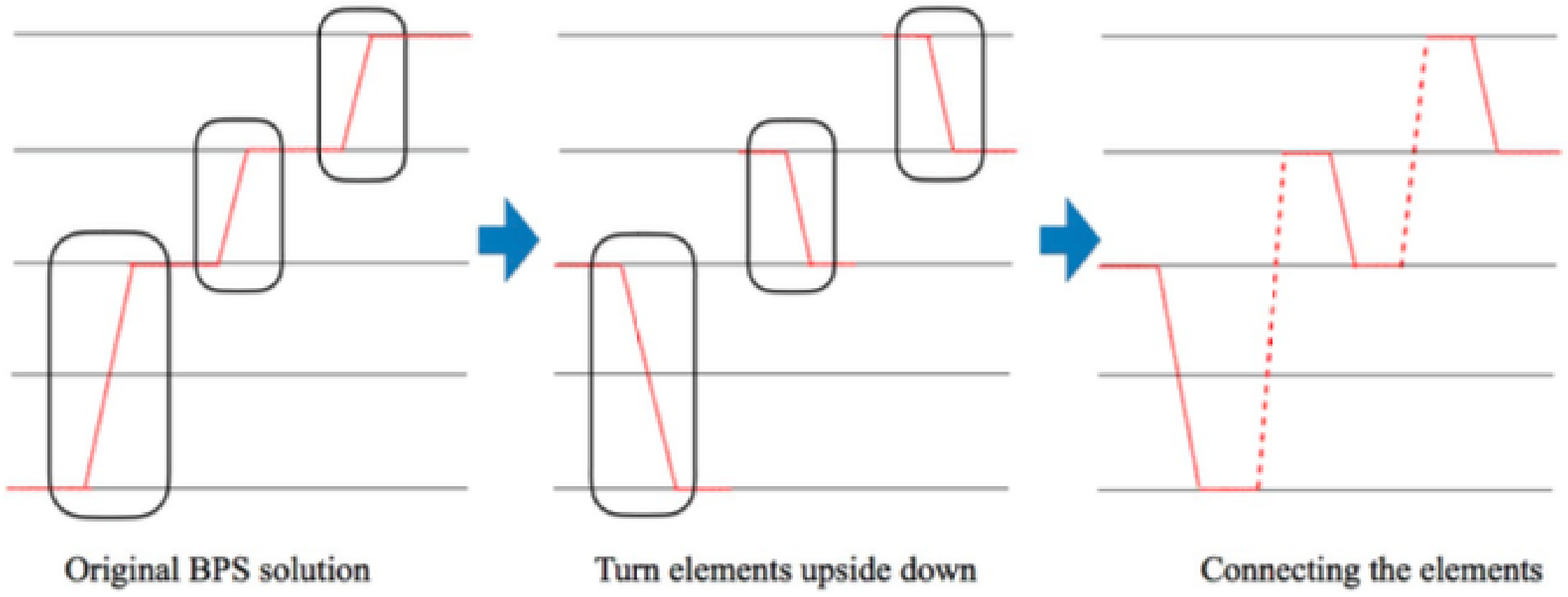}
\end{center}
\caption{Construction of the non-BPS exact solution from the BPS 
solution with one double fractional instanton and two fractional 
instantons for ${\mathbb C}P^{4}$ model. }
\label{nonbps_fig7}
\end{figure}

\begin{figure}[htbp]
\begin{center}
\includegraphics[width=0.95\textwidth]{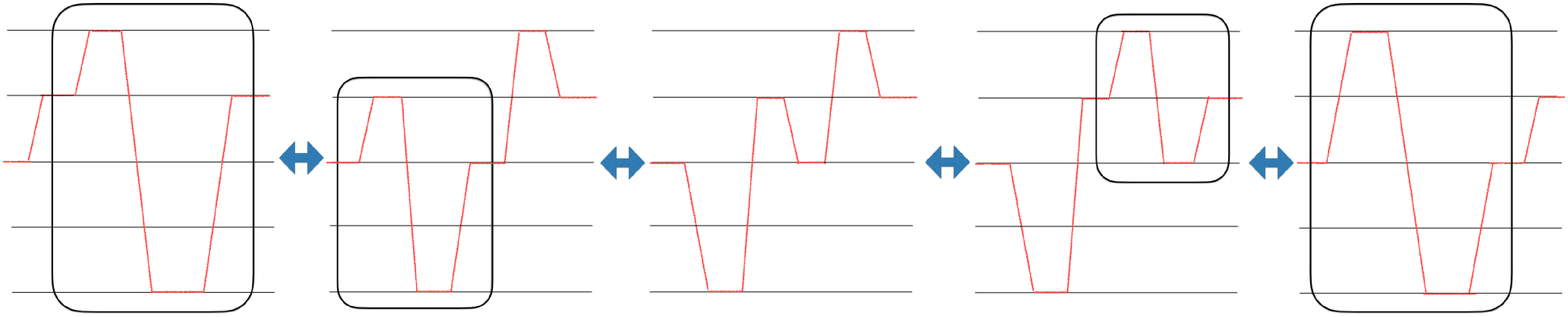}
\end{center}
\caption{Possible configurations are depicted for the non-BPS 
exact solution constructed from the BPS solution with one 
double fractional instanton and two fractional instantons for ${\mathbb C}P^{4}$ model. 
}
\label{nonbps_fig8}
\end{figure}

In Figs.~\ref{nonbps_fig7} and \ref{nonbps_fig8}, we show 
the construction starting from the BPS solution with one 
double fractional instanton and two fractional instantons. 
In this case, the configuration contains several zero-instanton-number 
parts, thus we again have five possible configurations as shown in 
Fig.~\ref{nonbps_fig8}.

\section{Balance of forces in the non-BPS exact solution}
\label{sec:balance}

To understand the reason why no force is acting on various 
constituents of the non-BPS exact solution, we first analyze 
in terms of two-body forces between solitons in various 
configurations and will reveal the presence of three-body forces. 
We consider the ${\mathbb C}P^{2}$ model for simplicity.

\subsection{A bion configuration :  $\bar {\cal I}{\cal I}$
}
We begin with the bion configuration in Fig.~\ref{bion} 
as the simplest non-BPS configuration. 
The effective interaction potential in 
Eq.~(\ref{eq:interation_pot}) between the two constituents 
for large separation $R$ is attractive for the relative 
phase $|\theta_b|<\pi/2$, but is repulsive for $|\theta_b|>\pi/2$. 
Consequently the total action at $\theta_b=\pi/2$ is flat 
along the $R$ direction for large separations, satisfying 
a necessary condition to be an exact solution. 
However, the total action has a positive $\theta_b$ derivative, 
and the bion cannot be a stationary point of the action. 
We need to remedy the strong $\theta_b$ dependence to achieve 
the balance of force. 

\begin{figure}[htbp]
\begin{center}
\includegraphics[width=0.8\textwidth]{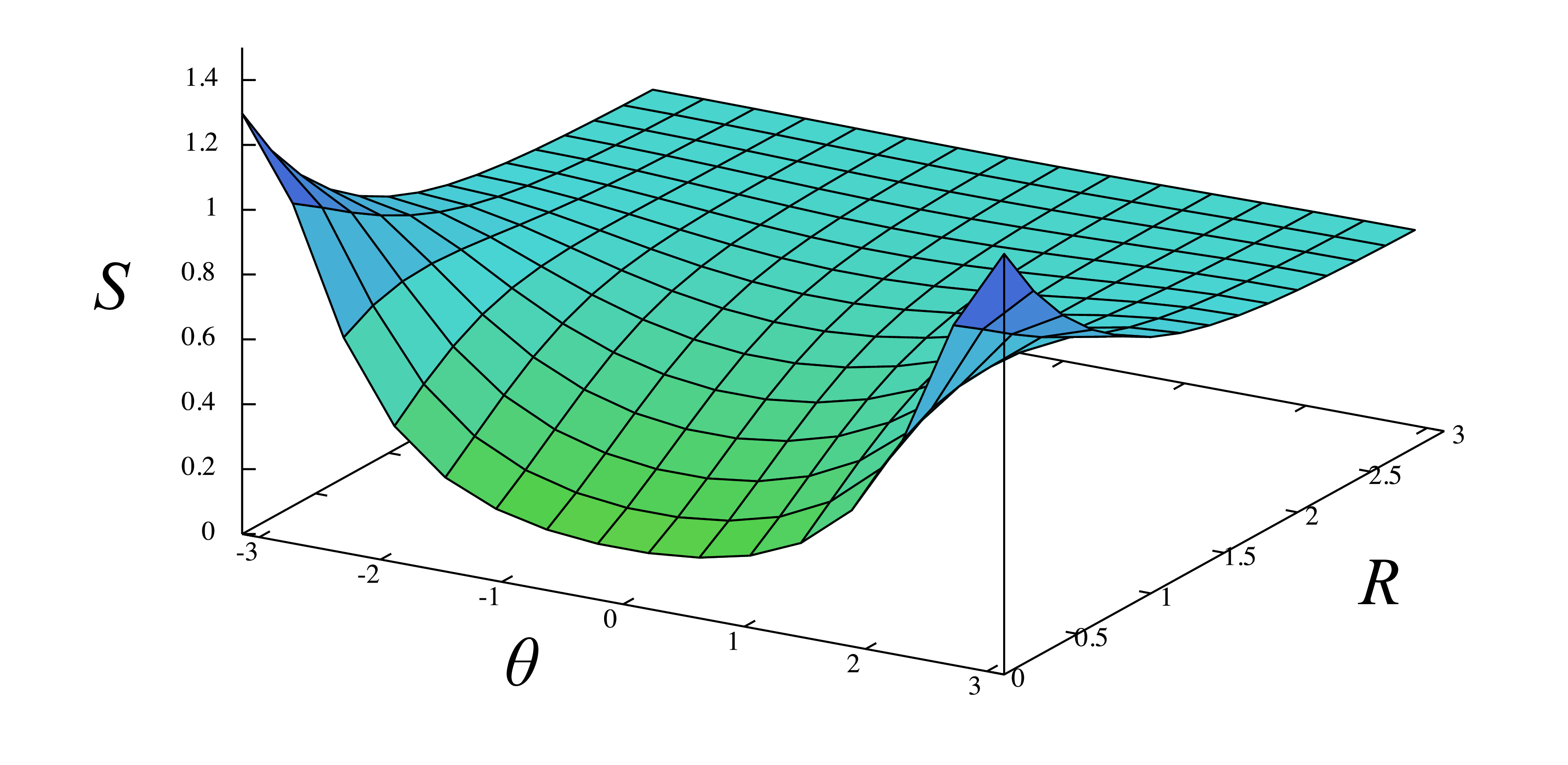}
\end{center}
\caption{The total action of bion configuration (\ref{eq:bion}) 
as a function of the separation $R$ and the relative phase 
$\theta_{b}\equiv \theta$ in ${\mathbb C}P^{2}$ model.}
\label{bionS}
\end{figure}

These features of the two body force is clearly visible in 
Fig.~\ref{bionS}, where we depict the total action of the bion 
in Eq.~(\ref{eq:bion}) as a function of the separation $R$ 
and the relative phase $\theta_b$.

\subsection{A fractional anti-instanton and a bion : 
$\bar {\cal I} \bar {\cal I} {\cal I}$}

\begin{figure}[htbp]
\begin{center}
\includegraphics[width=0.4\textwidth]{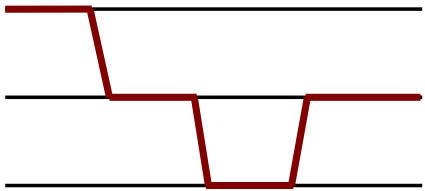}
\end{center}
\caption{The configuration of fractional anti-instanton + 
bion in (\ref{eq:bI-bion}) is depicted for ${\mathbb C}P^{2}$ model.}
\label{bI-bion}
\end{figure}

In order to eliminate the strong dependence on the relative 
phase, we next consider the addition of a fractional 
anti-instanton to the bion configuration from the small $x_{1}$ 
side as shown in Fig.~\ref{bI-bion}.

The configuration within our ansatz is given by
\begin{equation}
\omega_{\bar {\cal I}\bar {\cal I}{\cal I}} = 
\begin{pmatrix}
a e^{i\theta_a}e^{-{2\pi\over{3}}z},\,\,&
1+b e^{i\theta_b} e^{-{2\pi\over{3}}(z+\bar z)} 
,\,\,&
g e^{i\theta_g}e^{-{2\pi\over{3}}(z+2\bar{z})}
\end{pmatrix}\,,
\label{eq:bI-bion}
\end{equation}
with $a,b,g\in {\mathbb R}$ and $-\pi< \theta_a,\theta_b,\theta_g \leq \pi$. 
The action and topological charge densities are independent of
$\theta_a$ and $\theta_g$. 
The separation of two fractional anti-instantons in the left is 
$R_{\bar{\cal I}\bar{\cal I}}=\frac{3}{2\pi}\log\frac{b^2}{ag}$ and 
the separation between the middle fractional anti-instanton 
and the instanton is $R_{\rm bion}=\frac{3}{2\pi}\log\frac{a^2}{b}$. 
For large separations $R_{\bar{\cal I}\bar{\cal I}}$ and $R_{\rm bion}$, 
the total action depends only on $R_{\rm bion}$ and the 
phase $\theta_b$, and is independent of $R_{\bar{\cal I}\bar{\cal I}}$, 
reflecting the absence of static forces between fractional 
instantons. 
Therefore we obtain no improvement by adding the fractional 
anti-instanton, as long as they are well separated.

\begin{figure}[htbp]
\begin{center}
\includegraphics[width=0.4\textwidth]{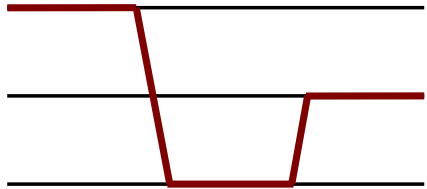}
\end{center}
\caption{The configuration (\ref{2I-1bI}) of a double 
fractional anti-instanton + a fractional instanton.}
\label{2I-1bIfig}
\end{figure}

If we let $b\to 0$ in Eq.~(\ref{eq:bI-bion}), the 
left-most fractional instanton is compressed into the 
middle fractional instanton. 
In the completely compressed limit $b=0$, we obtain the 
unnormalized vector field $\omega$ as 
\begin{equation}
\omega_{(\bar {\cal I}\bar {\cal I}){\cal I}} = 
\begin{pmatrix}
a e^{i\theta_a}e^{-{2\pi\over{3}}z},\,\,&
1,\,\,&
g e^{i\theta_g}e^{-{2\pi\over{3}}(z+2\bar{z})}
\end{pmatrix}\,.
\label{2I-1bI}
\end{equation}
The configuration contains one fully-compressed double 
fractional anti-instanton $(\bar {\cal I}\bar {\cal I})$ 
and one fractional instanton ${\cal I}$ 
as shown in Fig.~\ref{2I-1bIfig} ($Q=1/3$). 
The vacuum transitions occur $(0,0,1)\to(1,0,0)\to (0,1,0)$, 
as $x_1$ varies from $x_1=-\infty$ to $x_1=\infty$. 
The separation between the two constituents is given by 
$R=(3/(4\pi))\log(a^3/g))$. 
We emphasize that this configuration is independent 
of all the phase parameters $\theta_a, \theta_b, \theta_g$, 
since the action density and topological charge density is 
independent of phases $\theta_a, \theta_g$ for each component, 
and the phase $\theta_b$ disappears in the limit of $b\to 0$. 
Therefore we succeeded to eliminate the strong phase dependence 
of the effective interaction potential by just compressing 
another fractional anti-instanton with the fractional anti-instanton 
constituent of the bion. 
We can understand the disappearance of the phase dependence of 
the effective interaction potential when the two fractional 
instanton become closer (compressed), as a result of strong 
deformation of the internal structure of the compressed 
double fractional anti-instantons $(\bar {\cal I}\bar {\cal I})$ 
compared to two individual fractional anti-instantons. 
For large values of separation $R$, the phase dependent two-body 
forces like in Eq.~(\ref{eq:interation_pot}) is no longer acting.

\begin{figure}[htbp]
\begin{center}
\includegraphics[width=0.85\textwidth]{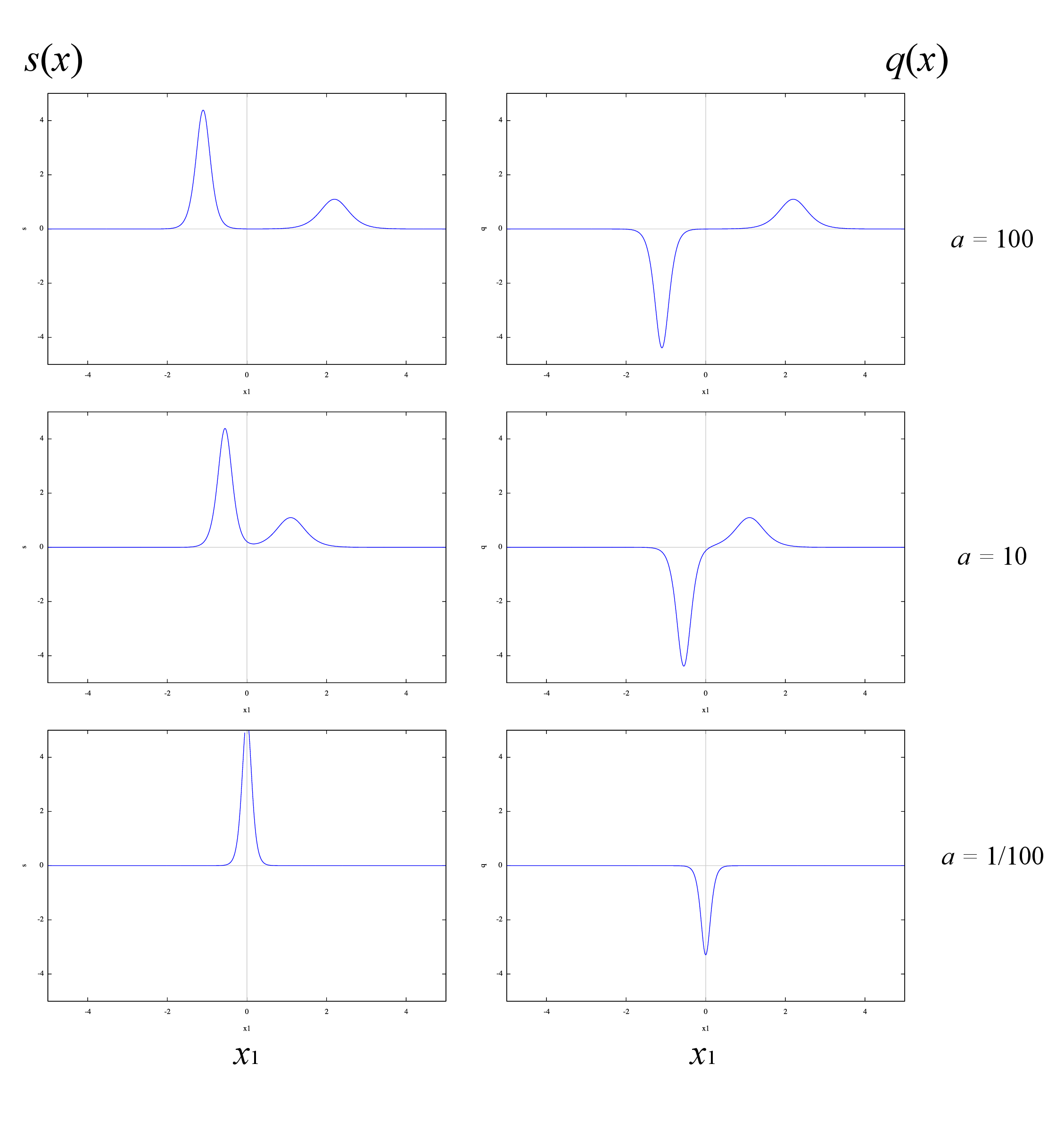}
\end{center}
\caption{The action density (left) and topological charge 
density (right) of the configuration ${(\bar {\cal I}\bar {\cal I}){\cal I}}$ 
in Eq.~(\ref{2I-1bI}) for three values of the parameter $a=100,10,1/100$ 
(from top to bottom) with $g=1$ fixed.
}
\label{2I-1bIden}
\end{figure}

However, the approximation with two-body forces between the 
compressed double fractional anti-instanton and a fractional 
instanton is no longer valid, 
if we compress these two constituents further by letting 
$a$ small with $g$ (the center of mass position) fixed. 
In Fig.~\ref{2I-1bIden}, we depict the action density (left) 
and topological charge density (right) for three values of 
the parameter 
$a=1/100,1/10,100$ (from top to bottom) with $g=1$ fixed.
We can clearly see that 
the total action is not equal to an absolute value of the topological 
charge, even when the two constituents are compressed ($a=100$).

\begin{figure}[htbp]
\begin{center}
\includegraphics[width=0.7\textwidth]{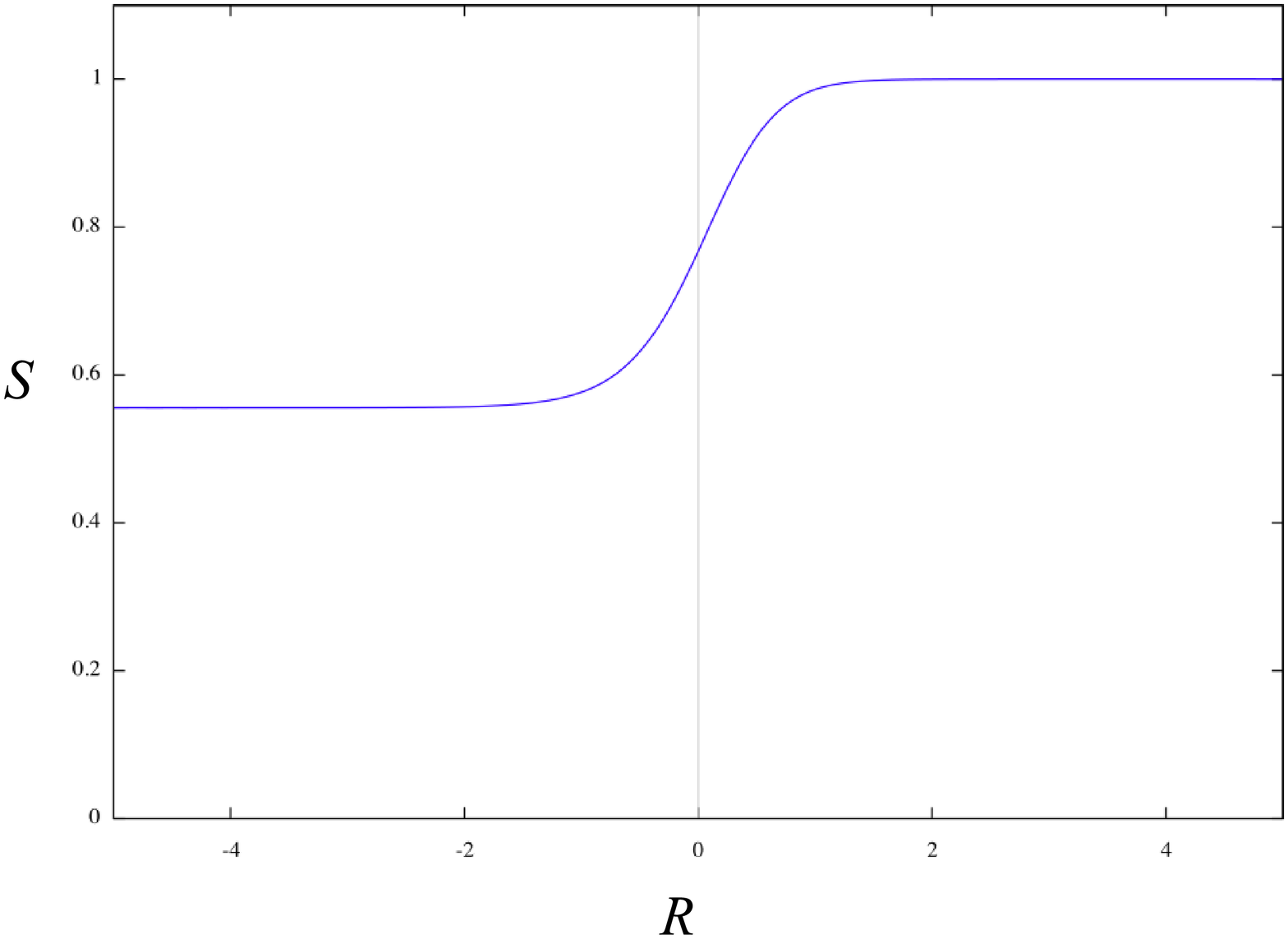}
\end{center}
\caption{The total action of the configuration 
${(\bar {\cal I}\bar {\cal I}){\cal I}}$ 
in Eq.~(\ref{2I-1bI}) as a function of the separation 
$R=(3/4\pi)\log(a^3/g)$.}
\label{2I-1bIS}
\end{figure}

In Fig.~\ref{2I-1bIS}, we depict the total action as a function 
of the separation $R=(3/4\pi)\log(a^{3}/ g)$. 
The smaller $a$ corresponds to the smaller separation $R$ 
between the constituents. 
The action starts from $S=1$ and approaches a composite soliton 
of compressed double fractional anti-instanton and an instanton
with the action $S=5/9$ and $Q=1/3$.
Therefore the configuration experiences a (weak) 
attractive force at small separations $R$. 
We need more modifications of the configuration to 
achieve the balance of force.

\subsection{An additional fractional instanton 
: ${\cal I}(\bar{\cal I}\bar{\cal I}){\cal I}$
}
\label{sc:pseudoSolution}

To see the importance of considering approximation beyond the 
two-body forces, let us modify the configuration in 
Eq.~(\ref{2I-1bI}) by adding one more fractional anti-instanton 
to the left of the compressed double fractional instanton, 
which can be parametrized as 
\begin{equation}
\omega_{{\cal I}(\bar{\cal I}\bar{\cal I}){\cal I}} = 
\begin{pmatrix}
a e^{-{2\pi\over{3}}z},\,\,&
1+d e^{i(\theta+\pi)}
e^{-{4\pi\over{3}}(z+\bar{z})},\,\,&
g e^{-{2\pi\over{3}}(z+2\bar{z})}
\end{pmatrix}\,.
\label{2I-1bImod2}
\end{equation}
We omit the irrelevant phase factor in the first component, and 
redefine the relative phase in the second component as 
$\theta \to \theta+\pi$, for later convenience. 
The configuration ${\cal I}(\bar{\cal I}\bar{\cal I}){\cal I}$ 
 consists of a compressed double fractional anti-instanton 
$(\bar{\cal I}\bar{\cal I})$ 
sandwiched between fractional instantons and looks very similar to 
Fig.~\ref{nonbpscp2}(a), one of configurations contained in the 
simplest non-BPS exact solution (\ref{eq:nonbps}) with $S=4/3,Q=0$.
The action density depends on the relative phase $\theta$ in 
the second component, the separation $R_{1}=(3/4\pi)\log (g^{3} /(a d^{2}))$ 
between $(\bar{\cal I}\bar{\cal I})$ and the left fractional instanton 
and the separation $R_{2}=(3/4\pi)\log(a^{3}/g)$ between 
$(\bar{\cal I}\bar{\cal I})$ and the right fractional instanton.

If we restrict the three parameters $a, d, g$ in terms of 
two parameters $l_1, l_2$ as 
$a=2l_1/l_{2},\,\,d=l_{1}^2,\,\,g=-2l_{1}^{2}/l_{2}$, we obtain 
\begin{equation}
\omega_{{\cal I}(\bar{\cal I}\bar{\cal I}){\cal I}} = 
\begin{pmatrix}
\frac{2l_{1}}{l_2}e^{-{2\pi\over{3}}z},\,\,&
1-l_{1}^2 e^{i\theta}e^{-{4\pi\over{3}}(z+\bar{z})},\,\,&
-\frac{2l_{1}^{2}}{l_{2}} e^{-{2\pi\over{3}}(z+2\bar{z})}
\end{pmatrix}\,. 
\label{2I-1bImod2-2}
\end{equation}
If the phase $\theta$ vanishes, the configuration becomes 
equivalent to dominant terms of the non-BPS exact solution 
(\ref{eq:nonbps}) in a parameter region $2l_1 \gg l_2^2$, 
representing one of the 
configurations in Fig.~\ref{nonbpscp2}(a). 
We see that two separations are equal : 
$R=R_{1}=R_{2}=(3/4\pi)\log(4l_{1}/l_{2}^{2})$. 
Thus, the energy density is 
symmetric under the reflection around the middle compressed 
double fractional instanton. 
We call this symmetry as {\it reflection symmetry}. 
If the reflection symmetry is broken, the balance of force 
fails, as we will see later. 
Thus, the reflection symmetry is one of the essential properties 
of the non-BPS exact solutions.

\begin{figure}[htbp]
\begin{center}
\includegraphics[width=0.8\textwidth]{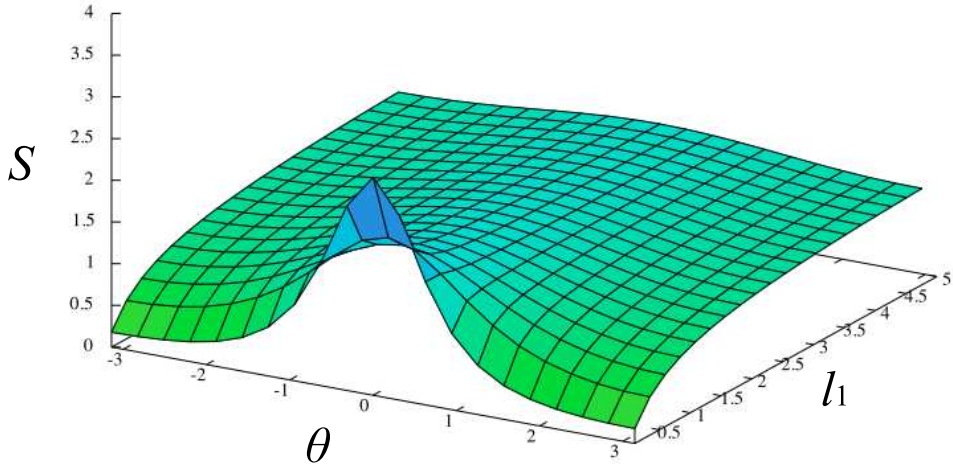}
\end{center}
\caption{The total action of the configuration 
 ${\cal I}(\bar{\cal I}\bar{\cal I}){\cal I}$ in 
Eq.~(\ref{2I-1bImod2-2}) as a function of the phase parameter 
$\theta$ and the separation parameter $l_{1}$ with $l_{2}=1$ 
fixed: $R=(3/4\pi)\log(4l_{1}/l_{2}^{2})$. 
}
\label{Non-solution}
\end{figure}

Fig.~\ref{Non-solution} shows the total action of 
(\ref{2I-1bImod2-2}) as a function of the phase parameter 
$\theta$ and the parameter $l_{1}$ with $l_{2}=1$ 
fixed, which gives the separation $R=(3/4\pi)\log(4l_{1}/l_{2}^{2})$. 
For $\theta=\pi$, in which the two terms in the second 
component in (\ref{2I-1bImod2-2}) have the same sign, 
the action decreases most rapidly as $l_{1}$ gets 
smaller ($R$ gets smaller), thus the effective force is attractive. 
For $\theta=0$, in which the two terms in the second component 
have the opposite sign, the action increases as $l_{1}$ gets smaller 
($R$ gets smaller), thus the effective force is repulsive. 
If one wants to achieve the balance of force to obtain a solution 
of field equations, it is a fatal flaw to have an attractive 
interaction, since the configuration tends to decay into vacuum 
in our case of topologically trivial sector. 
This is in accordance with the sign of the two terms of the 
second component in the non-BPS exact solution (\ref{eq:nonbps}).

It is important to realize that the force we have discussed here 
is a three-body force involving three constituent solitons 
: $\theta$ is the relative phase 
between the two terms corresponding to the left-most and 
right-most vacua. 
If one focuses on the two constituents, for example, the left 
anti-instanton and central double instanton, 
they have attractive force 
for any $\theta$ (only at small separations), as we have seen. 
Moreover, the reflection symmetry is also important, since the 
two-body force between constituents with shorter separation 
dominates to give the attractive force 
whenever the reflection symmetry is broken. 
This is why the reflection symmetry is also essential in the 
non-BPS exact solution.

Now, the only question is how the non-BPS exact solution 
suppresses the repulsive force resulting in the total action 
$S=4/3$ 
for any values of $R$ at $\theta=0$ and the vanishing 
derivative with respect to $\theta$. 
Compared to the present case, the non-BPS exact solution 
(\ref{eq:nonbps}) has the other terms, which produce the 
flipping-partner configuration for $l_{2}^2  > 2l_{1}$ as shown 
in Fig.~\ref{nonbpscp2}.
We have to conclude that these terms for the flipping partner 
also work to avoid the increase of the total action 
for the parameter range $2l_{1} > l_{2}^{2}$, even though these 
terms appear to be irrelevant at a glance.
Thus, the existence of flipping partners is also essential 
in the non-BPS exact solution.

\subsection{Essential properties of non-BPS exact solutions}

To sum up, we list essential properties of the non-BPS exact 
solution (\ref{eq:nonbps}), which give the constant total 
action without annihilation into vacuum in the entire moduli 
space. 

\begin{enumerate}

\item
{\bf Relative sign}

The terms in the solution have the appropriate relative signs : 
these signs serve to suppress the dominant two-body attractive 
forces, and to provide many-body repulsive forces suppressing 
a decay into vacuum. 

\item
{\bf Reflection symmetry}

The reflection symmetry around the middle compressed double 
fractional instanton prevents the dominance of attractive 
two-body forces. 

\item
{\bf Flipping partner}

The non-BPS exact solution makes a transition between two 
seemingly different configurations (flipping partners), 
as moduli parameters are varied across the point where all the 
constituents get closer. 
They never annihilate each other even though no topological 
quantum number guarantees the flatness of the total action. 
This transition is essential to provide the many-body 
forces to cancel the two-body and three-body forces in 
configurations without the flipping partners.

\end{enumerate}

All the configurations we have discussed including the non-BPS 
exact solutions correspond to some special cases of the 
generic two-bion ansatz in (\ref{eq:2bion}). 
We emphasize that this ansatz provides the multi-instanton 
computation needed to achieve the resurgence. 
Therefore, the non-BPS exact solutions contribute to a part 
of the multi-instanton moduli integral relevant to the resurgence 
theory.



\section{Local and Global stability }
\label{sec:NM}

It has been pointed out that the non-BPS exact solutions have 
unstable modes \cite{Din:1980jg}. 
In this section, we analyze the local stability of the non-BPS exact 
solution in Eq.~(\ref{eq:nonbps}) to understand the physical 
meaning of unstable negative modes within the parametrization 
of our general ansatz in Eq.~(\ref{eq:2bion}). 
We focus only on the simplest non-BPS solution in the ${\mathbb C}P^{2}$ model.
As we see in Sec.~\ref{sc:ansatz}, this ansatz has nine 
dimensional space of relevant parameters : four relative 
phases and five separations. 
However, different sets of parameters are relevant to analyze 
the stability of four different configurations in 
Fig.~\ref{G2bion}(a), (b), (c), and (d).  
We need to concentrate on two types of configurations 
(b) and (c), since only these two (flipping 
partners) arise in the vicinity of 
the non-BPS exact solution. 
The configuration in Fig.~\ref{G2bion}(b) depends on only 
two phases $b$ and $d$, and three separations $R_{21}', 
R_{32}', R_{43}'$. 
Another configuration in Fig.~\ref{G2bion}(c) can be 
obtained by the reflection symmetry. 
Therefore we need to consider only the five dimensional 
parameter space. 
The other two relative phases $(a, c)$ and $(f, g)$ are only 
relevant to examine the two (anti-)bion configurations in 
Fig.~\ref{G2bion}(a) and (d), which requires large 
deformations from the non-BPS exact solution. 
We also examine globally other possible 
saddle points within our ansatz, to which the exact 
solution can reach along 
these negative modes. 


\subsection{Phase negative modes I}

\begin{figure}[htbp]
\begin{center}
\includegraphics[width=0.7\textwidth]{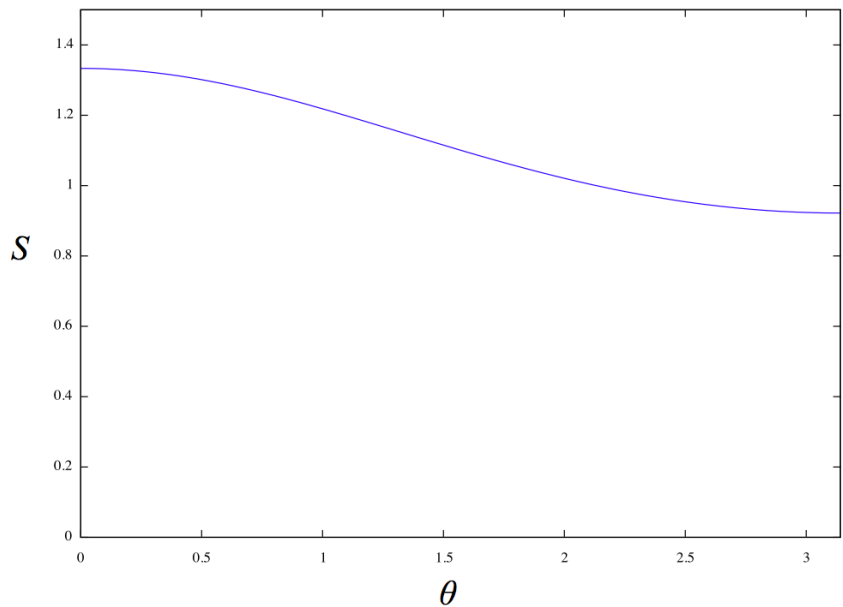}
\end{center}
\caption{ 
The total action of the configuration in Eq.~(\ref{mod00}) 
as a function of $\theta$ with 
$l_{1}=l_{2}=1$ fixed.} 
\label{thetaNM}
\end{figure}

In this subsection we violate the first of the essential properties 
({\it relative sign}) of non-BPS solutions by introducing a 
phase $\theta$ to a term in the second component of the exact 
solution in Eq.~(\ref{eq:embedding}), 
keeping the common separation $R$ 
for left and right separations (reflection-symmetry)  
\begin{equation}
\omega = 
\begin{pmatrix}
\frac{2l_{1}}{l_2}e^{-{2\pi\over{3}}z}
+l_{1}l_{2}e^{-{2\pi\over{3}}(2z+\bar{z})},\,\, &
1- e^{i\theta}l_{1}^{2}e^{-{4\pi\over{3}}(z+\bar{z})},\,\,&
-l_{2}e^{-{2\pi\over{3}}\bar{z}}
-\frac{2l_{1}^{2}}{l_2}e^{-{2\pi\over{3}}(z+2\bar{z})}
\end{pmatrix}\,.
\label{mod00}
\end{equation}
We here omit the common phases $\theta_{1},\theta_{2}$ of 
the two terms in the first and second components which are 
included in (\ref{eq:nonbps}), since the above configurations 
are independent of them. 
The argument in Sec.\ref{sc:pseudoSolution} suggests that the total 
action decreases due to the attractive force between the 
constituents if we flip the relative sign ($\theta=0 \to \pi$) 
between the second component in (\ref{eq:nonbps}). 
\begin{figure}[htbp]
\begin{center}
\includegraphics[width=0.8\textwidth]{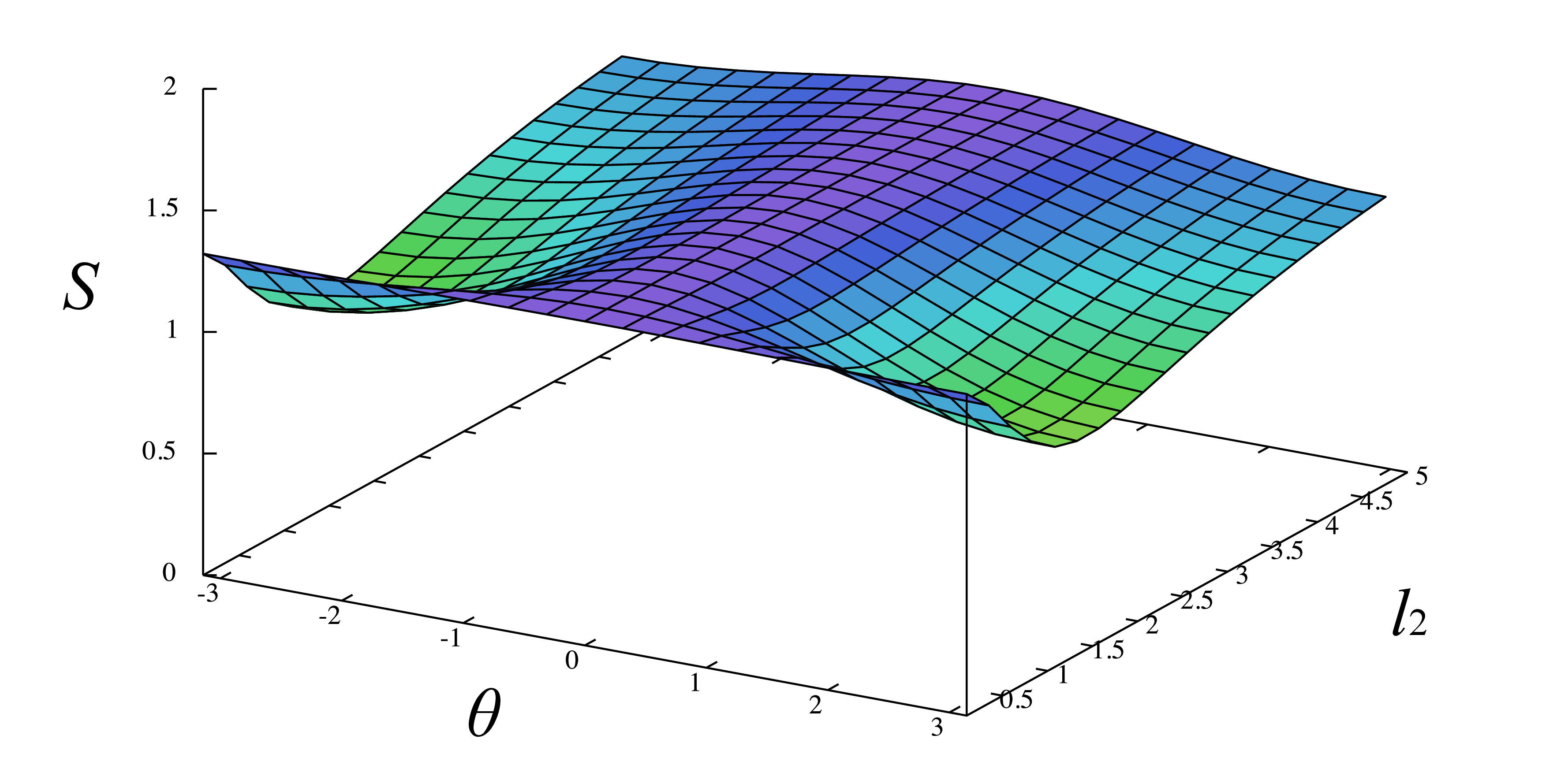}
\end{center}
\caption{
The total action for the configuration in Eq.~(\ref{mod00}) 
as a function of $\theta$ and $l_{2}$ with $l_{1}=1$ fixed 
(we note $R=(3/4\pi)\log(l_{2}^{2}/l_{1})$ for $l_{2}^{2} > 2l_{1}$).}
\label{RthetaNM}
\end{figure}
In Fig.~\ref{RthetaNM}, we show the total action as a function 
of the phase $\theta$ and $l_{2}$ with $l_{1}=1$ fixed 
(we note $R=(3/4\pi)\log(l_{2}^{2}/l_{1})$ for $l_{2}^{2} > 2l_{1}$).
We also display the total action as a function of $\theta$ with 
the fixed separation $l_{1}=1,l_{2}=1$ in Fig.~\ref{thetaNM}. 
We find that the total action decreases when we change 
$\theta$ from $0$ to $\pi$, indicating a negative mode along 
$theta$ direction. 
For the large separation $2l_{1}\gg l_{2}^{2}$ or $2l_{1}\ll l_{2}^{2}$,
the total action dependence on $\theta$ becomes small 
(but exists), since the magnitude of the two-body force decreases 
for small separations.

\subsection{Negative modes for asymmetric separation}

In this subsection we violate the second of the essential 
properties ({\it reflection symmetry}) of the non-BPS exact 
solution by introducing two multiplicative factors 
$\gamma, \gamma' \in {\mathbb R}$ 
in order to change the left and right separations 
\begin{equation}
\omega = 
\begin{pmatrix}
\frac{2l_{1}}{l_2}e^{-{2\pi\over{3}}z}
+l_{1}l_{2}e^{-{2\pi\over{3}}(2z+\bar{z})},\,\, &
\gamma' -\gamma l_{1}^{2}e^{-{4\pi\over{3}}(z+\bar{z})},\,\,&
-l_{2}e^{-{2\pi\over{3}}\bar{z}}
-\frac{2l_{1}^{2}}{l_2}e^{-{2\pi\over{3}}(z+2\bar{z})}
\end{pmatrix}\,.
\label{mod01}
\end{equation}
For $\gamma, \gamma' \not=1$, the configurations are no longer 
solutions.

In the parameter region $2l_{1}>l_{2}^{2}$, we have a 
configuration similar to that in Fig.~\ref{nonbpscp2}(a). 
However, the separation 
\begin{equation}
R_L=\frac{3}{4\pi}\log(\frac{4l_1}{\gamma^2l_2^2})\,,
\label{eq:leftSeparation}
\end{equation}
between the left instanton and the middle compressed double 
anti-instanton is decreased by $\gamma > 1$. 
Similarly, the separation 
\begin{equation}
R_R=\frac{3}{4\pi}\log(\frac{4l_1}{{\gamma'}^2l_2^2})\,,
\end{equation}
between the middle compressed double anti-instanton and the right 
instanton is decreased by $\gamma' > 1$.

\begin{figure}[htbp]
\begin{center}
\includegraphics[width=0.85\textwidth]{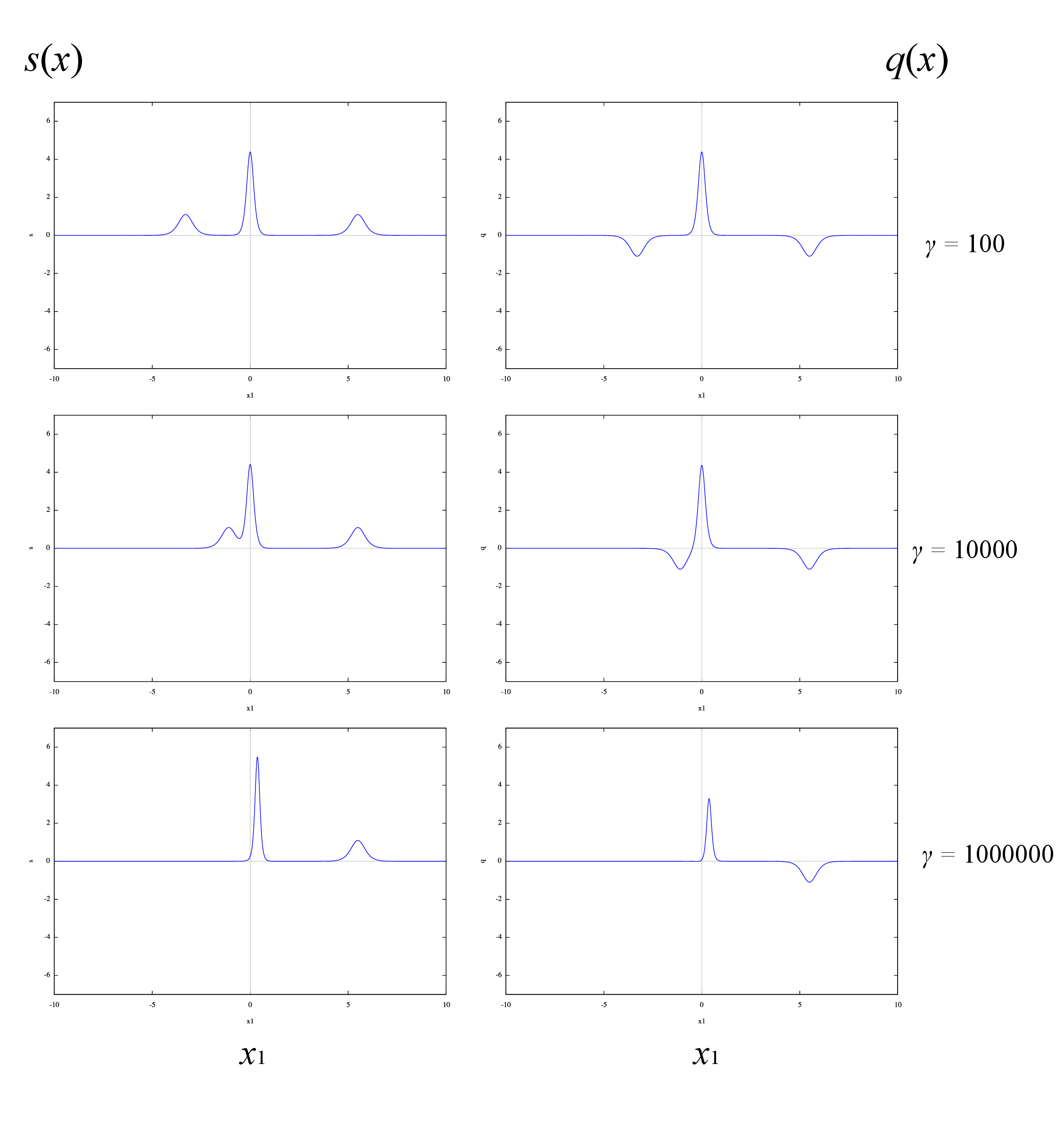}
\end{center}
\caption{
The action density and topological charge density 
of the configuration in Eq.~(\ref{mod01}) are 
shown in left and right columns, respectively. 
From top to bottom, densities are shown for three values 
of the parameter $\gamma=100,10000,1000000$, with $\gamma'=1$ 
and $l_{1}=1$ $l_{2}=100000$ fixed.}
\label{snm1}
\end{figure}

In the parameter region $2l_{1}<l_{2}^{2}$, we have another 
configuration similar to that in Fig.~\ref{nonbpscp2}(b). 
The separation 
\begin{equation}
R'_L=\frac{3}{4\pi}\log(\frac{l_2^2}{\gamma^2l_1})\,,
\end{equation}
between the left anti-instanton and the middle compressed double 
instanton is decreased by $\gamma > 1$, and 
the separation 
\begin{equation}
R'_R=\frac{3}{4\pi}\log(\frac{l_2^2}{{\gamma'}^2l_1})\,,
\end{equation}
between the middle compressed double instanton and the right 
anti-instanton is decreased by $\gamma' > 1$.

Fig.~\ref{snm1} shows the action and topological charge 
densities of the configuration in Eq.~(\ref{mod01}) 
with $\gamma>1, \gamma'=1$ and $2l_{1}<l_{2}^{2}$. 
When the right separation $R_{R}'$ is large and fixed, we 
can regard the configuration as the addition of noninteracting 
fractional anti-instanton to the configuration of compressed 
double fractional instanton and an anti-instanton similar to 
that in Eq.~(\ref{2I-1bI}). 
Therefore, the total action, which is originally $S=4/3$ for 
the non-BPS exact solution (\ref{eq:nonbps}), decreases and 
ends up with 
\begin{equation}
S=5/9 + 1/3 = 8/9 \,,
\end{equation}
\begin{figure}[htbp]
\begin{center}
\includegraphics[width=0.6\textwidth]{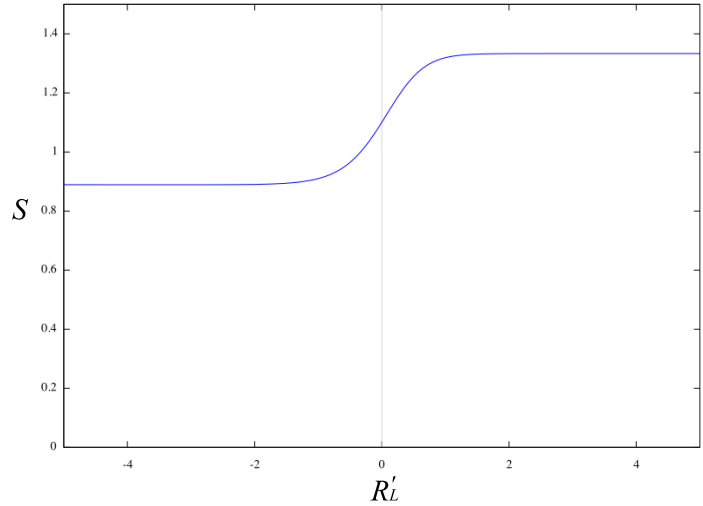}
\end{center}
\caption{Total action of configuration in Eq.~(\ref{mod01}) 
as a function of 
$R_L'= {3\over{4\pi}}\log{l_{2}^{2}\over{\gamma^{2} l_{1}}}$ 
with $\gamma' =1$, $l_{1}=1$, $l_{2}=10000$ fixed.
The non-BPS solution with $\gamma=1$ correspond to $R_L' \sim 4.4$}
\label{snm-S}
\end{figure}
for the $\gamma\to \infty$ limit, where the left fractional 
anti-instanton is fully compressed with the double fractional 
instanton.  
The value $5/9$ of the action comes from the compressed part 
of fractional anti-instanton + double fractional 
instanton as found in Fig.~\ref{2I-1bIS}.

Figure \ref{snm-S} shows the total action as a function of 
separation $R_L$ in Eq.~(\ref{eq:leftSeparation}) 
between the left anti-instanton and the middle double fractional 
instanton. 
Thus we find that increasing $\gamma$ with $\gamma'=1$ fixed 
is a negative mode. 
Moreover, the increase of $\gamma$ leads to a configuration 
of fully compressed fractional instanton + double fractional 
anti-instanton together with a (almost) noninteracting 
fractional instanton to their right.

\begin{figure}[htbp]
\begin{center}
\includegraphics[width=0.7\textwidth]{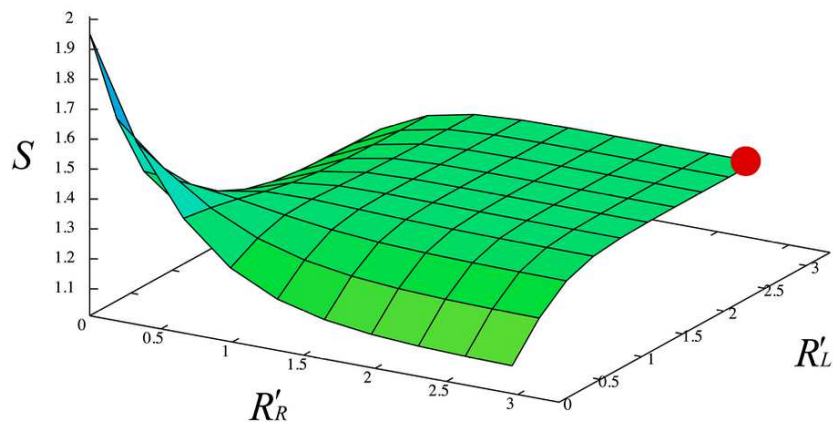}
\end{center}
\caption{The total action as a function of 
$R_{R}' ={3\over{4\pi}}\log{l_{2}^{2}\over{{\gamma'}^{2} l_{1}}}$ 
and $R_{L}' ={3\over{4\pi}}\log{l_{2}^{2}\over{\gamma^{2} l_{1}}}$ 
for (\ref{mod03}) with $\theta=0$.
In the calculation, we fix $l_{1}=1$ and $l_{2}=1000$, and vary $\gamma, \gamma'$.
The non-BPS solution (which means $\gamma=\gamma'=1$) corresponds to 
$R_{R}'=R_{L}'\sim3.3$, the red point at the corner.}
\label{Snm2d-t0}
\end{figure}

Similarly, increasing $\gamma'$ with $\gamma$ fixed 
is found to be a negative mode, leading to a configuration 
of fully compressed fractional double fractional 
anti-instanton + instanton together with a (almost) noninteracting 
fractional instanton to their left.

We next consider the deformation combining (\ref{mod00}) and (\ref{mod01}),
\begin{align}
\omega = 
\begin{pmatrix}
\frac{2l_{1}}{l_2}e^{-{2\pi\over{3}}z}
+l_{1}l_{2}e^{-{2\pi\over{3}}(2z+\bar{z})},\,\, &
\gamma' -\gamma e^{i\theta} l_{1}^{2}e^{-{4\pi\over{3}}(z+\bar{z})},\,\,&
-l_{2}e^{-{2\pi\over{3}}\bar{z}}
-\frac{2l_{1}^{2}}{l_2}e^{-{2\pi\over{3}}(z+2\bar{z})}
\end{pmatrix}\,.
\label{mod03}
\end{align}
As we have discussed in the previous section, this 
configuration behaves differently with $\theta=0$ and $\theta=\pi$.
In both cases, the increase of $\gamma>1$ with $\gamma' = 1$ or
the increase of $\gamma' >1$ with $\gamma=1$ decreases the 
total action from $S=4/3$ to $S=8/9$ since the double fractional 
instanton and the fractional anti-instanton 
at each side have attractive interaction.
However, things change if one increases both $\gamma$ and 
$\gamma'$ with $\gamma=\gamma'$. 

For $\theta=0$, the three instanton constituents have 
effectively repulsive interaction as we have seen in 
Fig.~\ref{Non-solution}. 
Thus the total action increases as one increases $\gamma=\gamma' >1$.
In Fig.~\ref{Snm2d-t0}, we depict the total action as a function 
of $R_{R}'={3\over{4\pi}}\log{l_{2}^{2}\over{{\gamma'}^{2} l_{1}}}$ 
and $R_{L}'={3\over{4\pi}}\log{l_{2}^{2}\over{\gamma^{2} l_{1}}}$.
In the calculation, we fix $l_{1}=1$ and $l_{2}=1000$, and vary $\gamma, \gamma'$.
Along $R_{R}'=R_{L}'\,\,\to\, \,0$, the action increases.

\begin{figure}[htbp]
\begin{center}
\includegraphics[width=0.8\textwidth]{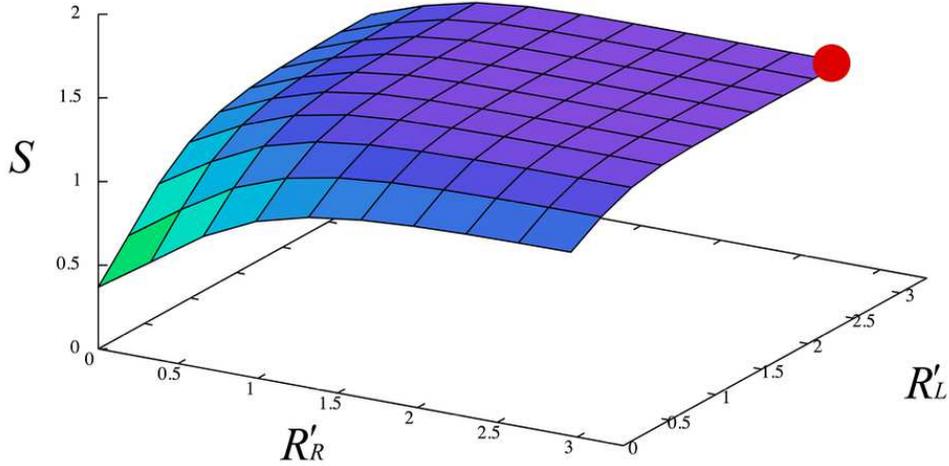}
\end{center}
\caption{The total action of Eq.~(\ref{mod03}) as a function of 
$R_{R}'={3\over{4\pi}}\log{l_{2}^{2}\over{{\gamma'}^{2} l_{1}}}$ 
and $R_{L}'={3\over{4\pi}}\log{l_{2}^{2}\over{\gamma^{2} l_{1}}}$ 
for (\ref{mod03}) with $\theta=\pi$.
In the calculation, we fix $l_{1}=1$ and $l_{2}=1000$, and vary $\gamma, \gamma'$.
The red point corresponds to $\gamma=\gamma' =1$.}
\label{Snm2d-tpi}
\end{figure}
\begin{figure}[htbp]
\begin{center}
\includegraphics[width=0.6\textwidth]{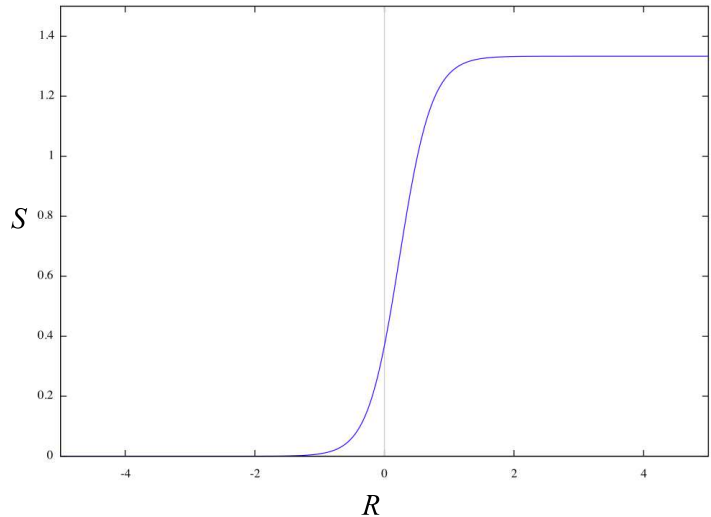}
\end{center}
\caption{The total action of Eq.~(\ref{mod03}) as a function of 
$R=R_{R}' =R_{L}'$ keeping $\theta=\pi$ with 
$l_{1}=1$, $l_{2}=100000$ fixed.}
\label{Snm-S}
\end{figure}

\begin{figure}[htbp]
\begin{center}
\includegraphics[width=0.85\textwidth]{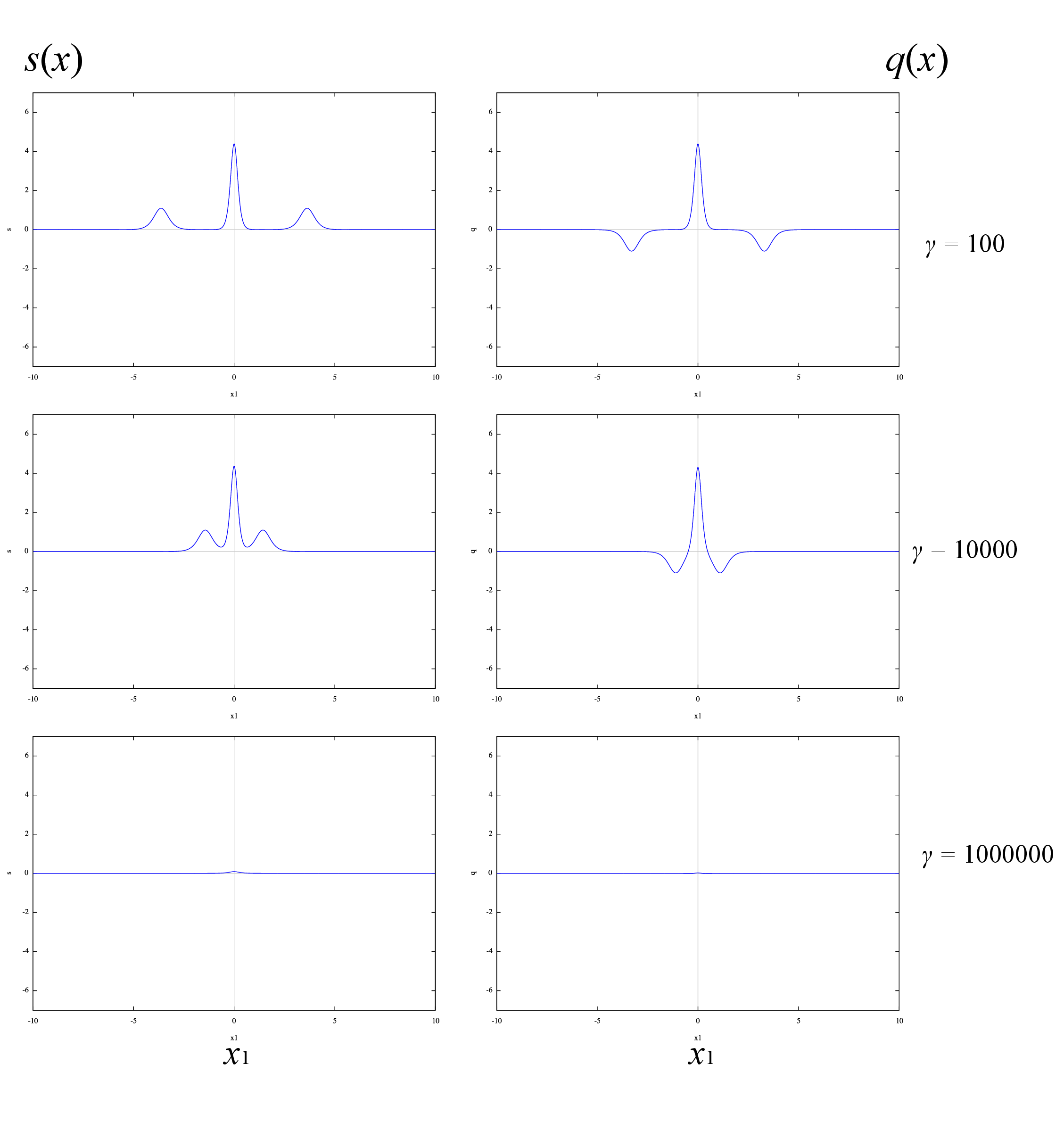}
\end{center}
\caption{
Action density (left) and topological charge density (right) 
of Eq.~(\ref{mod03}) 
for three values of parameter $\gamma=\gamma'=100,10000,1000000$ 
(from top to bottom) with $l_{1}=1$, $l_{2}=100000$, $\theta=\pi$ fixed.}
\label{Snm}
\end{figure}

For $\theta=\pi$, the three instanton constituents have 
effectively attractive interaction. 
Thus the total action decreases as one increases $\gamma=\gamma' >1$.
In Fig.~\ref{Snm2d-tpi}, we depict the total action as a function 
of $R_{R}'$ and $R_{L}'$, where we fix $l_{1}=1$ and $l_{2}=1000$, and vary $\gamma, \gamma'$.
In any direction of decreasing $R_{R}',R_{L}'$ (increasing $\gamma, \gamma'$) 
from the original configuration, the action decreases.
Figure \ref{Snm-S} shows the total action as a function 
of $R=R_{R}'=R_{L}'$, showing clearly that the total action 
decreases to $S=0$ towards $R=R_{R}'=R_{L}'\to -\infty$.
 (This corresponds to the curve obtained as a section along 
$R_{R}'=R_{L}'$ in Fig.~\ref{Snm2d-tpi} although we change the parameter value in the two figures.)
In Fig.~\ref{Snm}, corresponding to (\ref{mod03}) with 
$2l_{1}<l_{2}^{2}$,
we depict how the instanton constituents meet and 
are compressed when $\gamma=\gamma'$ increases.

\subsection{Number and directions of negative modes}

Before discussing the directions associated to the 
remaining parameters, we quantify the mass squared matrix of 
fluctuations around the non-BPS exact solution to find the 
number and directions of negative modes described by 
 $\gamma, \gamma', \theta$ in Eq.~(\ref{mod03}). 
As a natural coordinates of the parameter space, we take 
$R_{R}',R_{L}', \theta$, since they are associated to the flat 
metric for kinetic terms at least for large separations. 
For small separations, their physical meaning are admittedly 
less clear. 
The mass squared matrix is given by the $3\times 3$ matrix 
of second-derivatives. 
Diagonalization of the matrix gives the number of mass squared 
eigenvalues and the direction of eigenmodes.

We consider the three sets of the parameters in the non-BPS 
solution $l_{2}=100,\,10,\,1$ with $l_{1}=1$ fixed, 
which correspond to the cases that the separation between 
the double fractional instanton and the fractional 
anti-instantons at the both side are 
$R\sim2.20,\,1.10,\,0$.

For $R=2.20$ ($l_{2}=100,\,\,\,l_{1}=1$), the separation is 
relatively large. 
The second-derivative matrix at $\gamma=1,\gamma'=1,\theta=0$ 
is numerically calculated as 
\begin{equation}
{\partial^2 S\over{\partial (R_{R}', R_{L}', \theta)^{2}}} \approx
\begin{pmatrix}
-9.11 & 9.17 &  O(10^{-3}) \\
9.17 & -9.11 &  O(10^{-3})\\
O(10^{-3}) & O(10^{-3}) & -2.11
\end{pmatrix}
\times 10^{-4}\,,
\label{2dMat100}
\end{equation}
where the label $R_{R}', R_{L}', \theta$ stands 
for the ordering of rows and columns. 
By denoting the unit vectors as ${\bf e}_{R}, {\bf e}_{L}, 
{\bf e}_{\theta}$ respectively, we approximately obtain the eigenvalues 
and the eigenvectors as 
\begin{align}
\sim\,\,&6.00\times 10^{-6}:\,\,\,\, {\bf e}_{R} + {\bf e}_{L}  \\
\sim\,\,-&1.83\times 10^{-3}:\,\,\,\, {\bf e}_{R} - {\bf e}_{L} \\
\sim\,\,-&2.11 \times 10^{-4}:\,\,\,\,{\bf e}_{\theta}\,.
\end{align}
We find that there are two negative modes in 
${\bf e}_{R} - {\bf e}_{L}$ and ${\bf e}_{\theta}$ directions 
in this case.

For $R=1.10$ ($l_{2}=10,\,\,\,l_{1}=1$), the separation is 
relatively small and the constituents are about to crash.
The second-derivative matrix at $\gamma=1,\gamma'=1,\theta=0$ is numerically calculated as  
\begin{equation}
{\partial^2 S\over{\partial (R_{R}', R_{L}', \theta)^{2}}} \approx
\begin{pmatrix}
-0.0606 & 0.0852 & O(10^{-6}) \\
0.0852 & -0.0606 & O(10^{-6})\\
O(10^{-6}) & O(10^{-6}) & -0.0204
\end{pmatrix}\,.
\label{2dMat10}
\end{equation}
Then the eigenvalues and the eigenvectors are approximately given by
\begin{align}
\sim\,\,&0.0246:\,\,\,\, {\bf e}_{R} + {\bf e}_{L}  \\
\sim\,\,-&0.1458:\,\,\,\, {\bf e}_{R} - {\bf e}_{L} \\
\sim\,\,-&0.0204 :\,\,\,\,{\bf e}_{\theta}\,.
\end{align}
In this case too, we find that there are two negative modes in 
${\bf e}_{R} - {\bf e}_{L}$ and ${\bf e}_{\theta}$ directions.

For $R=0$ ($l_{2}=1,\,\,\,l_{1}=1$), the separation is zero 
and the constituents crash.
The second-derivative matrix at $\gamma=1,\gamma'=1,\theta=0$ is numerically calculated as  
\begin{equation}
{\partial^2 S\over{\partial (R_{R}', R_{L}', \theta)^{2}}} \approx
\begin{pmatrix}
0.664 & 0.639& O(10^{-5}) \\
0.639 & 0.664 & O(10^{-5})\\
O(10^{-5}) & O(10^{-5}) & -0.267
\end{pmatrix}\,.
\label{2dMat1}
\end{equation}
Then the eigenvalues and the eigenvectors are approximately given by
\begin{align}
\sim\,\,&1.303 :\,\,\,\, {\bf e}_{R} + {\bf e}_{L}  \\
\sim\,\,&0.025 :\,\,\,\, {\bf e}_{R} - {\bf e}_{L} \\
\sim\,\,-&0.267 :\,\,\,\,{\bf e}_{\theta}\,.
\end{align}
In this case, there is one negative mode in 
${\bf e}_{\theta}$ directions.
We should mention that coordinates $R_{R}'$ and $R_{L}'$ do 
not have a simple physical meaning as separations at 
small $R$ region such as $R=0$. 

These results show that the number of negative modes depends 
on the parameter region. 
However, the relative phase fluctuation always gives a negative 
mode.


\subsection{
Splitting of 
two bions 
}


To analyze 
the stability 
of the non-BPS exact solution, 
we still need to consider two more parameters : a separation 
parameter corresponding to the splitting of the middle compressed 
double fractional (anti-)instanton, and its associated phase. 
The splitting can be described by the following ansatz containing 
a new term with the parameter $\delta e^{i\theta'}$ in the 
second component 
\begin{equation}
\omega = 
\begin{pmatrix}
\frac{2l_{1}}{l_2}e^{-{2\pi\over{3}}z}
+l_{1}l_{2}e^{-{2\pi\over{3}}(2z+\bar{z})},\,\, &
1 +\delta e^{i\theta'}e^{-{2\pi\over{3}}(z+\bar{z})}
- l_{1}^{2}e^{-{4\pi\over{3}}(z+\bar{z})},\,\,&
-l_{2}e^{-{2\pi\over{3}}\bar{z}}
-\frac{2l_{1}^{2}}{l_2}e^{-{2\pi\over{3}}(z+2\bar{z})}
\end{pmatrix}\,.
\label{mod1}
\end{equation}
with $\delta\geq0,\,\,\,\,0\leq\theta'<2\pi$.
For $\delta\not=0$, the configuration is no longer a solution. 
In the limit of the exact solution ($\delta=0$), the phase 
$\theta'$ obviously is ill-defined and loses a physical meaning. 
Therefore we consider here only deformations due to these new 
parameters $\delta, \theta'$ and postpone the analysis combined 
with other parameters to the next subsection.

A nonzero $\delta$ splits the double fractional 
(anti-)instanton in the middle of Fig.~\ref{nonbpscp2}
into two fractional (anti-)instantons, and leads to the 
two bion configurations as shown in Fig.~\ref{G2bion}(b) 
and (c).

We first study the case of large separations between 
all consitituents, where the two-body forces \cite{Misumi:2014jua} 
in Eq.~(\ref{eq:interation_pot}) between fractional instanton 
and anti-instanton is applicable. 
Since two-body forces are attractive (repulsive) for the relative 
phase smaller (larger) than $\pi/2$, we find that the effective 
forces between the fractional constituents in both two bions 
in Fig.~\ref{G2bion}(b) and (c) become attractive for any values 
of $\delta$, only at $\theta'=\pi/2$. 
This is because some of the terms in (\ref{mod1}) have relative 
signs, in such a way that one of the constituent pairs has 
an attractive interaction while the other has a repulsive 
interaction for $\theta'\not=\pi/2$. 
One should note that two fractional (anti-)instantons experience 
no static force because they are mutually BPS.

For example, let us consider the parameter region $2l_{1}<l_{2}^{2}$, 
corresponding to Fig.~\ref{G2bion}(b). 
The fractional anti-instanton in the left end is located 
at $\frac{3}{2\pi}\log\frac{l_1}{l_2}$, whereas the left 
fractional instanton emerging from the middle double instanton 
is located at  $\frac{3}{2\pi}\log\frac{l_1l_2}{\delta}$. 
They have a relative phase $e^{i\pi}/e^{i\theta'}=e^{i(\pi-\theta')}$, 
and compose a left bion. 
Thus the constituents have an attractive force for 
$|\theta'|>\pi/2$, and repulsive for 
$|\theta'|<\pi/2$. 
Similarly, the fractional instanton and anti-instanton in the right bion have 
a relative phase $e^{i\theta'}$, and exhibit an attractive 
force for $|\theta'|<\pi/2$, and repulsive for $|\theta'|>\pi/2$.
If $|\theta'|=\pi/2$ and the separation is small, 
which is beyond the scope of the two-body force approximation,
the strong attractive force between the constituents of bions emerges as shown in Fig.~\ref{bionS}. 
Therefore attractive forces are the strongest 
around $\theta'=\pm\pi/2$ since constituents of both 
bions are attractive for small separation only at these parameters.

We note that, for $2l_{1}>l_{2}^{2}$, the situation is similar.

\begin{figure}[htbp]
\begin{center}
\includegraphics[width=0.48\textwidth]{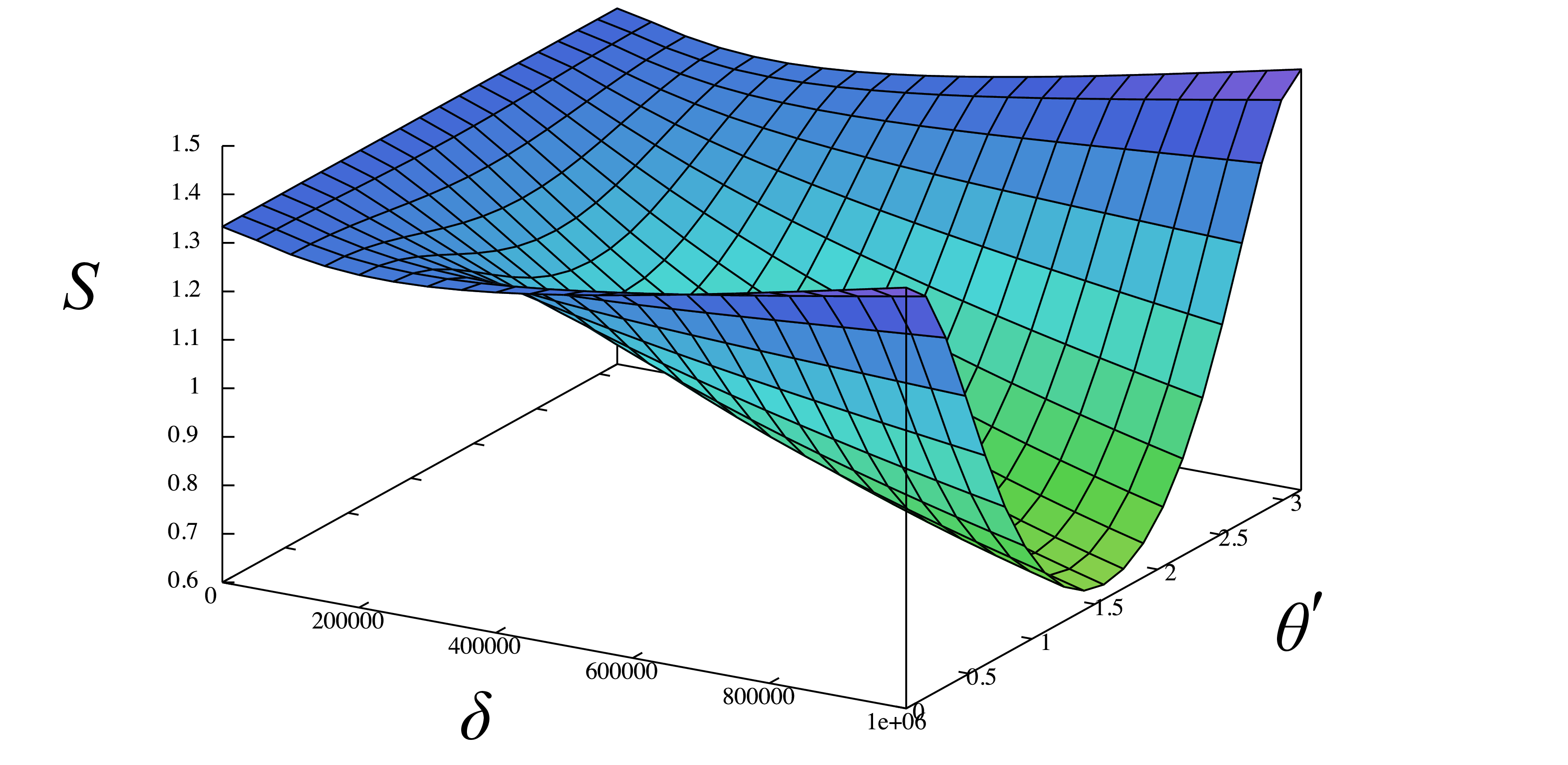}
\includegraphics[width=0.48\textwidth]{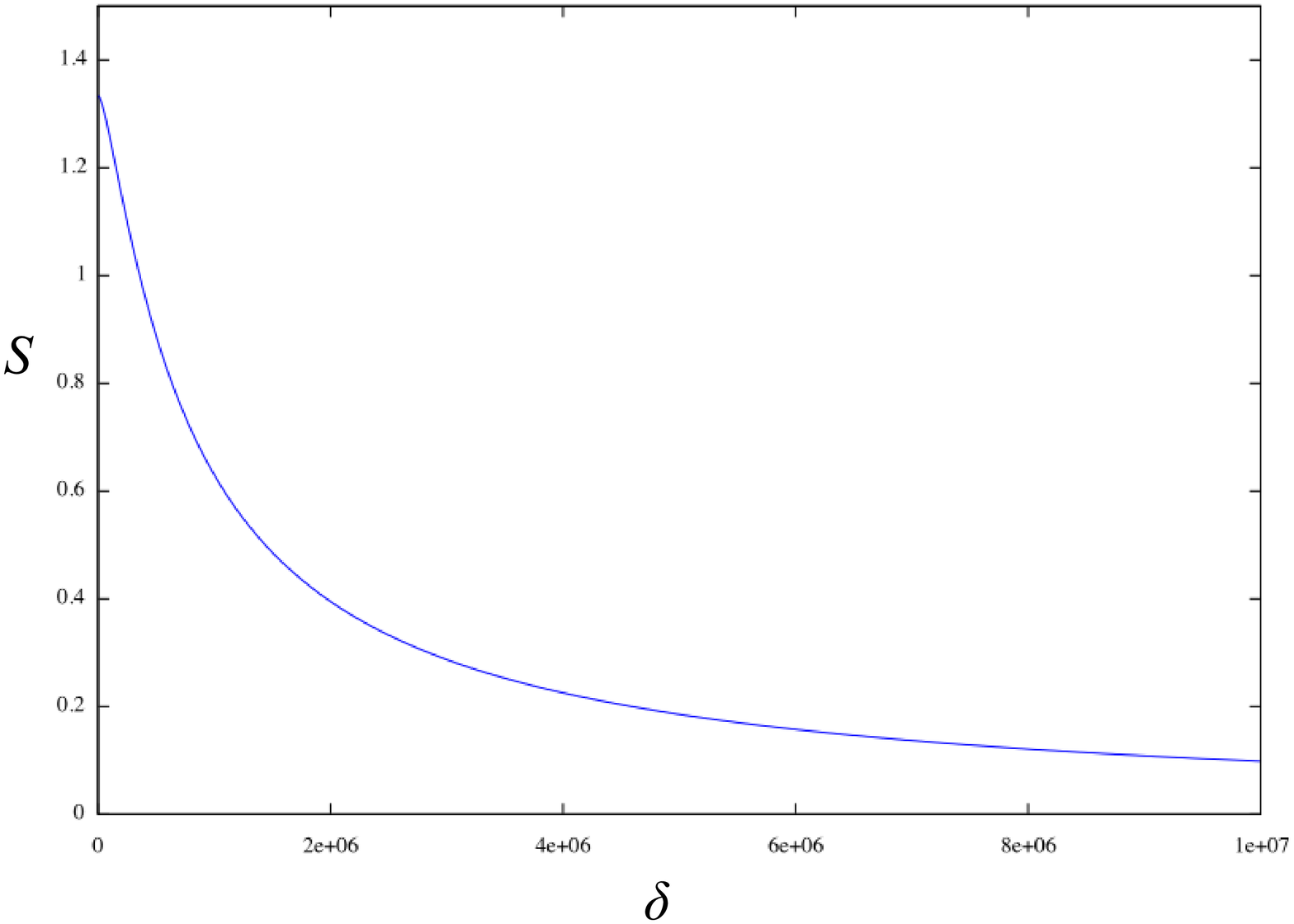}
\end{center}
\caption{The total action as a function of $\delta$ and $\theta'$ 
in the modified configuration (\ref{mod1}) for $l_{1}=1, l_{2}=100$ (left). 
The total action as a function of $\delta$ with $\theta'=\pi/2$ fixed 
is also depicted (right).} 
\label{3d-action}
\end{figure}

\begin{figure}[htbp]
\begin{center}
\includegraphics[width=0.48\textwidth]{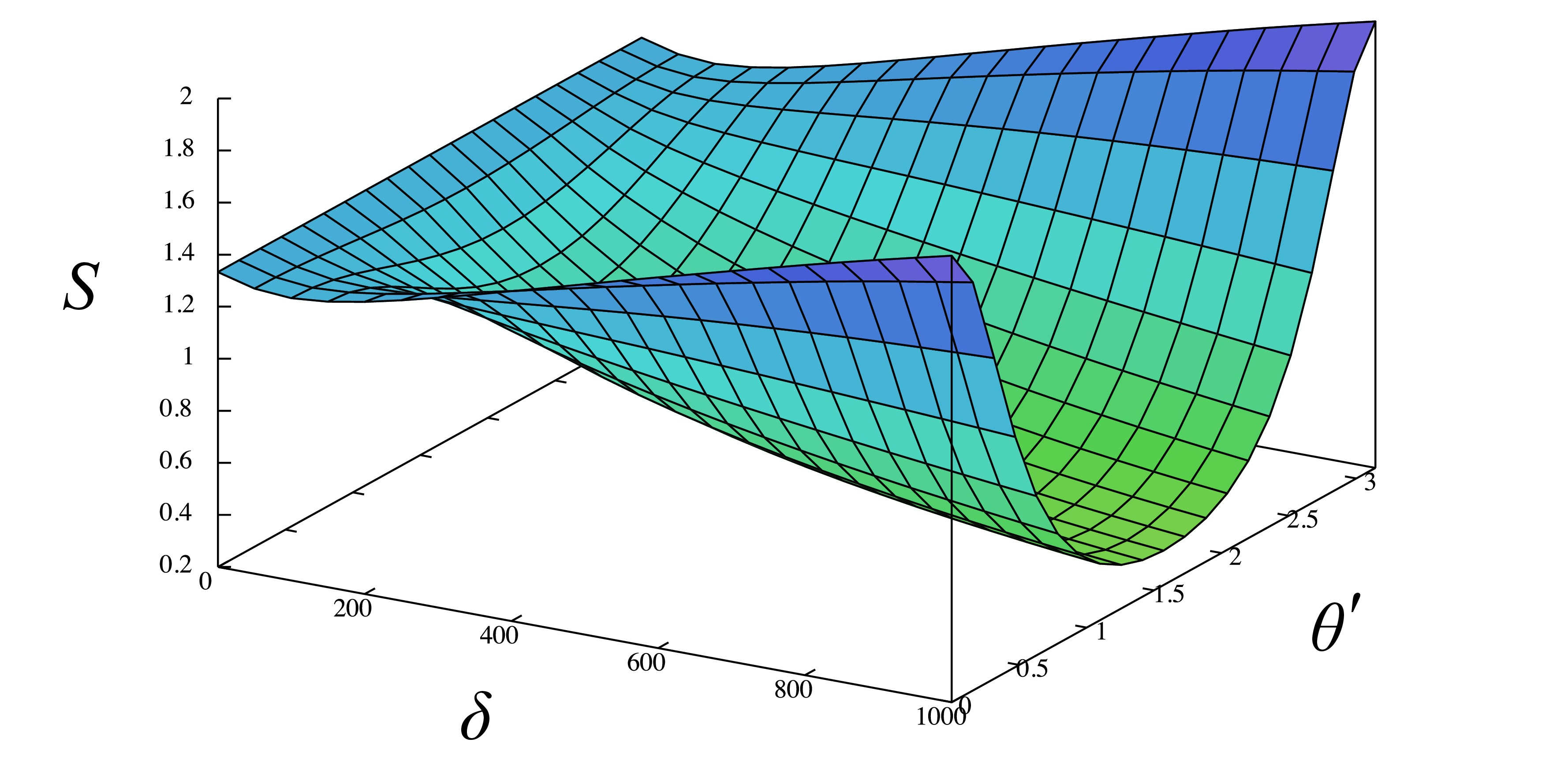}
\includegraphics[width=0.48\textwidth]{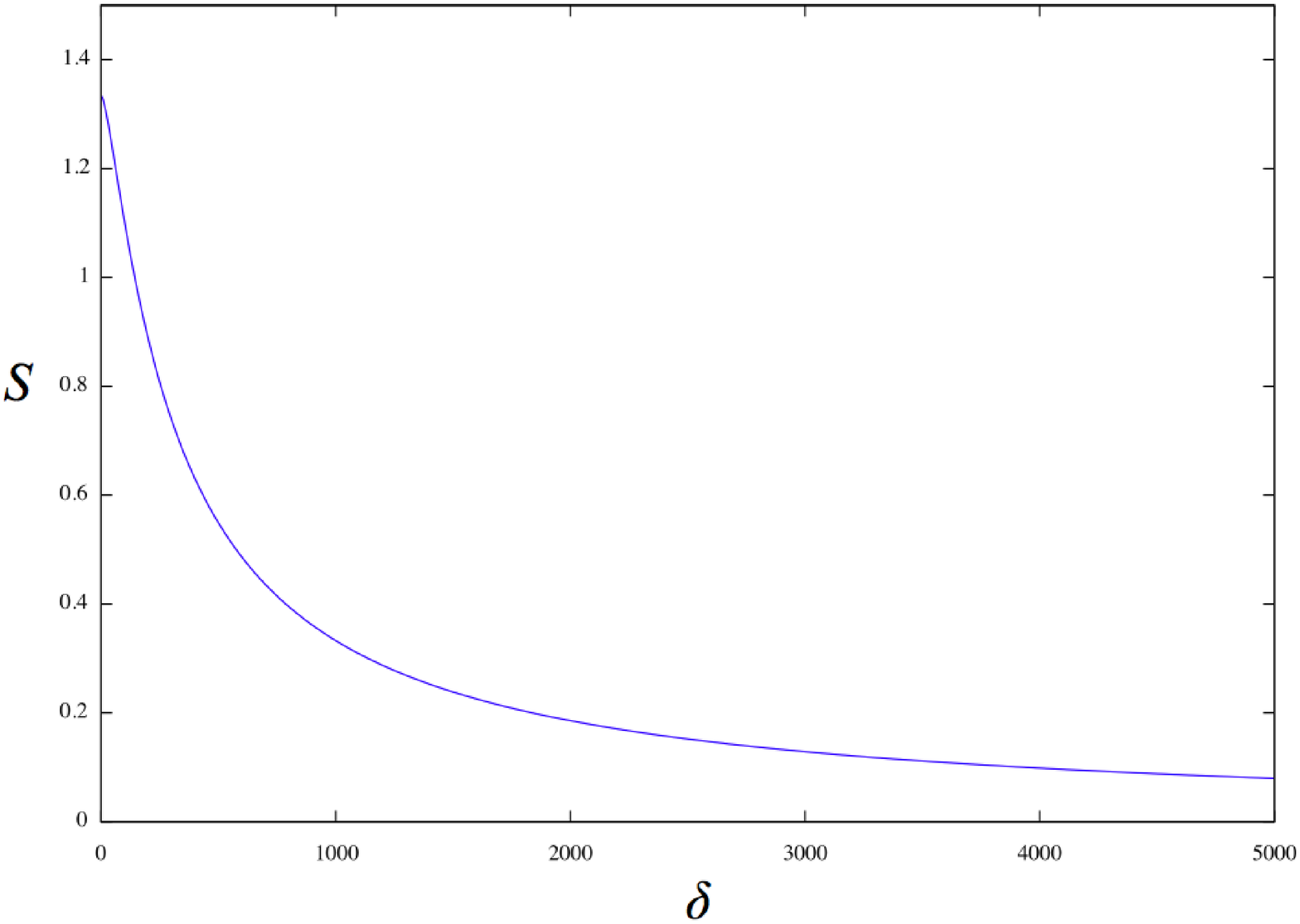}
\end{center}
\caption{The total action as a function of $\delta$ and $\theta'$ 
in the modified configuration (\ref{mod1}) for $l_{1}=1, l_{2}=1/100$ (left). 
The total action as a function of $\delta$ with $\theta'=\pi/2$ 
fixed is also depicted (right).} 
\label{3d-action2}
\end{figure}

In Fig~\ref{3d-action} and Fig.~\ref{3d-action2}, 
the total action is depicted as a function of 
 $\delta$ and $\theta'$ for fixed $l_{1}$ and $l_{2}$.
One can see that, for $\theta'=0,\pi$, the total action decreases 
at first as $\delta$ gets nonzero, 
then at some point it takes a turn and increases.
It means that, for these values of $\theta'$, the effective 
force between the fractional instanton and anti-instanton is 
changed from attractive to repulsive ones at some value of $\delta$.
Here we used the parameter $\delta$ instead of $R$ to describe 
the length of the middle vacuum region between the two fractional 
instantons, since the metric for the two BPS solitons is 
typically cigar-like and $\delta$ is more appropriate at 
small separations \cite{Tong:2002hi}. 

\begin{figure}[htbp]
\begin{center}
\includegraphics[width=0.32\textwidth]{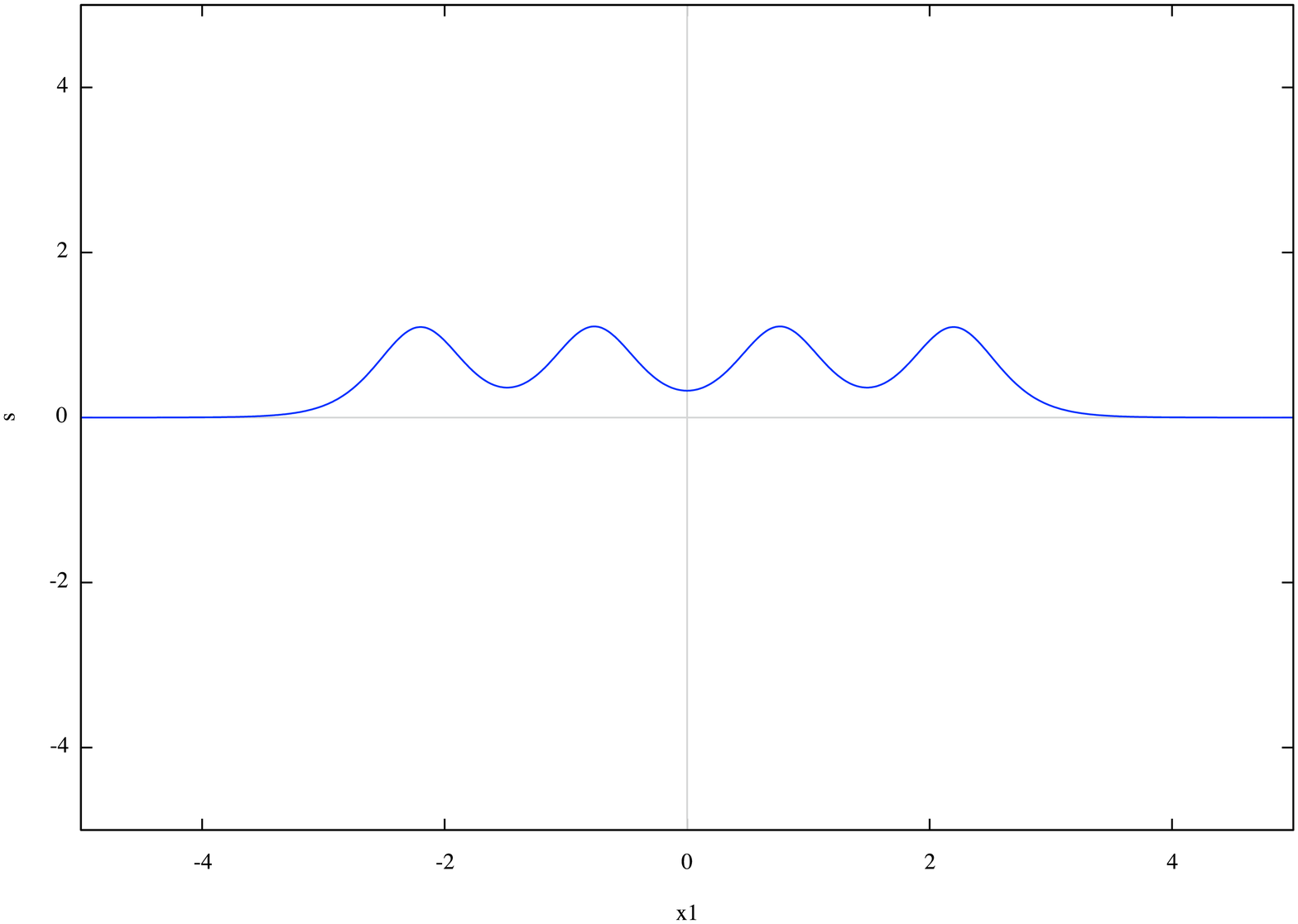}
\includegraphics[width=0.32\textwidth]{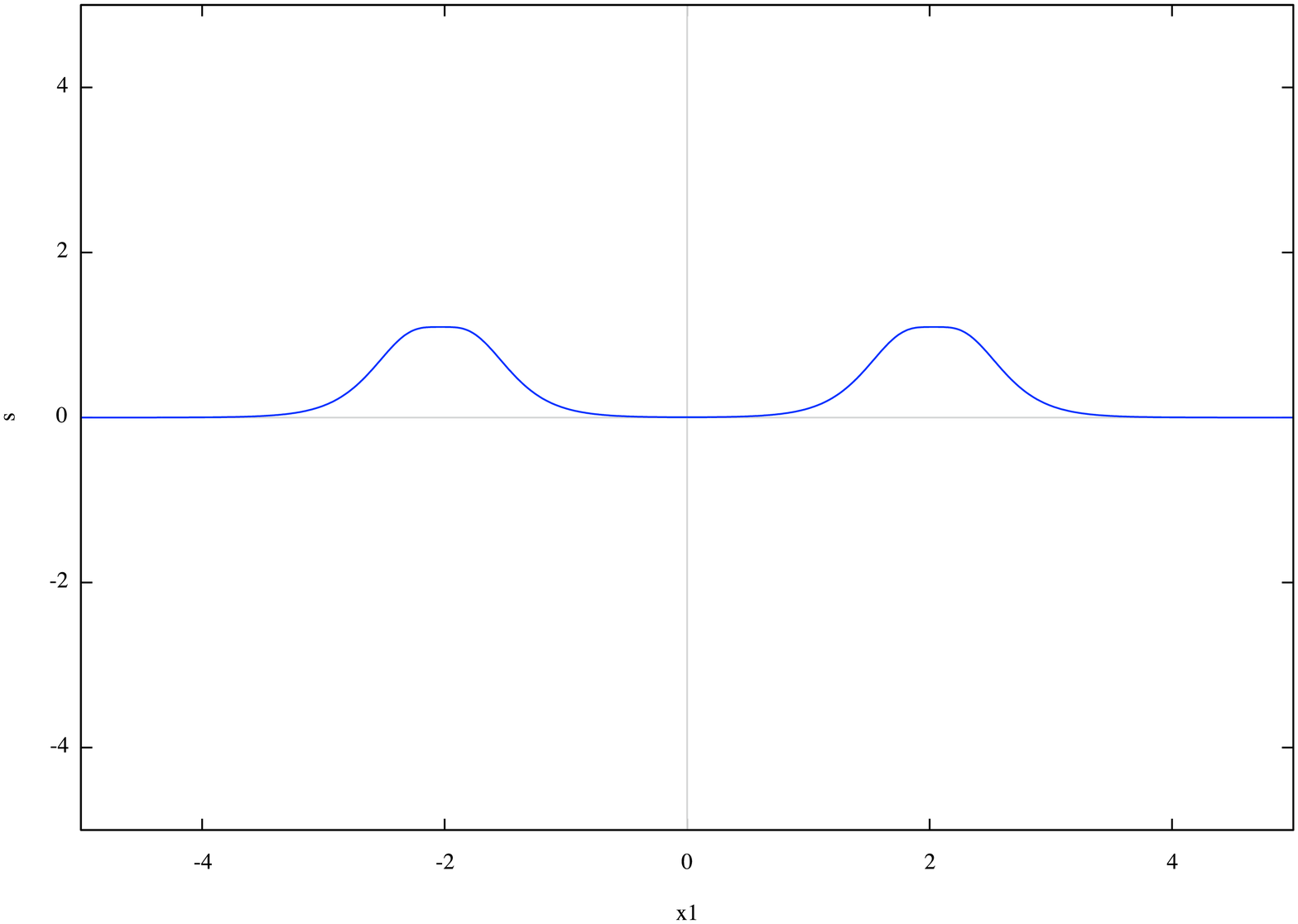}
\includegraphics[width=0.32\textwidth]{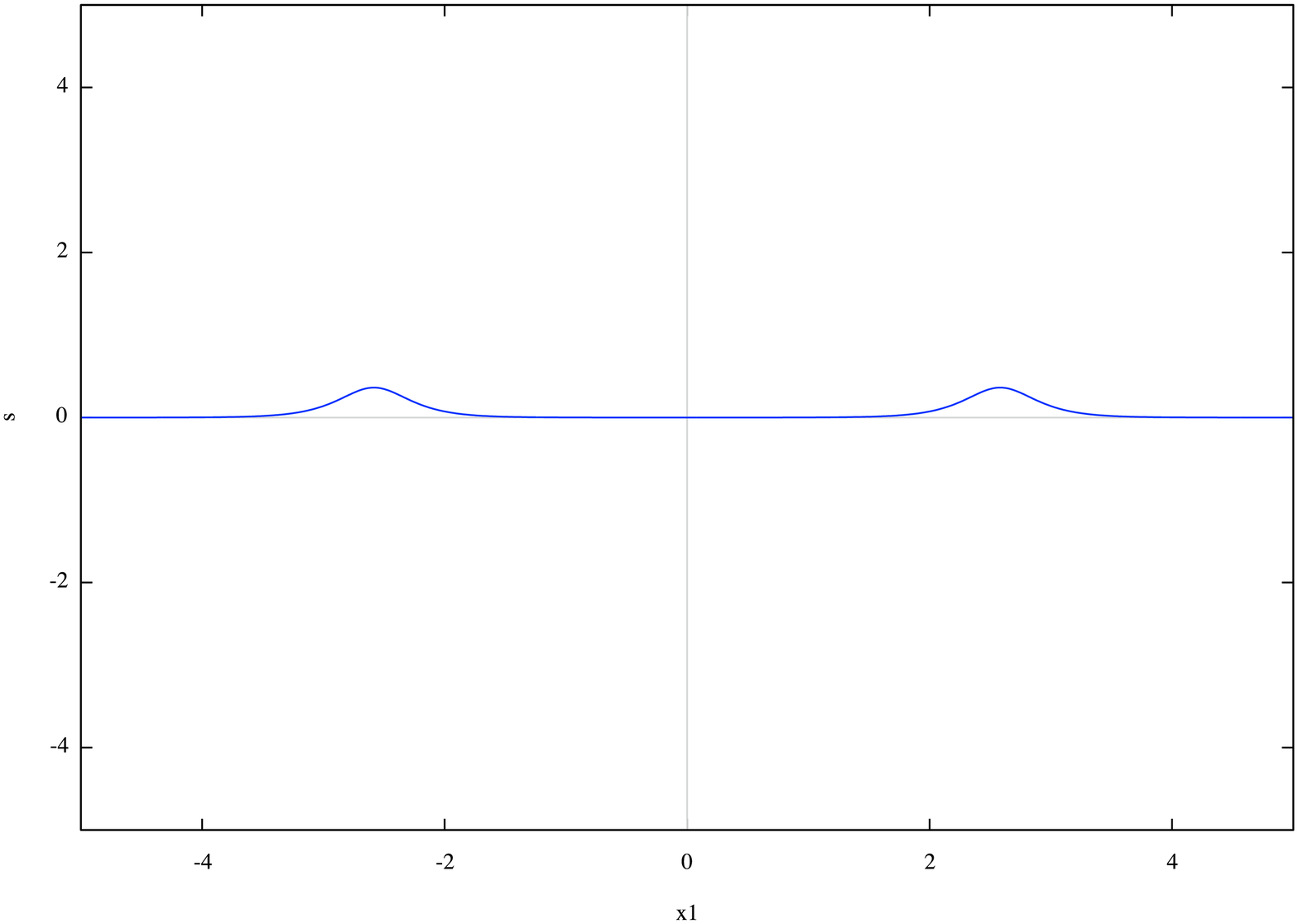}
\end{center}
\caption{For Eq.~\ref{mod1}, action densities for $\delta=5\times10^4$(left), $5\times10^5$(middle), $5\times 10^6$(right) with $\theta'=\pi/2$ fixed are depicted for $l_{1}=1, l_{2}=100$ . }
\label{s-pi_2}
\end{figure}

Our observation shows that, for example, if we fix $\theta'=\pi/2$, 
the parameter $\delta$ is identified as one of the unstable modes, 
which connects the non-BPS solution and the two bion configuration. 
We show how the action density changes with $\delta$ for 
$\theta=\pi/2$ in Fig.~\ref{s-pi_2}. 
In both left and right bions, the fractional constituents 
gradually annihilate into vacuum as they approach each other.

\subsection{Combined effects of deformation parameters 
}

The final step of analyzing the (global) stability of the non-BPS 
exact solution is to consider the five relevant deformation 
parameters at the same time. 
As a most significant effect, we consider the case where 
the strong phase dependence of two-body forces become most 
visible. 
Let us consider the following modification of the non-BPS 
exact solution 
\begin{equation}
\omega = 
\begin{pmatrix}
\frac{2l_{1}}{l_2}e^{-{2\pi\over{3}}z}
+l_{1}l_{2}e^{-{2\pi\over{3}}(2z+\bar{z})},\,\, &
e^{i\theta} -\delta e^{-{2\pi\over{3}}(z+\bar{z})}
- l_{1}^{2}e^{-{4\pi\over{3}}(z+\bar{z})},\,\,&
-l_{2}e^{-{2\pi\over{3}}\bar{z}}
-\frac{2l_{1}^{2}}{l_2}e^{-{2\pi\over{3}}(z+2\bar{z})}
\end{pmatrix}\,,
\label{mod2}
\end{equation}
with $\delta\geq0,\,\,\,\,0\leq\theta<2\pi$. 
We have chosen the phase of the 
the second term of the second component as 
$ -\delta e^{-{2\pi\over{3}}(z+\bar{z})}$, so that 
the strong phase dependence of two-body forces 
in Eq.~(\ref{eq:interation_pot}) of both bions can collaborate 
together to act as attractive forces when $\theta=\pi$.

\begin{figure}[htbp]
\begin{center}
\includegraphics[width=0.6\textwidth]{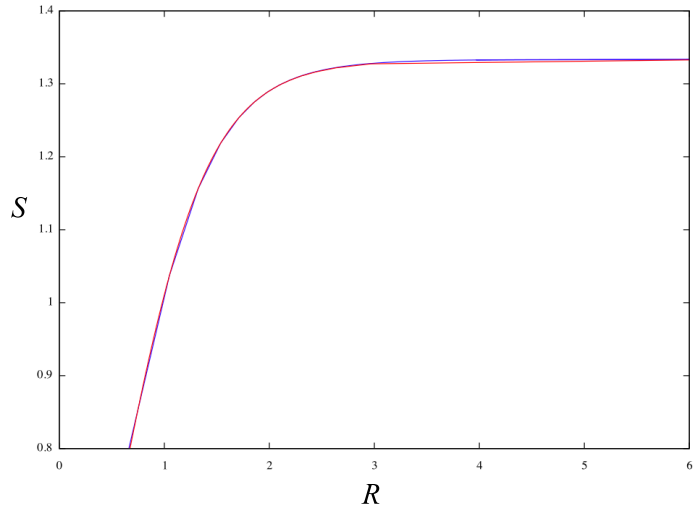}
\end{center}
\caption{The total action (red) as a function of the symmetric 
separation $R={3\over{2\pi}}\log{l_{2}^{3}\over\delta}$ 
of the bions at both right and left sides Eq.~(\ref{mod2}). 
The results with $l_{1}=1, l_{2}=100$, 
$\theta=\pi$ (red) is compared with the sum of total actions of 
left and right bions (blue).}
\label{2-nega}
\end{figure}

We depict the total action as a function of the symmetric 
separation $R={3\over{2\pi}}\log{l_{2}^{3}\over\delta}$ 
of the bions at both right and left sides 
with $\theta=\pi$ fixed in Fig.~\ref{2-nega}, and 
compare it to the sum of the total action of each individual 
left and right bion. 
Up to numerical errors, these two plots are consistent.
It indicates that the configuration (\ref{mod2}) is seen as almost two-bion
configurations, and there is no interaction between the two bions.

We note that the bion parts have the attractive interaction.
It means that, if we calculate their amplitudes by integrating moduli integrals, we find the imaginary ambiguities, which will be cancelled out combined with the perturbative calculations around the appropriate background.
Again, we conclude that the non-BPS solutions can be seen as a special limit of the configurations relevant to the resurgence theory.


\section{Non-BPS solutions in Grassmann sigma models}
\label{sec:Gr}

In this section, we briefly discussextensions to the Grassmann 
sigma models 
$Gr_{N_{F},N_{C}} \simeq {SU(N)/SU(N_F-N_C)\times SU(N_C) \times U(1)}$ 
with twisted boundary conditions.
For this case, we have several ways of constructing non-BPS 
solutions \cite{Din:1981bx, Din:1983fj}.
We introduce two simple classes of non-BPS solutions in the 
Grassmann sigma models.

\subsection{Simple Extension of ${\mathbb C}P^{N-1}$ Projection}

We introduce one simple class of non-BPS solutions in the Grassmann model \cite{Din:1981bx}. 
which is given by the following projection,
\begin{equation}
H_{0}^{\rm nonBPS}=
\begin{pmatrix}
Z_{+}^{k_{1}}\omega \\
Z_{+}^{k_{2}}\omega \\
\cdot \\
\cdot \\
Z_{+}^{k_{N_{c}}}\omega
\end{pmatrix}\,,
\hspace{2cm}
\omega\equiv\omega(z)\,,
\end{equation}
with
\begin{equation}
Z_{+}:\,\,\omega\,\to\, Z_{+}\omega\equiv \partial_{z}\omega-{(\partial_{z}\omega)\omega^{\dag}\over{\omega\omega^{\dag}}}\omega\,,
\end{equation}
and 
\begin{equation}
0\leq k_{1} < k_{2}<\cdot\cdot\cdot<k_{N_{c}}\leq N_{F}-1\,.
\end{equation}
$\omega(z)$ is a arbitrary holomorphic $N_{F}$ vector.
This projection makes $N_{c}$ $N_{F}$-vectors orthogonal.
$H_{0}$ is a moduli matrix, which is related to the physical scalar $H$ 
by $H=(H_{0}H_{0}^{\dag})^{-1/2} H_{0}$ for this case
since $H_{0}H_{0}^{\dag}$ is diagonal.

\begin{figure}[htbp]
\begin{center}
\includegraphics[width=0.4\textwidth]{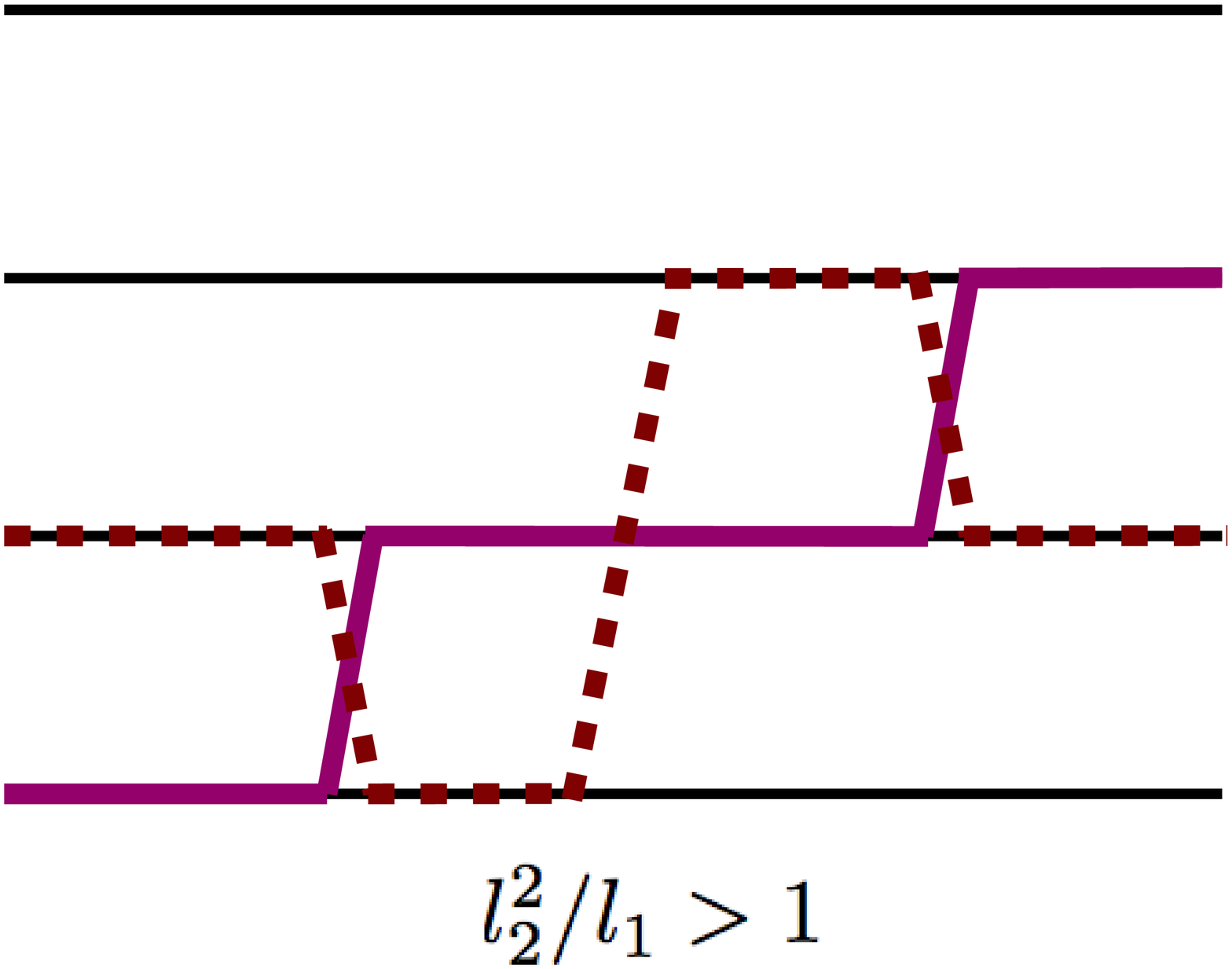}
\includegraphics[width=0.4\textwidth]{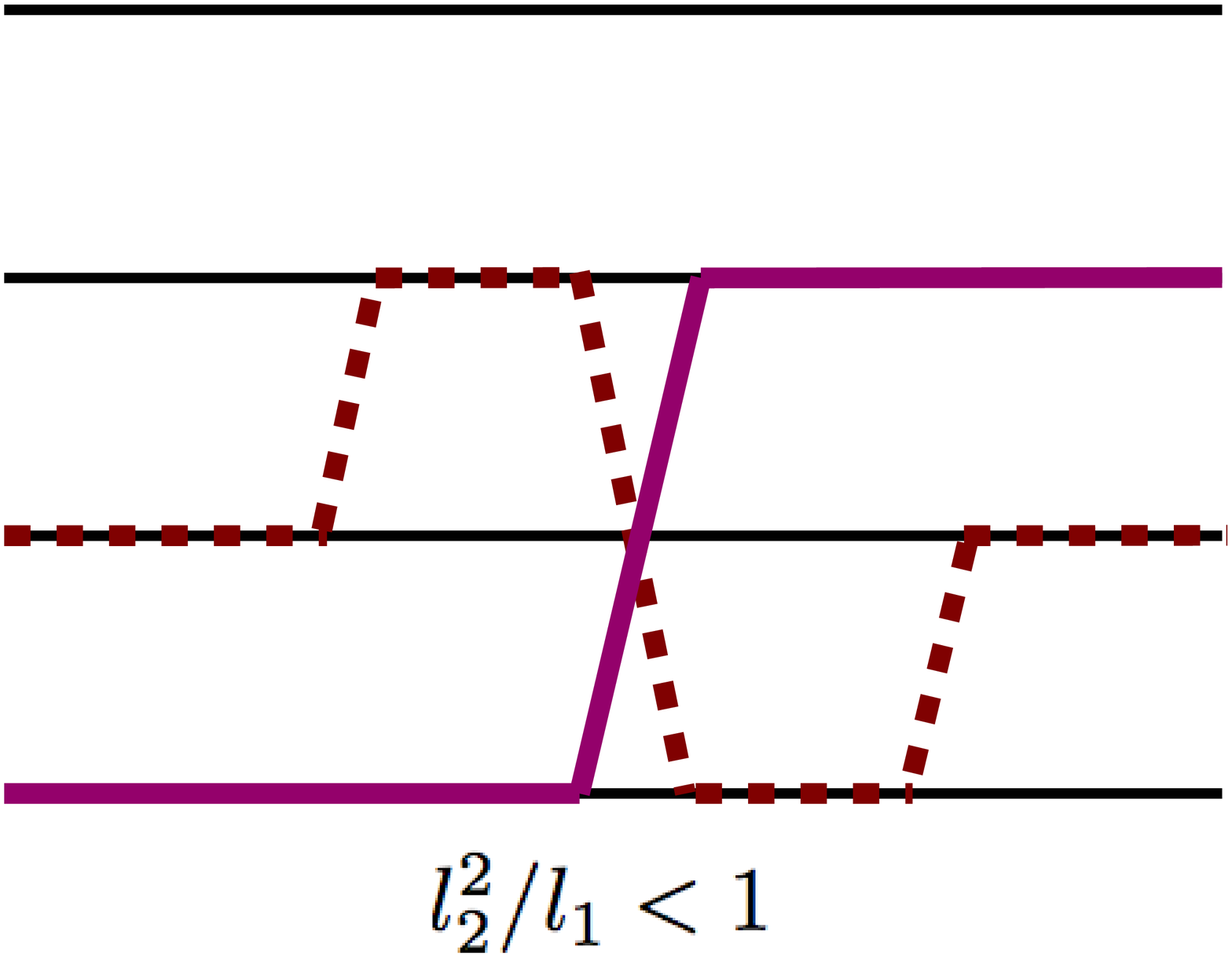}
\end{center}
\caption{The flipping partner configurations in the Non-BPS 
exact solution in $Gr_{4,2}$ model.
Two different colored lines stand for two colors.}
\label{42}
\end{figure}

We apply this projection to $Gr_{4,2}$.
We consider the holomorphic vector
\begin{equation}
\omega = 
\begin{pmatrix}
l_{1}e^{i\theta_{1}}e^{-\pi z},\,\,&
l_{2}e^{i\theta_{2}}e^{-{\pi\over{2}}z},\,\,&
1,\, \, &
0
\end{pmatrix}\,.
\end{equation}
It is nothing but the BPS solution of the ${\mathbb C}P^{3}$ model with $S=1/2,\,Q=1/2$.
For example, we apply the projection with $k_{1}=0, k_{2}=1$ on the
holomorphic vector $\omega$,
then one of non-BPS solutions in $Gr_{4,2}$ is given by 
\begin{align}
\small
H_{0}^{\rm nonBPS}\,=\,
\begin{pmatrix}
l_{1}e^{-\pi z}&
l_{2}e^{-{\pi\over{2}}z}&
1&
 0\\
 l_{1}l_{2}^{2}e^{-{\pi\over{2}}(2z+\bar{z})}+2l_{1}e^{-{\pi\over{2}}z}\,\, &
-l_{1}^{2}l_{2}e^{-\pi(z+\bar{z})}+l_{2}\,\,&
-2l_{1}^{2}e^{-{\pi\over{2}}(z+2\bar{z})}-l_{2}^{2}e^{-{\pi\over{2}}\bar{z}}&
0
\end{pmatrix}\,.
\end{align}
We here drop the phase variables $\theta_{1}, \theta_{2}$ since they do not contribute to
the action and topological charge densities.
This solution is shown in Fig.~\ref{42}.
The total action of this solution is $S=3/2$ while the total topological charge is $Q=1/2$.
It is notable that positions of constituent instantons
in different color lines coincide in the configurations of the solution. 

When we apply the projection with $k_{1}=1, k_{2}=2$ on the
same holomorphic vector $\omega$,
we obtain a non-BPS solutions in $Gr_{4,2}$
\begin{align}
\small
H_{0}^{\rm nonBPS}\,=\,
\begin{pmatrix}
 l_{1}l_{2}^{2}e^{-{\pi\over{2}}(2z+\bar{z})}+2l_{1}e^{-{\pi\over{2}}z}\,\, &
-l_{1}^{2}l_{2}e^{-\pi(z+\bar{z})}+l_{2}\,\,&
-2l_{1}^{2}e^{-{\pi\over{2}}(z+2\bar{z})}-l_{2}^{2}e^{-{\pi\over{2}}\bar{z}}&
0\\
l_{2}&
2l_{1}e^{-{\pi\over{2}}\bar{z}}&
-l_{1}l_{2}e^{-\pi \bar{z}}&
 0\\
\end{pmatrix}\,.
\end{align}
The total action of this solution is $S=3/2$ while the total topological charge is $Q=-1/2$.
This solution also has the transition between two 
seemingly distinct configurations in a similar way to the case of Fig.~\ref{42}.

\subsection{Projection Operation on Physical Scalars}

As another class of non-BPS solutions, we introduce the class derived from
the following projection on the physical scalar $H^{\rm b}$ for a BPS solution \cite{Din:1981bx},
\begin{equation}
Z_{+}:\,\,H^{\rm b}\,\to\, Z_{+}H^{\rm b}\equiv \partial_{z}H^{\rm b}
-(\partial_{z}H^{\rm b})(H^{\rm b})^{\dag}\left(H^{\rm b}(H^{\rm b})^\dag\right)^{-1} H^{\rm b}\,.
\end{equation} 
Then, the scalar field $H^{\rm nonBPS}$ of the non-BPS solution from this projection is given as 
$H^{\rm nonBPS}=(Z_{+}H^{\rm b}(Z_{+}H^{\rm b})^{\dag})^{-1/2} Z_{+}H^{\rm b}$.

\begin{figure}[htbp]
\begin{center}
\includegraphics[width=0.4\textwidth]{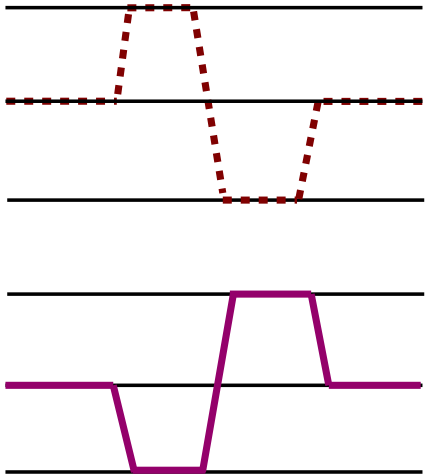}
\end{center}
\caption{A Non-BPS 
exact solution in $Gr_{6,2}$ model for $l_1\ll l_2^2, l_3\gg l_4^2$.}
\label{62}
\end{figure}

We here only concentrate on the simplest case, where the lines of different colors
do not run on the same flavor vacua.
We apply this projection to a BPS solution in $Gr_{6,2}$
with the moduli matrix
\begin{equation}
H_{0}^{\rm b}\,=
\begin{pmatrix}
l_{1}e^{i\theta_{1}}e^{-{2\pi\over{3}}z}&
l_{2}e^{i\theta_{2}}e^{-{\pi\over{3}}z}&
1 &
0 &
0 &
0\\
0&
0&
0&
l_{3}e^{i\theta_{3}}e^{-{2\pi\over{3}}z}&
l_{4}e^{i\theta_{4}}e^{-{\pi\over{3}}z}&
1
\end{pmatrix}
\end{equation}
with $S=2/3,\,\,Q=2/3$.
The physical scalar $H^{\rm b}$ is derived as 
$H^{\rm b}=(H_{0}^{\rm b}(H_{0}^{\rm b})^{\dag})^{-1/2} H_{0}^{\rm b}$ for this case.
Then, the non-BPS solution derived from the above projection is 
\begin{align}
H^{\rm nonBPS}\,=\, 
\begin{pmatrix}
 & 
 {\omega^{(1)}\over |\omega^{(1)}|} &
 &
 0&
 0&
 0\\
 0&
 0&
 0&
 &
 {\omega^{(2)}\over |\omega^{(2)}|} &
 &
\end{pmatrix}\,,
\end{align}
with
\begin{equation}
\omega^{(1)} = 
\begin{pmatrix}
l_{1}l_{2}^{2}e^{-{\pi\over{3}}(2z+\bar{z})}
+2l_{1}e^{-{\pi\over{3}}z},\,\, &
-l_{1}^{2}l_{2}e^{-{2\pi\over{3}}(z+\bar{z})}+l_{2},\,\,&
-2l_{1}^{2}e^{-{\pi\over{3}}(z+2\bar{z})}
-l_{2}^{2}e^{-{\pi\over{3}}\bar{z}}
\end{pmatrix}\,,
\end{equation}
and 
\begin{equation}
\omega^{(2)} = 
\begin{pmatrix}
l_{3}l_{4}^{2}e^{-{\pi\over{3}}(2z+\bar{z})}
+2l_{3}e^{-{\pi\over{3}}z},\,\, &
-l_{3}^{2}l_{4}e^{-{2\pi\over{3}}(z+\bar{z})}+l_{4},\,\,&
-2l_{3}^{2}e^{-{\pi\over{3}}(z+2\bar{z})}
-l_{4}^{2}e^{-{\pi\over{3}}\bar{z}}
\end{pmatrix}\,.
\end{equation}
This solution is shown in Fig.~\ref{62} for $l_1\ll l_2^2, l_3\gg l_4^2$.
The total action of this solution is $S=4/3$ while the total topological charge is $Q=0$.

More generic forms of non-BPS solutions in Grassmann sigma model
are discussed in \cite{Din:1983fj}.
We will discuss the properties of these non-BPS solutions and their relation to bions in
Grassmann sigma model in the future works.


\section{Summary and Discussion}
\label{sec:SD}

We can summarize the present work as follows. 
Firstly, we have studied the non-BPS 
exact solutions in terms of ansatz for arbitrarily many 
fractional instantons, which gives solutions of field equations 
for asymptotically large separations of constituent fractional 
instantons. 
Since the ansatz serves as a basis to obtain all the 
multi-instanton contributions as integrals over the (quasi-)moduli, 
we find that the non-BPS exact solutions are also included as 
a part of multi-instanton contributions, which play an 
important role in the resurgence theory. 
Secondly, we have studied the balance of forces that assures 
the non-BPS configuration to be an exact solution. 
The interaction between a fractional instanton and a fractional anti-instanton 
is attractive (repulsive) depending on the value of the relative 
phase modulus $\theta$. 
From this strong dependence of two-body forces, 
we found three essential properties of non-BPS exact solutions: 
(i) the appropriate relative sign in two terms of the same component, 
(ii) the reflection symmetry around the middle compressed double 
fractional instanton, and (iii) the transition between two 
seemingly different configurations called flipping partners 
as moduli parameters vary.
Thirdly, we have found a generic pattern of flipping partners 
that arises in the non-BPS exact solutions, and conjecture a 
diagrammatic procedure to find possible configurations in 
various other exact solutions. 
Fourthly, we have analyzed the local and global instabilities 
of the non-BPS exact solution in terms of the ansatz, and found 
the physical meaning of deformations that give negative modes and 
flow to other configurations of possible global saddle points. 
We  have discussed the simple cases of non-BPS solutions 
in Grassmann sigma models based on Din-Zakrzewski projection method.
The detailed study of properties of solutions will be a subject 
of future works.


The $U(N)$ Yang-Mills theory 
coupled with 
$N$  Higgs fields in the fundamental representation 
admits a non-Abelian vortex solution, 
around which 
${\mathbb C}P^{N-1}$ zero modes are localized 
\cite{Hanany:2003hp,Auzzi:2003fs,Eto:2005yh}. 
BPS lump or instantons in the ${\mathbb C}P^{N-1}$ model 
on the vortex represent Yang-Mills instantons in the bulk point of view
\cite{Eto:2004rz}. 
This correspondence was established in the BPS case.
The non-BPS solutions in the ${\mathbb C}P^{N-1}$ model 
that we discussed 
should correspond to  non-BPS solutions in the Yang-Mills theory.
We can consider un-Higgsing by taking the decoupling limit of 
the Higgs fields. Then 
we may be able to obtain non-BPS multi-fractional-instanton 
or bion solutions in the Yang-Mills theory, 
that may shed light on 
the resurgence in the Yang-Mills theory.


\begin{acknowledgments}
We are grateful to Gerald Dunne for let the authors pay attention to
the present subjects and thank the organizers for giving us a chance 
to discuss with him in ``KEK Theory Workshop December 2015". 
This work is supported in part 
by  the Japan Society for the 
Promotion of Science (JSPS) 
Grant-in-Aid for Scientific Research
(KAKENHI) Grant Numbers 
(16K17677 (T.\ M.),  25400268 (M.\ N.), 16H03984 (M.\ N.) and 
25400241 (N.\ S.)).
The work of M.N. is also supported in part 
by a Grant-in-Aid for Scientific Research on Innovative Areas
``Topological Materials Science"
(KAKENHI Grant No. 15H05855) and 
``Nuclear Matter in neutron Stars investigated by experiments and
astronomical observations"
(KAKENHI Grant No. 15H00841) 
from the Ministry of Education, Culture, Sports, Science,
and Technology (MEXT) of Japan.
This work is also supported by
MEXT-Supported Program for the Strategic Research Foundation
at Private Universities ``Topological Science" (Grant No. S1511006).
\end{acknowledgments}



\end{document}